\documentclass[twocolumn,apj,numberedappendix,iop]{openjournal}

\usepackage[T1]{fontenc}

\usepackage{graphicx}	
\usepackage{amsmath}	
\usepackage{newtxtext,newtxmath}    
\usepackage[dvipsnames]{xcolor}  
\usepackage[colorlinks,linkcolor=blue,citecolor=blue,urlcolor=blue ]{hyperref}
\usepackage{amssymb}	
\usepackage{dsfont}
\usepackage{float}
\usepackage{ulem}
\usepackage{soul}
\usepackage{enumitem}
\usepackage{lineno}

\usepackage{orcidlink}

\usepackage{natbib}

\usepackage{fontawesome}

\usepackage[export]{adjustbox}

\usepackage{arydshln}
\usepackage{macros_vdb} 

\begin{document}


\title{BASILISK IV. No $S_8$ Tension with Satellite Kinematics}

\shorttitle{BASILISK IV. No $S_8$ Tension with Satellite Kinematics}
\shortauthors{Mitra et al.}

\author{Kaustav Mitra$^1$\orcidlink{0000-0001-8073-4554}}
\author{Frank C. van den Bosch$^1$\orcidlink{0000-0003-3236-2068}}   
\author{Josephine Baggen$^1$\orcidlink{0009-0005-2295-7246}}
\author{Johannes U. Lange$^2$\orcidlink{0000-0002-2450-1366}}

\affiliation{$^1$Department of Astronomy, Yale University, PO. Box 208101, New Haven, CT 06520-8101}
\affiliation{$^2$Department of Physics, American University, 4400 Massachusetts Avenue NW, Washington, DC 20016}

\email{kaustav.mitra@yale.edu}

\label{firstpage}


\begin{abstract} 
  We develop a novel technique to probe the $S_8$ tension, using information from the smallest scales of galaxy redshift survey data. Specifically, we use \Basilisc, a Bayesian hierarchical tool for forward modeling the kinematics and abundance of satellite galaxies extracted from spectroscopic data, to first constrain the galaxy-halo connection precisely and accurately. We then demand self-consistency in that the galaxy-halo connection predicts the correct galaxy luminosity function, which constrains the halo mass function and thereby cosmology. Crucially, the method accounts for baryonic effects and is free of halo assembly bias issues. We validate the method against realistic SDSS-like mock data, demonstrating unbiased recovery of the input cosmology. Applying it to the SDSS-DR7, we infer that $\Omega_{\rmm} = 0.324 \pm 0.012$, $\sigma_8 = 0.775 \pm 0.063$ and $S_8 \equiv \sigma_8 \sqrt{\Omega_{\rmm}/0.3}  = 0.81 \pm 0.05$, in perfect agreement with the cosmic microwave background constraints from \Planck. The most stringent constraint is with regard to the parameter combination $\sigma_8 (\Omega_\rmm/0.3)^2$, which we infer to be $0.91 \pm 0.05$. Hence, unlike many low-redshift analyses of large-scale structure data, we find no indication of $S_8$ tension. We demonstrate that these results are robust to reasonable variation in the implementation of baryonification used to model the host halo's gravitational potentials in response to baryonic processes. We also highlight the importance of correctly modeling the satellite radial profile in any analysis involving small-scale information. Finally, we underscore the hidden potential of this methodology for constraining baryonic physics using data from ongoing and upcoming surveys.
\end{abstract} 




\section{Introduction}
\label{sec:intro}

The $\Lambda$CDM concordance cosmology is highly successful in accounting for a diverse range of observations across different epochs in the history of the Universe and across many scales. However, with the advent of improved observational capabilities, a number of challenges have emerged \citep[see][for comprehensive reviews]{Abdalla.etal.2022, Perivolaropoulos.&.Skara.2022}. One of these has become known as the $S_8$ tension, which refers to the fact that constraints on the strength of matter clustering inferred from various probes of the low-redshift (low-$z$) Universe, are in tension with constraints inferred from the primary anisotropies of the cosmic microwave background (CMB) as measured by the \Planc satellite. In particular, the low-$z$ probes suggest a lower value for both $\Omega_\rmm$, the fraction of the energy density in matter, and $\sigma_8$, the amplitude of matter fluctuations on scales of $8 \mpch$. It is expressed most often in terms of tension with respect to the composite parameter $S_8 = \sigma_8\,(\Omega_\rmm/0.3)^{1/2}$. The Planck Collaboration \citep[][]{Planck.18} has tightly constrained $S_8$ to be $0.834 \pm 0.016$ (TT,TE,EE + lowE), which has been confirmed by an independent CMB analysis based on data from the {\it Atacama Cosmology Telesope} (ACT) and the {\it Wilkinson Microwave Anisotropy Probe} (WMAP) by \citet{Aiola.etal.20} which finds $S_8 = 0.840 \pm 0.030$. This is about 5-10 percent higher than the value preferred by the low-$z$ data. Most importantly, the $S_8$ tension manifests in various different observational probes, including, but not limited to, galaxy cluster counts \citep{Kilbinger.etal.2013, Hikage.etal.2019, Costanzi.etal.2021, Asgari.etal.2021}, projected and redshift-space galaxy clustering \citep{Macaulay.etal.2013, Li.etal.2016, Gil-Marin.etal.2017, Philcox.etal.2022}, weak gravitational lensing \citep{Joudaki.etal.2017, Hildebrandt.etal.2020, Joudaki.etal.2020, Asgari.etal.2021, Loureiro.etal.2022}, and combinations thereof \citep{Lange.etal.19c, Lange.etal.23, Wibking.etal.20, Heymans.etal.2021, Abbott.etal.2022, Yuan.etal.22}.  Another manifestation of the $S_8$ tension is the ``lensing is low'' problem, which refers to the fact that the galaxy-galaxy lensing signal is low compared to predictions based on the galaxy clustering signal, assuming \Planck's best-fit $\Lambda$CDM cosmology \citep[][]{Leauthaud.etal.17, Yuan.etal.20, Lange.etal.21, Amon.etal.23}. 

For completeness, we point out that more than a decade ago various analyses already inferred similar values for $\Omega_\rmm$ and $\sigma_8$ as the more recent studies cited above. This includes, among others, studies based on peculiar velocities inferred from redshift-space distortions in the two-point correlation function \citep[e.g.,][]{Yang.etal.04, Tinker.etal.06, Reid.etal.14}, studies that combined clustering with constraints on the mass-to-light ratios of galaxy clusters \citep[e.g.,][]{vdBosch.etal.03b, Tinker.etal.05}, and the combined analysis of clustering plus galaxy-galaxy lensing \citep[e.g.,][]{Cacciato.etal.09, Cacciato.etal.13, Mandelbaum.etal.13}. However, at the time these studies were published, these constraints were consistent with the then best-fitting CMB constraints of the WMAP mission \citep[][]{Komatsu.etal.2011} and thus did not signal any tension.

Although no individual analysis can claim a $>5\sigma$ discrepancy with the CMB results, there is an ever-growing list of studies, based on different methods and different data sets, that all reveal a similar $S_8$ tension. This has prompted numerous authors to explore beyond-standard-$\Lambda$CDM models including, but not limited to, modifications of dark energy \citep{Kunz.etal.2015, Allali.etal.2021, Akarsu.etal.2021, Sola.etal.2021, Heisenberg.etal.2023}, non-zero neutrino mass \citep[][]{Chudaykin.etal.2022, Poulin.etal.2018}, invoking various interactions in the dark sector \citep[][]{Kumar.etal.2019, Archidiacono.etal.2019, diValentino.etal.2020, Lucca.2021}, or departures from general relativity (GR) that weaken gravity in the low-$z$ universe \citep{Nesseris.etal.2017, Kazantzidis.&.Perivolaropoulos.2021, Skara.&.Perivolaropoulos.2020, Nguyen.etal.2023}. 

However, before concluding that the $S_8$ tension requires a revision of the $\Lambda$CDM concordance cosmology, it must be ruled out that it arises from some kind of observational bias or from systematic errors in the theoretical modeling. In fact, each of the different methods used to constrain $\Omega_\rmm$ and/or $\sigma_8$ has its own shortcomings and/or challenges. For example, cosmic shear and galaxy-galaxy lensing rely on accurate shear measurements that are subject to systematics that manifest as a multiplicative bias that scales linearly with $\sigma_8$ \citep[][]{Heymans.etal.06, Huff.Mandelbaum.17, Sheldon.etal.20}. In addition, gravitational lensing analyses typically rely on photometric redshifts, which can also introduce systematic errors in cosmological inference. Another significant challenge for cosmic shear measurements is how best to mitigate the impact of intrinsic alignments \citep[][]{Hirata.Seljak.04, Troxel.Ishak.14, Joachimi.etal.15}, which can induce large systematic errors if not accounted for \citep[][]{Bridle.King.07, Yao.etal.17}.  

Methods relying on galaxy clustering, especially those probing the quasi-linear and non-linear scales, are sensitive to halo assembly bias, the fact that the clustering of dark matter halos depends on their assembly history \citep{Gao.etal.05, Gao.White.07, Dalal.etal.08, Li.etal.2008, Lacerna.Padilla.11}, and several secondary halo properties such as the concentration parameter, spin parameter, and halo shape \citep{Wechsler.etal.06, Faltenbacher.White.10, Villarreal.etal.17, Salcedo.etal.18}. As a result, the clustering strength of galaxies is not a clean indicator of halo mass, and failing to account for this can significantly bias the results \citep[][]{Croton.etal.2007, Blanton.Berlind.07, Zentner.etal.14, Beltz-Mohrmann.etal.2020}. Although various authors have developed methods to incorporate assembly bias in their modeling \citep[e.g.,][]{Hearin.etal.16, Lehmann.etal.17, Contreras.etal.2023}, the true challenge is to ensure that the model is sufficiently general.

Another issue, one that plagues virtually all methods that have been used to uncover the $S_8$ tension, is `baryonic effects', that is, the effects that processes related to galaxy formation, such as cooling, star formation, and feedback, have on the matter power spectrum. Ideally, one would forward model these baryonic effects using large cosmological hydrodynamical simulations. However, such simulations rely heavily on uncertain subgrid models to account for processes below the resolution limit and are extremely expensive to run. Despite remarkable progress \citep{Villaescusa-Navarro.etal.2021, Villaescusa-Navarro.etal.2023} in the last few years, it is still difficult, if not impossible, to marginalize over all the uncertainties involved in  hydrodynamical-simulation-based modeling.  Therefore, most analyses rely instead on halo-occupation models to populate dark matter halos with galaxies, either using dark-matter only (DMO) simulations or using semi-analytical techniques that rely on Press-Schechter theory. The most commonly used halo occupation models are the Halo-Occupation Distribution \citep[HOD;][]{Jing.etal.98, Seljak.00, Scoccimarro.etal.01, Berlind.etal.03, Zheng.etal.05}, which specifies the number of central and satellite galaxies in a halo above a luminosity threshold as simple functions of the host halo mass, and the Conditional Luminosity Function \citep[CLF;][]{Yang.etal.03, vdBosch.etal.03, Cooray.06}, which specifies the luminosity distributions of central and satellite galaxies as a function of their host halo mass. Although halo-occupation modeling is a powerful tool to marginalize over uncertain galaxy formation physics, the models are not self-consistent in that they typically do not account for the way galaxy formation modifies the matter distribution with respect to what is inferred from a DMO simulations, which can significantly impact any cosmological inference \citep[][]{Zentner.etal.13, Springel.etal.2018, Arico.etal.2021, Beltz-Mohrmann.Berlind.2021, Ayromlou.etal.2023}. Since baryonic effects are most pronounced on small scales, one can partially, but not entirely, mitigate their impact by restricting the cosmological analysis to larger scales, as is often done. This has the added benefit that larger scales are more linear and thus easier to model. However, this is far from ideal, since the data extracted from cosmological surveys are typically most precise on scales of a few Mpc, where linear theory is no longer applicable. Furthermore, several studies have shown that small-scale data contain a wealth of information that should not be underutilized \citep[][]{Yuan.etal.22, Lange.etal.23}.

In this paper, we present a novel approach of probing the $S_8$ tension using observables that (i) are insensitive to halo assembly bias, (ii) fully exploit the data on the smallest, fully nonlinear scales, and (iii) allow for a straightforward correction for baryonic effects. The method uses a two-step approach. First, a Bayesian hierarchical method, called \Basilisk \citep[][]{vdBosch.etal.19, Mitra.etal.24},  is used to model the abundance and kinematics of satellite galaxies extracted from the Sloan Digital Sky Survey \citep[SDSS;][]{York.etal.00}. This yields extremely tight constraints on the galaxy-halo connection (characterized using the CLF) that are largely independent of cosmology. In the next step, these constraints are used to model the galaxy luminosity function, which is directly related to the halo mass function and is therefore strongly cosmology dependent. The technique is conceptually similar to the idea of using kinematics to estimate cluster masses and then using the cluster mass function to constrain cosmology. The main difference is that here we use data that cover the entire halo mass range from $\sim {\rm few} \times 10^{11} \Msunh$ to that of the most massive clusters.

By using the full line-of-sight velocity information of a large sample of tens of thousands of satellite galaxies, \Basilisk successfully breaks the mass-anisotropy degeneracy that often hampers dynamical mass estimates \citep[][]{Binney.Mamon.82}, simultaneously constraining both halo mass and the orbital anisotropy of satellite galaxies, while properly accounting for scatter in the galaxy-halo connection \citep{Mitra.etal.24}. Baryonic effects are taken into account and marginalized over, using the model recently developed by \citet[][hereafter B25]{Baggen.etal.25}. 

This paper is organized as follows. Section~\ref{sec:methods} presents a detailed description of our methodology, including our sample selection (Sections~\ref{sec:selection}-\ref{sec:datasample}), a description of the model used to characterize the galaxy-halo connection (Sections~\ref{sec:ghc}-\ref{sec:satellite_profile}), a model for how baryons modify the halo potential compared to a dark-matter only scenario (Section~\ref{sec:method_baryon}), and the method used for modeling the cosmological likelihood (Section~\ref{sec:evidence_model}). In Section~\ref{sec:mock_test} we test and validate our methodology against realistic SDSS-like mock data, demonstrating that it yields unbiased cosmological constraints. In Section~\ref{sec:results} we apply our machinery to SDSS DR-7 data, which intriguingly yields cosmological constraints in perfect agreement with \Planck, without any sign of a $S_8$ tension. Section~\ref{sec:discussion} discusses how our results depend on uncertainties related to baryonic effects, in particular the poorly known halo mass dependence of the fraction of baryonic mass that is ejected from the halo due to feedback processes. We also underscore the importance of correctly modeling the radial profile of satellite galaxies by demonstrating that standard assumptions often made in the literature can result in a strongly biased cosmological inference. Finally, Section~\ref{sec:summary} summarizes our findings.

\section{Methodology}
\label{sec:methods}

Our method for constraining the cosmological parameters, $\Omega_\rmm$ and $\sigma_8$, is a two-step process. First, we use \Basilisk to model the abundance and kinematics of satellite galaxies in the SDSS, which yields tight constraints on the galaxy-halo connection. Crucially, we demonstrate that these constraints are virtually independent of the assumed cosmology. Next, using the posterior on the galaxy-halo connection thus obtained, we predict the galaxy luminosity function (LF) for different combinations of $\Omega_\rmm$ and $\sigma_8$. By comparing these predictions with the actual observed LF, we are thus able to constrain cosmology.

\Basilisc, developed by \citet[][hereafter Paper~I]{vdBosch.etal.19} and subsequently improved by \citet[][hereafter Paper~II]{Mitra.etal.24} and \citet[][hereafter Paper~III]{Mitra.etal.25}, is a Bayesian hierarchical inference formalism that uses the abundance and kinematics of satellite galaxies to constrain the galaxy halo connection. Crucially, unlike previous methods, \Basilisk does not resort to stacking the kinematics of satellite galaxies in bins of central luminosity and does not make use of summary statistics, such as satellite velocity dispersion. Rather, it leaves the data in its raw form and computes the corresponding total likelihood. By modeling the full projected phase-space distribution of the satellite galaxies, one can simultaneously solve for halo mass and orbital velocity anisotropy of the satellites. Due to the Bayesian hierarchical scheme, \Basilisk naturally accounts for the scatter in the galaxy-halo connection and accurately constrains it. In addition, \Basilisk can be applied to flux-limited, rather than volume-limited samples, greatly enhancing the amount and dynamic range of the data. Moreover, in a typical galaxy redshift survey such as the SDSS used here, the vast majority ($\sim 90\%$) of central galaxies do not have a spectroscopically detected satellite around it -- an immense wealth of information that is typically discarded in traditional satellite kinematics analyses. \Basilisk fully utilizes this information by computing the likelihoods for each of those ``lonely'' centrals to have no detectable satellite given the properties of the central and the survey's flux limit.

In the following subsections, we describe the various ingredients of our methodology. The description of \Basilisk is kept brief in order to avoid redundancy with Papers~II and III, to which we refer the interested reader for more details. 

\subsection{Selecting centrals and satellites}
\label{sec:selection}

The first step in analyzing the abundance and kinematics of satellite galaxies is to select a sample of centrals and their associated satellites from a redshift survey. Unfortunately, this selection is never perfect; one undoubtedly ends up selecting some bright satellites as potential centrals (we refer to these as `impurities') and not every galaxy selected as a potential satellite actually resides in the same host dark matter halo as the corresponding central (those that do not are referred to as `interlopers'). We therefore use the terms `primaries' and `secondaries' to refer to galaxies that are selected as potential centrals and satellites, respectively.

A galaxy at redshift $z$ is considered a primary if it is the brightest galaxy in a conical volume of opening angle $\theta_{\rm ap}^{\rm pri} \equiv \Rh/d_\rmA(z)$ centered on the galaxy in question, and extending along the line of sight from $z-(\Delta z)^{\rm pri}$ to $z+(\Delta z)^{\rm pri}$. Here $d_\rmA(z)$ is the angular diameter distance at redshift $z$, and $(\Delta z)^{\rm pri} = (\dVh/c) \, (1+z)$. The parameters $\Rh$ and $\dVh$ specify the primary selection cone.  All galaxies fainter than the primary and located inside a similar cone, but defined by $\Rs$ and $\dVs$, centered on the primary are identified as its secondaries.  The four parameters $\Rh$, $\dVh$, $\Rs$, and $\dVs$ scale with the luminosity of the galaxy on which the cone is centered, which is tuned to optimize the completeness and purity of the sample. We emphasize that we have extensively tested that \Basilisc's inference is robust and insensitive to the exact details of the selection, primarily because the detailed selection effects are forward modeled in the Bayesian hierarchical framework. See \paperII for details.

\subsection{Data sample}
\label{sec:datasample}
\begin{figure}
\centering
\includegraphics[width=0.49\textwidth]{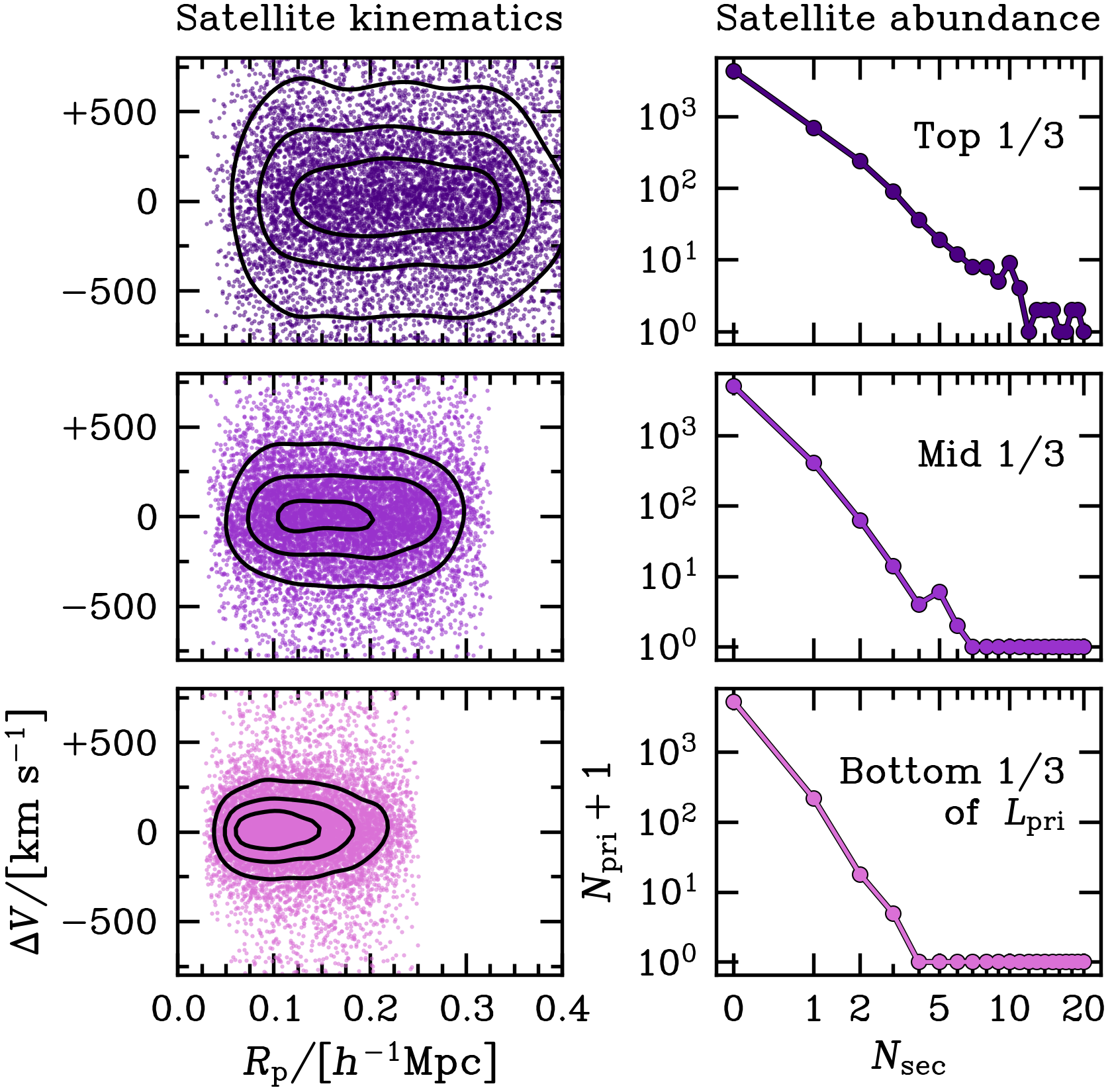}
\caption{The SDSS data used by \Basilisk to constrain the galaxy-halo connection. The panels in the left-hand column show the projected phase-space distribution of secondaries, while the right-hand panels show their multiplicity distributions.  To highlight the dependence on the luminosity of the primary, the data is split in three consecutive bins of $\Lpri$ that contain equal numbers of primaries, with the brightest (faintest) subsample shown at the top (bottom). Note, the velocity dispersion of the secondaries is higher for the more luminous primaries, which also contain, on average, more secondaries, implying that they reside in more massive halos.}
\label{fig:data}
\end{figure}
The analysis presented in this paper uses the same set of primaries and secondaries selected from the SDSS (and thus the same $\Rh$, $\dVh$, $\Rs$, and $\dVs$) as used in \papII. Primaries are restricted to have luminosities in the range $9.504 \leq \log[\Lpri/(h^{-2}\Lsun)] \leq 11.104$ and redshifts in the range $0.034 \leq z \leq 0.184$. Using the selection method described above, this yields $\sim 165,000$ primaries, of which $N_+=18,373$ primaries have at least one secondary. The total number of secondaries, and thus the total number of primary-secondary pairs for which kinematic data is available, is $30,431$. For each primary-secondary pair in the sample, we use their projected separation, 
\begin{equation}\label{Rpdef}
\Rp = d_\rmA(\zpri) \, \vartheta
\end{equation}
and their line-of-sight velocity difference,
\begin{equation}\label{dvdef}
\dV = c \, \frac{(\zsec - \zpri)}{1 + \zpri}
\end{equation}
Here, $\zpri$ and $\zsec$ are the observed redshifts of the primary and secondary, respectively, $c$ is the speed of light, and $\vartheta$ is the angular separation between the primary and secondary in the sky.

As input, \Basilisk takes two data vectors. The first is the satelite kinematics (SK) data vector
\begin{equation}\label{dataprim}
\bD_{\rm SK} = \bigcup\limits_{i=1}^{N_{+}} \, \left( \{\dVij, \Rij | j=1,...,\Nsi \} | \Lci, \zci, \Nsi\right)
\end{equation}
where the union is over all $N_{+}$ primaries with at least one secondary. Here $\Nsi$ is the number of secondaries associated with primary $i$, and it is made explicit that $\Lci$, $\zci$, and $\Nsi$ are only treated as {\it conditionals} for the data $\{\dVij, \Rij | j=1,...,\Nsi \}$ (see \paperII for a justification of this approach). The left-hand column of Fig.~\ref{fig:data} shows the projected phase-space distributions of all secondaries, split into three luminosity bins that have equal numbers of primary-secondary pairs. Several trends are worth noting. First, the typical $\dV$ is larger around more luminous primaries, indicating that these reside in more massive halos. Secondly, secondaries extend to larger $\Rp$ around brighter primaries. This simply reflects the fact that the conical volume used to select secondaries is wider around brighter primaries. Finally, the dearth of secondaries at small $\Rp$ is due to the fact that we remove all secondaries that are within $55$ arcseconds of their primary to mitigate the impact of fiber collisions (see \paperII and Section~\ref{sec:method_baryon}).

The second data vector used by \Basilisk is 
\begin{equation}\label{dataNs}
\bDNs = \bigcup\limits_{i=1}^{N_{\rm Ns}} \left( \Nsi | \Lci, \zci \right)
\end{equation}
which specifies the number of secondaries associated with each primary. Here, the union is over a subset of $N_{\rm Ns} \simeq \calO(N_{+})$ primaries that are randomly selected from the entire $\Npri = N_0 + N_{+}$ sample of primaries, independent of how many secondaries they have (that is, including primaries with zero secondaries). The reason for using a random subset rather than all $\Npri$ primaries is to limit the computational cost, as discussed in \papII. The right-hand column of Fig.~\ref{fig:data} shows the multiplicity distributions for three bins in $\Lpri$ that have equal numbers of primaries. As is evident, brighter primaries have, on average, more secondaries. This reflects both the larger volume of their secondary selection cone and the fact that they reside in more massive halos that have more satellites above a given flux limit. Together with the kinematic information shown in the left-hand columns, these are the data that we use to constrain the galaxy-halo connection.  However, we emphasize that we forward model the entire data in raw and unbinned format; the binning in Fig.~\ref{fig:data} is only for the sake of visualization.

\subsection{Sampling the posterior}
\label{sec:sampler}

\Basilisk uses an affine invariant ensemble sampler \citep[][]{Goodman.Weare.10} to constrain the posterior distribution, 
\begin{equation}\label{posterior_Basilisk}
P(\btheta | \bD) \propto \calL(\bD | \btheta) \, P(\btheta)
\end{equation}
Here $\bD = \bD_{\rm SK} + \bDNs$ is the total data vector\footnote{Note that this differs from \papII, where we also folded in the galaxy luminosity function (LF) as additional constraints. Here we first ignore the LF to constrain the galaxy-halo connection, which we subsequently use to model the LF with the goal of constraining the cosmological parameters.}, $\btheta$ is the vector that describes our model parameters (see below), $P(\btheta)$ is the prior probability distribution on the model parameters, and $\calL(\bD|\btheta)$ is the likelihood of the data given the model. The latter consists of two parts: the likelihood $\calL_{\rm SK}$ for the satellite kinematics data $\bD_{\rm SK}$ and the likelihood $\calL_{\rm Ns}$ for the numbers of secondaries as described by the data vector $\bDNs$. 

A detailed description of how these two different likelihood terms are computed can be found in Appendix~\ref{App:AppA} and in \papII. Briefly, we assume that satellite galaxies are virialized steady-state tracers of the gravitational potential in which they orbit, and we use the second and fourth order velocity moments of the collisionless Boltzmann equation to compute the probability $P(\Delta V|\Rp,M)$ for the line-of-sight velocity difference $\Delta V$ at a projected separation $\Rp$ in a (spherically symmetric) halo of mass $M$. In computing $\calL_{\rm SK}$ we fully marginalize over halo mass and allow for (mass- and radius-independent) velocity anisotropy. By capturing the full projected phase-space distribution, \Basilisk successfully breaks the so-called ``mass-anisotropy degeneracy'' \citep[][]{Binney.Mamon.82}, and is able to simultaneously constrain the galaxy-halo connection of central and satellite galaxies, including their scatter, as well as the radial profile and orbital velocity anisotropy of satellite galaxies. The effects of interlopers and impurities are taken into account using a detailed, data-driven forward modeling approach.

\subsection{The galaxy halo connection}
\label{sec:ghc}

The galaxy occupation statistics of dark matter halos are modeled using the conditional luminosity function (CLF), $\Phi(L|M,z) \, \rmd L$, which specifies the average number of galaxies with luminosities in the range $[L-\rmd L/2, \, L + \rmd L/2]$ that reside in a halo of virial mass $M$ at redshift $z$ \citep[][]{Yang.etal.03, vdBosch.etal.03}. In particular, we write
\begin{equation}
    \Phi(L|M,z) = \Phi_\rmc(L|M) + \Phi_\rms(L|M)
\end{equation}
Here, the subscripts `c' and `s' refer to central and satellite, respectively, and we assume that the CLF is independent of the redshift over the range considered ($0.02 \leq z \leq 0.2$).

The CLF of centrals is parametrized using a log-normal distribution,
\begin{equation}\label{CLFcen}
\Phi_\rmc (L | M) \rmd L = \frac{\log e}{\sqrt{2\pi \sigma_\rmc^2}} \exp \left[ -\left(\frac{\log L - \log\bar{L}_\rmc}{\sqrt{2} \sigma_\rmc} \right)^2\right] \frac{\rmd L}{L}
\end{equation}
The mass dependence of the median luminosity, $\bar{L}_\rmc$, that is, the central luminosity-halo mass relation, is parametrized by a double power law: 
\begin{equation}\label{averLc}
\bar{L}_\rmc (M) = L_0 \frac{(M / M_1)^{\gamma_1}}{(1 + M / M_1)^{\gamma_1 - \gamma_2}}
\end{equation}
which is characterized by four free parameters: a normalization, $L_0$, a characteristic halo mass, $M_1$, and two power-law slopes, $\gamma_1$ and $\gamma_2$. 

The mass dependence of the scatter is modeled using
\begin{equation}\label{scatter}
\sigma_\rmc(M) = \sigma_{13} + \sigma_\rmP \, \log M_{13}
\end{equation}
where $M_{13} = M/(10^{13}\Msunh)$. This has two additional free parameters, a normalization, $\sigma_{13}$, which specifies the intrinsic scatter in $\log\Lc$ in halos of mass $\Mh = 10^{13}\Msunh$, and a power law slope $\sigma_\rmP$. 

For the satellite CLF we adopt a modified Schechter function: 
\begin{equation}\label{satCLF}
\Phi_\rms (L | M) = \frac{\phi_\rms^*}{L_\rms^*} \left( \frac{L}{L_\rms^*} \right)^{\alpha_\rms} \exp \left[ - \left( \frac{L}{L_\rms^*} \right)^2 \right]
\end{equation}
Thus, the luminosity function of satellites in halos of a given mass follows a power law with slope $\alpha_\rms$ and with an exponential cutoff above critical luminosity $L_\rms^*(M)$. Throughout, we adopt
\begin{equation}\label{scatter2}
\alpha_\rms(M) = \alpha_{13} + \alpha_\rmP \,\log M_{13}
\end{equation}
and 
\begin{equation}
\log[L_\rms^*(M)] = \log[\bar{L}_{\rmc} (M)] + \Delta_{13} + \Delta_\rmP \, \log M_{13}
\end{equation}
Finally, the normalization $\phi_\rms^*(M)$ is parametrized by 
\begin{equation}\label{satCLFnorm}
\log \left[ \phi_\rms^*(M) \right] = b_0 + b_1 \log M_{12} + b_2 (\log M_{12})^2
\end{equation}
where $M_{12} = M/(10^{12}\Msunh)$. 

Hence $\Phi_\rmc(L|M)$ and $\Phi_\rms(L|M)$ are characterized by a total of 6 and 7 free parameters, respectively.  Note that this characterization of the CLF is similar to, but significantly more flexible than, that adopted in a number of previous studies \citep[][]{Cacciato.etal.09, Cacciato.etal.13, More.etal.09b, vdBosch.etal.13, Lange.etal.19a, Lange.etal.19b}. In particular, we have introduced a halo mass dependence for the faint-end slope, $\alpha_\rms$, and for the ratio $L_\rms^*/\bar{L}_{\rmc}$, which differs from the fiducial model used in Papers~I and~II.  As discussed in detail in \papIII, this extra freedom greatly improves the fits to the data, indicating that the galaxy-halo connection is more complicated than is often assumed. In Appendix~\ref{App:C} we also showcase the results for the other CLF parametrizations discussed in \papIII.

In addition to the 13 parameters that characterize the CLF, \Basilisk also includes 3 nuisance parameters used for interloper modeling, and one parameter, $\beta$, to characterize the orbital anisotropy of satellite galaxies within their host halos, bringing the grand total of free parameters that make up our model vector $\btheta$ to 17.

\subsection{Spatial distribution of satellites}
\label{sec:satellite_profile}

Throughout we assume that the radial distribution of satellite galaxies is given by a spherically symmetric generalized NFW profile:
\begin{equation}\label{nsatprof}
n_{\rm sat}(r|M,z) \propto \left( \frac{r}{\calR \, r_\rms} \right)^{-\gamma} \left( 1 + \frac{r}{\calR \, r_\rms} \right)^{\gamma - 3}
\end{equation}
Here $\calR$ and $\gamma$ are free parameters, and $r_\rms$ is the scale radius of the dark matter halo, which is related to the halo virial radius via the concentration parameter $\cvir = \rvir / r_\rms$. This profile has sufficient flexibility to adequately describe a wide range of radial profiles, from satellites being unbiased tracers of their dark matter halo, that is $(\gamma,\calR) = (1,1)$, to extended and cored profiles that resemble the radial profile of surviving subhalos in numerical simulations, for which $(\gamma,\calR) \sim (0,2)$ \citep[e.g.,][]{Diemand.etal.04, Springel.etal.08, Han.etal.16, Jiang.vdBosch.17}. This also brackets the range of observational constraints on the radial distribution of satellite galaxies in groups and clusters \citep[e.g.,][]{Carlberg.etal.97, vdMarel.etal.00, Lin.etal.04, Yang.etal.05a, Chen.08, More.etal.09b, Guo.etal.12a, Cacciato.etal.13, Watson.etal.10, Watson.etal.12}. 

Although \Basilisk uses a flexible model for the radial profile of satellite galaxies, $(\gamma, \calR)$ are not treated as free parameters in its MCMC inference. To make it feasible to compute the $\calO(10^6-10^7)$ likelihood evaluations required to adequately probe the posterior, it is essential that a large number of quantities that depend on $n_{\rm sat}(r|M,z)$ be precomputed and stored in memory. Therefore, instead of letting the radial profile vary in each MCMC step, we run \Basilisk on a grid of $(\gamma, \calR)$ values and use the best-fit values of $\gamma$ and $\calR$ for our subsequent analysis (see Papers~I and~II). For the cosmological analysis presented here, we do this separately for each set of cosmological parameter (see Section~\ref{sec:evidence_model}).
\begin{figure*}
\centering
\includegraphics[width=0.95\textwidth]{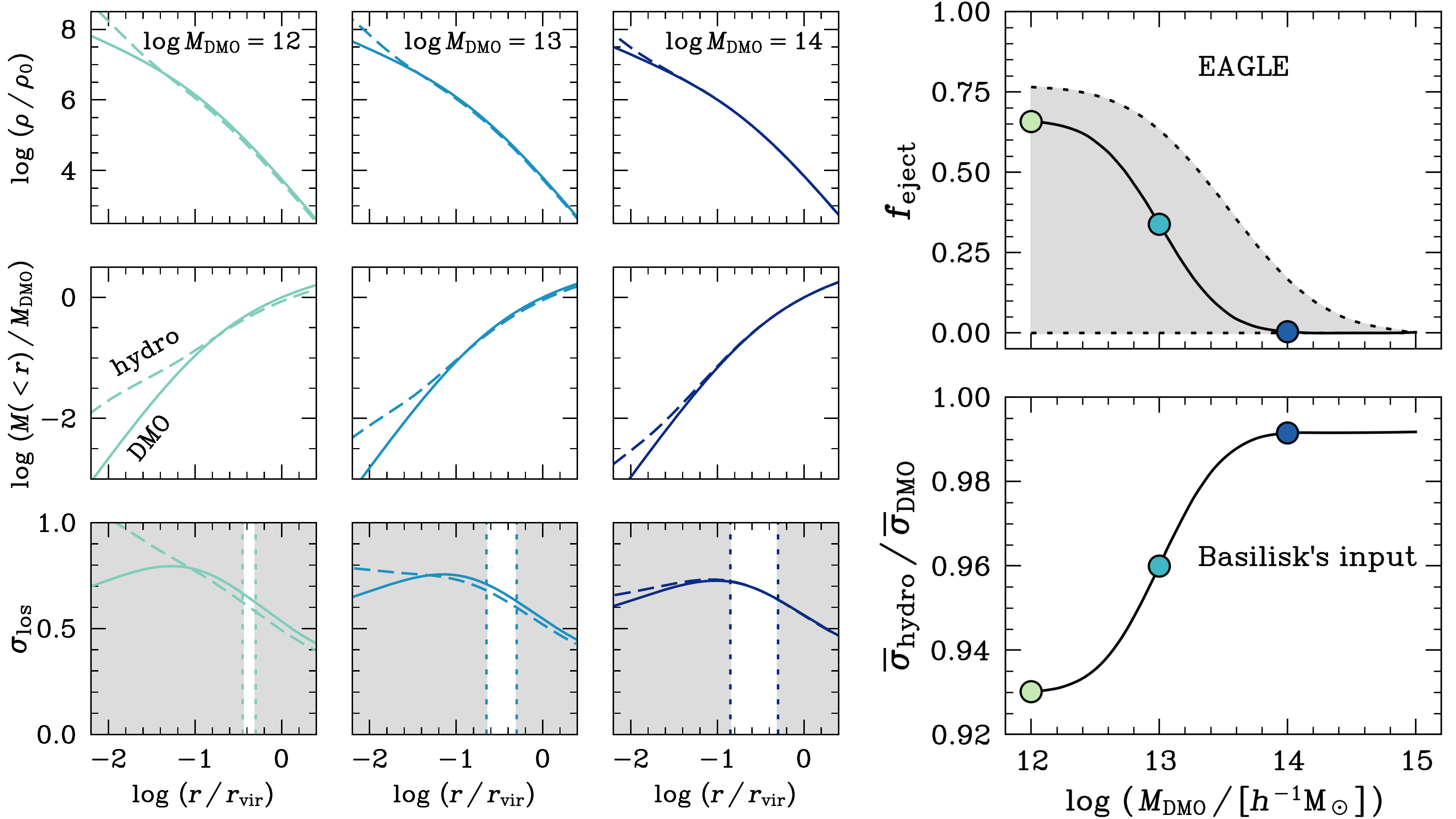}
\caption{The $3\times 3$ panels on the left show how baryonic effects impact satellite kinematics for three different halo masses (different columns). From top to bottom, the panels show radial profiles of total mass density, enclosed mass, and the resultant line-of-sight velocity dispersion of satellite galaxies modelled as a massless tracer population. Solid and dashed curves correspond to the dark matter only (DMO) case and the case in which baryonic effects have been taken into account, respectively. The gray shaded regions in the lower panels indicate where \Basilisk does not have data, due to secondary selection cuts; inner radii are excluded because of fiber collisions, while outer radii are excluded to minimize the impact of interlopers. The top right-hand panels plots $f_{\rm eject}$, the fraction of baryons ejected from halos in the \EAGLE simulation as a function of the DMO virial mass, which is the most relevant parameter controlling the impact of baryonic effects in our analysis of satellite kinematics. The gray-shaded region envelopes the range of $f_{\rm eject}(M)$ models considered in this study (see Section~\ref{sec:subsec:varying_baryon}). For the fiducial baryonic prescription of B25, which includes the adiabatic response of the dark matter, that ratio of interest $(\overline{\sigma}_{\rm hydro}/\, \overline{\sigma}_{\rm DMO})$ is shown in the bottom-right panel, labelled ``\Basilisc's input''. The coloured circles correspond to the 3 different halo masses shown in the panels on the left.}
\label{fig:baryon_model}
\end{figure*}

\subsection{Baryonic effects}
\label{sec:method_baryon}

A crucial ingredient for modeling satellite kinematics is the mass model for the gravitational potential in which the satellites orbit. All previous studies that used satellite kinematics to constrain the galaxy-halo connection, including Papers I, II \& III in this series and many others \citep[][]{Prada.etal.03, vdBosch.etal.04, Norberg.etal.08, More.etal.09b, More.etal.11, Wojtak.Mamon.13, Lange.etal.19b}, have assumed that these potential wells are made of dark matter only. In particular, in Papers I-III we assumed that satellite galaxies orbit in spherical dark matter halos with a NFW density profile. In reality, the density profiles of the host halos consist of both dark matter and baryons, and have been modified due to the combined effects of cooling, star formation and feedback processes. These `baryonic effects' can significantly modify the potential well of the host halo, and thus impact the kinematics of the satellite galaxies. 

The impact of baryonic effects on satellite kinematics was investigated in detail in the recent study by \citet[][hereafter B25]{Baggen.etal.25}, who compared the mass distributions of halos in the  hydrodynamical \EAGLE ('Evolution and Assembly of GaLaxies and their Environment') simulations \citep[][]{Schaye.etal.15, Crain.etal.15, McAlpine.etal.16} with their matched halos in the dark matter-only (DMO) run. If all baryons remain associated with their halo, and exactly follow the dark matter, the mass distributions in each pair of matched halos would have been identical. However, some of the baryonic matter cools and condenses at the halo center to form a central galaxy, which enhances the central mass distribution of the halo. At the same time, feedback processes can expel large fractions of baryons from the host halo, while some of the baryons remain bound to the halo either in gaseous form or as a stellar halo. The latter manifests as intracluster light in massive groups and clusters. Using Jeans modeling, B25 compared the line-of-sight kinematics of a massless tracer population (representing satellite galaxies) in halos affected by baryonic processes with the kinematics of an identical tracer population in the matched DMO halo. 

The key finding of B25 is that the ejection of baryons is by far the dominant effect. This is easy to understand; satellite kinematics mainly probes the total (enclosed) mass. If that is reduced because of the removal of (baryonic) matter, it impacts the kinematics across the entire system. Although the formation of the central galaxy can have a very significant effect on kinematics close to the center of the halo, where it typically causes a large boost in the enclosed mass compared to the DMO case, this tends to have an almost negligible effect on the analysis in \Basilisc. The reason is somewhat fortuitous. The SDSS is affected by fiber collisions, which causes incompleteness in the spectroscopic data on small scales. \Basilisk partially corrects for this, but this is only effective on scales larger than 55 arcseconds (see \paperI for details). On smaller scales the effect is too strong to allow for a simple correction, which is why all secondaries that are closer than 55 arcseconds from their primaries are removed from the data. By being blind to the kinematics of satellite galaxies at small halo-centric radii, we are also insensitive to how baryonic effects modify the halo potential on those scales. 

Fig.~\ref{fig:baryon_model} illustrates how baryonic effects impact satellite kinematics. The top row of the $3\times 3$ grid of panels compares the total density profiles of halos (including dark matter, stars, and gas) in the hydrodynamic runs (dashed lines) with those of the corresponding DMO halos (solid lines). Results are shown for three different mass scales, as indicated. The enclosed mass profiles corresponding to these density profiles are shown in the middle row of panels. The upturn of the dashed curves at small radii reflects the formation of the central galaxy. At larger radii, the reduced density in the hydrodynamic run is mainly a result of the ejection of baryons from the halo. The panel in the upper right corner of Fig.~\ref{fig:baryon_model} shows the halo mass dependence of the ejected baryonic mass fraction,
\begin{equation}\label{eqn:feject}
f_{\rm eject} \equiv 1 - \dfrac{M_{\rm bar}}{f_{\rm bar}  M_{\rm DMO}} = 1 - \dfrac{M_{\rm cen} + M_{\rm ICL} + M_{\rm gas}}{f_{\rm bar}  M_{\rm DMO}}  \,,
\end{equation}
in the \EAGLE simulation, as taken from B25. Here $M_{\rm bar}$ is the total baryonic mass inside the virial radius in the hydrodynamic run, which is the sum of the stellar mass of the central galaxy, $M_{\rm cen}$, the stellar mass of the stellar halo, $M_{\rm ICL}$, and the total gas mass inside the halo, $M_{\rm gas}$. The constant $f_{\rm bar}$ is the universal baryon fraction, and $M_{\rm DMO}$ is the total mass within the virial radius in the dark matter only run. Typically, $f_{\rm eject}$ increases with decreasing halo mass simply because more massive halos are better at retaining mass \citep[e.g.,][]{Dekel.Silk.86}. The gray-shaded region in the top-right panel envelopes the extreme cases of $f_{\rm eject}$ that we explore in our analysis (see Section~\ref{sec:subsec:varying_baryon}). 

The bottom rows of the $3\times 3$ grid of panels show the line-of-sight velocity dispersions of a massless tracer population whose density profile follows that of the dark matter. As expected, the larger enclosed mass in the central regions increases the central line of sight velocity dispersion, whereas the reduced mass of the system as a whole implies that $\sigma_{\rm los}$ in the halo outskirts is reduced. The gray-shaded regions indicate the radial ranges that are excluded in \Basilisc. On small scales, we exclude data because of the issue with fiber collisions. Note that 55 arcseconds typically corresponds to a larger fraction of the halo's extent in lower mass halos\footnote{At $z=0.1$, roughly the median redshift of the SDSS, 55 arcseconds corresponds to $\sim 100 \kpc$, which is a large fraction of the virial radius for low mass halos.} The data at large radii is excluded because of the selection criteria used to identify secondaries. The opening angle of the secondary selection cone is tuned to be approximately 40 percent of the halo virial radius (see \paperII for details). This is done to reduce the impact of interlopers, which become more prevalent at larger projected separations from the primary. Note how in the non-shaded radial interval baryonic processes mainly lower the line-of-sight velocity dispersion with respect to the DMO case, and with virtually no dependence on radius.  The lower right panel of Fig.~\ref{fig:baryon_model} summarizes these findings by plotting the average ratio $\overline{\sigma}_{\rm hydro}/\overline{\sigma}_{\rm DMO}$, where $\overline{\sigma}_{\rm hydro/DMO}$ is the integral of $\sigma_{\rm los}$ for the hydro/DMO cases over the range of projected radii covered by \Basilisc. As is evident, baryonic effects are quite negligible in massive halos but cause a slight reduction in $\sigma_{\rm los}$ in low-mass halos.  The change in $\overline{\sigma}_{\rm los}$ is not simply the result of mass loss due to feedback; there are other effects at play, such as the accumulation of mass at the halo center associated with the formation of the central galaxy, and the corresponding adiabatic contraction \citep[][]{Blumenthal.etal.86} of the dark matter halo, all of which are taken into account. Despite that, the ejection of baryonic matter is the dominant factor, and we find that the net baryonic effect is approximately captured by the following simple expression (see B25):
\begin{equation}\label{barcorr}
\overline{\sigma}_{\rm hydro} \, /\,\overline{\sigma}_{\rm DMO} \approx \sqrt{1 - f_{\rm bar} f_{\rm eject}}
\end{equation}
with $f_{\rm eject} = f_{\rm eject}(M_\rmh)$.

In \Basilisc, we compute the line-of-sight velocity dispersion, $\sigma_{\rm los}$, of satellite galaxies assuming that their host halos are made entirely of dark matter, which is assumed to follow a NFW density profile with a concentration parameter set by the concentration-mass relation of \citet{Diemer.Kravtsov.15}. However, we then correct $\sigma_{\rm los}$ for baryonic effects using the mass-dependent ratio shown in the lower right panel of Fig.~\ref{fig:baryon_model}, which we take as our fiducial model. We emphasize that this fiducial model is obtained by B25 by calibrating their model against the \EAGLE simulation. Given that different hydrodynamical simulations yield significantly different results, mainly because they all use different subgrid prescriptions to model the various feedback processes \citep[e.g.,][]{Vogelberger.etal.20, Beltz-Mohrmann.etal.2020, Strawn.etal.2024}, this correction for baryonic effects has significant uncertainties. In Section~\ref{sec:subsec:varying_baryon} we discuss the implications of these uncertainties on the cosmological constraints. 

\subsection{Cosmological evidence modeling}
\label{sec:evidence_model}

The main goal of this paper is to use \Basilisk to constrain cosmological parameters and in particular to see if the analysis of the abundance and kinematics of satellite galaxies in SDSS reveal a similar $S_8$ tension as other low-redshift probes. In order to do so, we proceed as follows. First, we use \Basilisk and the SDSS data on the abundances and kinematics of satellite galaxies to put tight constraints on the CLF, $\Phi(L|M)$. As we explicitly demonstrate in Section~\ref{sec:mock_test}, these constraints are virtually independent of the assumed cosmology. This follows from the fact that satellite kinematics is insensitive to the clustering of dark matter halos and with only a very weak dependence on the halo mass function. Rather, satellite kinematics mainly probes the detailed potential of the host halos, which in turn depends on halo mass and halo concentration. And since the concentration-mass relation of dark matter halos only has a weak cosmology dependence \citep[at least over the limited range of cosmologies considered here; ][]{Diemer.Kravtsov.15}, the posterior CLF constraints obtained with \Basilisk only reveal a very weak dependence on $\Omega_\rmm$ and $\sigma_8$. 

Having tight constraints on the galaxy-halo connection, as characterized by the CLF $\Phi(L|M)$, allows us to make precise predictions for the galaxy LF for a given cosmology, using
\begin{equation}\label{LFeq}
\Phi(L) = \int_0^{\infty} \Phi(L|M) \, n(M) \, \rmd M
\end{equation}
with $n(M)$ the cosmology-dependent halo mass function. In the absence of {\it any} prior on $\Phi(L|M)$, one can {\it always} perfectly fit $\Phi(L)$, for {\it any} cosmology. After all, $\Phi(L|M) = [\Phi(L)/n(M)] \, \delta_\rmD(M-M_\rmc)$, with $\delta_D(x)$ the Dirac delta function and $M_\rmc$ an arbitrary mass, clearly solves equation~(\ref{LFeq}). Obviously such a galaxy-halo connection, in which galaxies only reside (with the correct abundance) in halos of one particular mass $M_\rmc$, is clearly absurd, but it does elucidate that without any prior on the galaxy-halo connection, the LF puts no constraints at all on cosmological parameters.  However, using the tight constraints on $\Phi(L|M)$ coming from \Basilisk's posterior, we can use equation~(\ref{LFeq}) to make accurate predictions for the galaxy LF for a given cosmology. A comparison with the actual LF of SDSS galaxies then yields constraints on $\Omega_\rmm$ and $\sigma_8$, which are the two main cosmological parameters that impact the halo mass function. 

More formally, the posterior distribution for cosmology $\calC$, given the data $\bD$, can be written as
\begin{align}
    P(\calC|\bD) = \frac{\calZ(\bD|\calC) \, P(\calC)}{\calZ(\bD)}
\end{align}
Here $\calZ(\bD)$ is the marginal Bayesian evidence for the data, and $P(\calC)$ is the prior on cosmology ($\Omega_\rmm$ and $\sigma_8$), for which we use flat uninformed priors.  Since $\calZ(\bD)$ is independent of cosmology, the ratio $P(\calC) / \calZ(\bD)$ is merely a multiplicative constant that normalizes the posterior $P(\calC|\bD)$, and we have
\citep[][]{Lange.etal.19d}
\begin{equation} \label{evidence_cosmo}
P(\calC|\bD) \propto \calZ(\bD|\calC) = \int \rmd \calG \, P\left(\bD | \hat{\bD}(\calC,\calG)\right) \,  P(\calG) 
\end{equation}
where $\calG$ indicates the galaxy-halo connection, the prior of which is given by $P(\calG)$. In our analysis, $\bD$ is the LF data taken from the SDSS, and $\hat{\bD}(\calC,\calG)$ is the LF predicted for cosmology $\calC$ and the galaxy-halo connection $\calG$. For the prior $P(\calG)$, we use the robust cosmology-independent posterior given by equation~(\ref{posterior_Basilisk}) and obtained with \Basilisc. If we define
\begin{equation}\label{chi2LF}
\chi^2_{\rm LF} = (\hat{\bD}-\bD)^\rmT \, \bC^{-1}_{\rm LF} \, (\hat{\bD}-\bD)
\end{equation}
where $\bC_{\rm LF}$ is the covariance matrix of the LF data (see below), we can rewrite equation~(\ref{evidence_cosmo}) as
\begin{equation} \label{evidence_cosmo_new}
P(\calC|\bD) = N_{\rm MC}^{-1} \, \sum_{i=1}^{N_{\rm MC}} \exp\left[-\tfrac{1}{2} \chi_{\rm LF}^2\left(\bD | \hat{\bD}(\calC,\calG_i)\right) \right]
\end{equation}
Here, the summation is over the $N_{\rm MC}$ elements of the Monte Carlo Markov chain that samples Basilisk's posterior of the galaxy-halo connection, and we have used the fact that the density of MCMC elements in the parameter space of $\calG$ is a direct representation of $P(\calG)$. This method of computing the evidence can be somewhat noisy for small MCMC chains. Using extensive tests based on separate MCMC runs with a total of 1.25 million Markovian realizations each, we have explicitly verified that our evidence and the resulting cosmological constraints are not prone to such noise.

In practice, rather than using a continuous LF, $\Phi(L)$, we follow \cite{Lange.etal.19a, Lange.etal.19b} and \paperII and use the number density of galaxies in ten, 0.15~dex bins in luminosity, ranging from $10^{9.5}$ to $10^{11} \Lsunh$. These are computed using the corresponding volume-limited subsamples, carefully accounting for the SDSS DR-7 footprint. In what follows, we refer to the data vector representing these 10 number densities as $\bD_{\rm LF}$. The covariance matrix of these data, $\bC_{\rm LF}$, is computed using a jackknife estimator applied to maximally compact equal-area partitions \citep{Zhou.etal.21, Wang.etal.22}. We use 100 jackknife fields, and their distribution is similar to Fig.~2 in \citet{Wang.etal.22}. We also apply a Hartlap correction \citep[][]{Hartlap.etal.07} to the inverse of the covariance matrix (see \paperII for details). In order to model these number densities, we properly account for the evolution of the halo mass function and the luminosity dependence of the survey depth by integrating over the lightcone:
\begin{equation}\label{eqn:numdens}
n_{\rm gal}(L_1,L_2) = \int_{L_1}^{L_2} \rmd L \int_{z_{\rm min}}^{z_{\rm max}} \rmd z \dfrac{\rmd V}{\rmd z} \int_{0}^{\infty} \rmd M \, \Phi(L|M) \, n(M,z) 
\end{equation}
where $\rmd V/\rmd z$ is the comoving volume element per unit of solid angle. The minimum redshift $z_{\rm min}=0.02$ is set by our selection criteria, while the maximum redshift $z_{\rm max}(L)$ is defined as the redshift out to which a galaxy of luminosity $L$ makes the flux limit of the spectroscopic SDSS data (see Section~\ref{sec:results}).

The cosmological inference described above is a two-step process; we first use the abundance and kinematics of satellite galaxies to constrain the galaxy-halo connection, which in the next step is used as a prior when constraining cosmological parameters by fitting the LF. The reader may wonder why we don't simply use the combined data vector $\bD_{\rm SK} + \bDNs + \bD_{\rm LF}$ and use an MCMC method to simultaneously constrain the posterior of the CLF plus cosmological parameters. The reason for this is purely computational: \Basilisc's algorithm relies heavily on extensive pre-computation of necessary quantities that are cosmology dependent (see \paperI for details). Hence, performing a Markov Chain likelihood inference that includes $\Omega_\rmm$ and $\sigma_8$ as free parameters is computationally prohibitive. Rather, we make use of the fact that \Basilisc's inference regarding the galaxy-halo connection based on $\bD_{\rm SK} + \bDNs$ only has a very weak cosmology dependence. In particular, we run \Basilisk on the SDSS data for 9 different cosmologies (that is, 9 different combinations of $\Omega_\rmm$ and $\sigma_8$) that span a range of values that is much larger than what is considered consistent with the \Planc CMB constraints (see Sections~\ref{sec:mock_test} and \ref{sec:results} for details). We then combine these 9 posterior distributions (which are very similar to each other) into a single posterior distribution for the galaxy-halo connection. This combined posterior is subsequently used in the cosmological evidence modeling described above as the prior $P(\calG)$ for the CLF,  properly marginalized over cosmology.

Whenever using galaxy survey data for a cosmological analysis, one has to cope with the fact that the data used typically depend on a fiducial cosmology, $\calC^{\rm fid}$, adopted to convert the raw data (i.e., apparent magnitudes, angles on the sky, and redshifts) to physical quantities (i.e., luminosities and distances). When analyzing such data for any other cosmology, $\calC$, one needs to scale the data or the model accordingly \citep[][]{More.13}. In our analysis, the fiducial cosmology used to convert the data is the flat $\Lambda{\rm CDM}$ \Planc cosmology with $\Omega_\rmm = 1 - \Omega_\Lambda = 0.315$ and $h = H_0/(100\kmsmpc) = 0.6736$. When using any other cosmology, \Basilisk converts the projected separations $r_\rmp$ between two galaxies at redshift $z$ computed for $\calC$ by the ratio of angular diameter distances
\begin{equation}
r_\rmp^{\rm corr} = r_\rmp(\calC) \, \frac{d_\rmA(z,\calC^{\rm fid})}{d_\rmA(z,\calC)}
\end{equation}
before comparing it with the value of $r_\rmp$ measured assuming $\calC^{\rm fid}$. Similarly, when computing the number densities of galaxies for comparison with the $n_{\rm gal}(L_1,L_2)$ obtained from the SDSS data assuming our fiducial \Planc cosmology, we multiply $\Phi(L)$, given by equation~(\ref{LFeq}), with the ratio of the comoving volumes to which a galaxy of luminosity $L$ can be observed given the SDSS flux limit. Note that for convenience, we do not convert luminosities. Hence, galaxies are always `labeled' with the luminosities they would have for the fiducial \Planc cosmology $\calC^{\rm fid}$. Since we attach no physical meaning to $L$ in terms of galaxy formation physics but are only interested in the abundance of galaxies carrying the label $L$, this does not impact our cosmological inference.
\begin{figure*}
\centering
\includegraphics[width=0.8\textwidth]{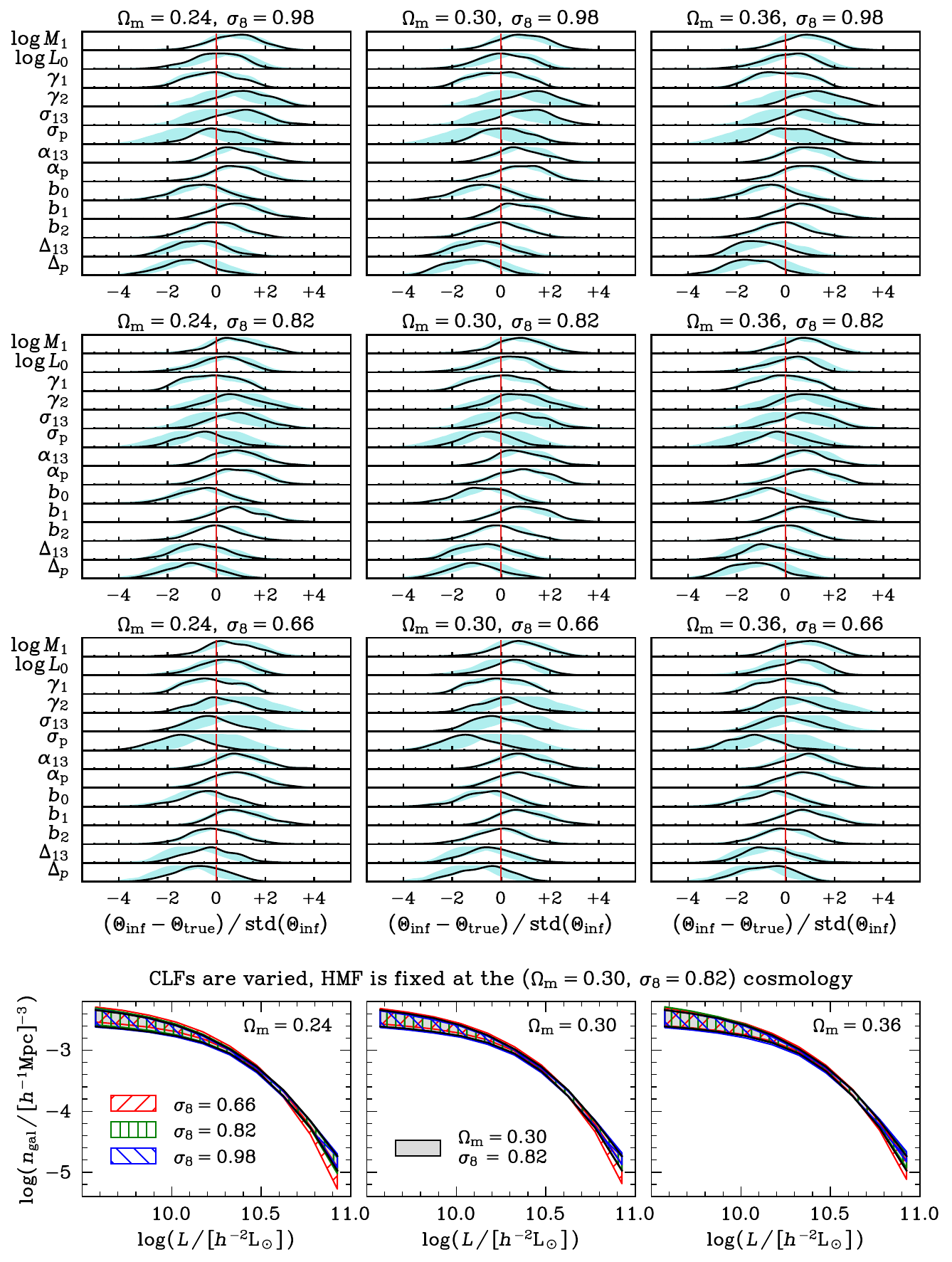}
\caption{\textit{Top 3 rows} show \Basilisc's inference of galaxy-halo connection parameters from the mock data discussed in the text for 9 different cosmologies. Each set of histograms shows the posterior distributions of the parameters for a specific cosmology, as indicated on the top of each panel. The cyan shaded regions show the amounts of variation of the posterior distributions across all 9 cosmologies shown here. All posterior distributions are shifted by the true value of the input parameter used to create the mock, and scaled by the standard deviation of the posterior distribution. All posterior inferences are statistically consistent with the truth (red vertical line in each panel), irrespective of the cosmology. The narrow width of the cyan shaded regions demonstrates that the inferred posteriors have only a very weak cosmology-dependence. As described in the text, the combined posterior of these 9 cosmologies serves as the prior in our cosmological inference procedure. \textit{Bottom row} shows the 90 percent confidence intervals of the predicted luminosity functions using the CLF posteriors for each of the 9 cosmologies, all computed using the halo mass function for $\Omega_\rmm = 0.30$ and $\sigma_8 = 0.82$. Each one is compared to the luminosity function predicted for the true cosmology (gray shaded region in each of these 3 panels), to demonstrate that they almost perfectly overlap.}
\label{fig:mock_param_compare}
\end{figure*}
%


\section{Test on mock data}
\label{sec:mock_test}

\begin{figure}
\centering
\includegraphics[width=0.45\textwidth]{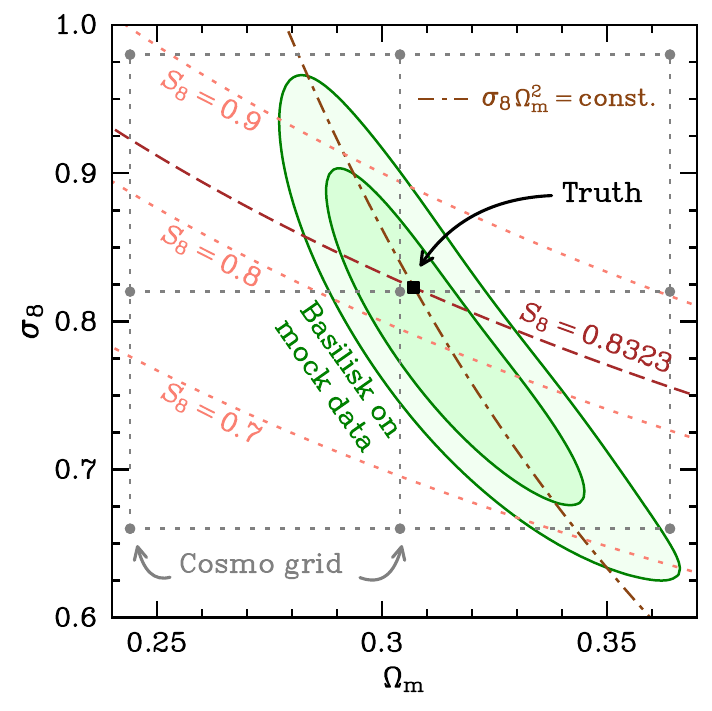}\caption{\Basilisc's cosmology inference based on mock data. The contours mark the 68 and 95 percent confidence intervals, and are in perfect agreement with the cosmology of the SMDPL simulation used to create the mock data, shown by the black square. The gray points on a $3\times 3$ grid shows the set of 9 cosmologies used to infer the combined $P(\calG)$, that acts as the prior in the cosmology inference. The brown dashed curve is the locus for constant $S_8 = 0.8323$, the true value in the SMDPL simulation. As is evident from the other loci of constant $S_8$ shown as light-brown dotten lines, the constraints on $S_8$ are of the order of $\sim 8$ (15) percent at 68 (95) percent confidence. Note, though, that \Basilisc's inference is degenerate along the direction $\sigma_8 \Omega_\rmm^2$ (brown dot-dashed curve), rather than $\sigma_8 \Omega_\rmm^{0.5}$. The accuracy with which our analysis constrains $\sigma_8 \Omega_\rmm^2$ is $\sim 6$ percent (68 percent confidence).}
\label{fig:cosmo_mock}
\end{figure}

Before applying our methodology to SDSS data, we first test it on realistic SDSS-like mock data. We use the same Tier-3 mock as used in Papers~I-III, to which we refer the reader for details. Briefly, the mock is constructed using the $z=0$ halo catalog of the high-resolution SMDPL simulation \citep[][]{Klypin.etal.16}. Each host halo in the catalog with mass $M_{\rm vir} \ge 3 \times 10^{10} \Msunh$ is populated with mock galaxies with luminosities $L \geq 10^{8.5} \Lsunh$ according to a particular fiducial CLF model. Each central galaxy is given the position and velocity of its halo core, defined as the region that encloses the innermost 10\% of the halo virial mass, while satellite galaxies are assigned the phase-space coordinates of the subhalos. Thus, the satellites do not follow a NFW profile, and have highly complex radius- and mass-dependent velocity anisotropies (see \papI). Crucially, no assumptions are made in the mock-making process about whether the satellites are virialised or relaxed massless tracers in their host halos. They only obey the Jeans equations in as much as the subhalos do in the SMDPL simulation. Once all halos have been populated with mock galaxies, we construct a mock SDSS survey that properly accounts for redshift space distortions, the SDSS flux limit and angular footprint, redshift errors, spectroscopic incompleteness, and fiber collisions. Using the resulting mock spectroscopic survey, we select primaries and secondaries using the selection cones described in Section \ref{sec:selection} excluding secondaries that are located within $55$ arcseconds of their primary. Finally, we use the mock data to compute the comoving abundances of galaxies in the ten luminosity bins described in Section \ref{sec:evidence_model} using the same method as used for the real SDSS data.

Next, we use \Basilisk to analyze the abundance and kinematics of these mock data for 9 different cosmologies that sample a $3\times 3$-grid in $(\Omega_\rmm, \sigma_8$)-space, with $\Omega_\rmm \in \{ 0.24, \, 0.30, \, 0.36\}$ and $\sigma_8 \in \{ 0.66, \, , 0.82, \, 0.98\}$. For each cosmology, we first find the best fit radial profile, $n_{\rm sat}(r|M,z)$, characterized by the parameters $\gamma$ and $\calR$ (see equation~[\ref{nsatprof}]), marginalized over all other model parameters, using a simple $\chi^2$ minimization algorithm. The best-fit parameters are $(\gamma,\calR) \simeq (0.0, 2.3)$ with no significant dependence on cosmology, indicating that $n_{\rm sat}(r)$ is cored and with a scale parameter that is roughly twice that of the dark matter. This is in good agreement with the radial distribution of subhalos in SMDPL (see \papII), indicating that \Basilisk accurately recovers the radial distribution of satellite galaxies. Next, for each cosmology we use \Basilisk to infer the full posterior for the 17 parameters that characterize the galaxy-halo connection (using broad uninformed priors for each), each time keeping $\gamma$ and $\calR$ fixed at their best-fit values for the cosmology in question. Note that since the mock is based on a dark-matter-only simulation, no baryonic corrections (as described in Section~\ref{sec:method_baryon}) are applied here.
 
The black histograms in Fig.~\ref{fig:mock_param_compare} show the inferred posteriors of the CLF parameters thus obtained for each of the nine different cosmologies, as indicated at the top of each panel. The posteriors are shown with respect to the true values of each parameter and scaled by the standard deviation of each posterior. The cyan-shaded regions show the range of variation between the $3\times 3$ cosmologies. Each of the inferred galaxy-halo connection parameters is consistent with their true values within $1 \sigma$, and with only a very weak dependence on the assumed cosmology, as evidenced by the narrow width of the cyan bands. For some parameters, notably $\sigma_{13}$ and $\sigma_\rmP$ that quantify the scatter in $L_\rmc(M)$, the cosmology dependence is slightly more pronounced, but in each case the dependence on $\Omega_\rmm$ and $\sigma_8$ is considerably smaller than the posterior width for any one cosmology, at least over the ranges of $\Omega_\rmm$ and $\sigma_8$ probed.

The posteriors being consistent with each other in the one-dimensional histograms does not necessarily prove their consistency in the multi-dimensional parameter space. However, what matters for our methodology is the consistency among their CLF posteriors in terms of their predicted luminosity function (LF) given a fixed halo mass function (HMF). To demonstrate this, the bottom row of Fig.~\ref{fig:mock_param_compare} compares the LF predictions for each of the CLF posteriors corresponding to the 9 different cosmologies, but each time using the HMF for the $(\Omega_\rmm = 0.30, \, \sigma_8 = 0.82)$ cosmology. This illustrates how much the LF varies due to changes in the CLF alone. The different panels correspond to CLFs inferred for different values of $\Omega_\rmm$, while the different hatched regions in each panel correspond to different $\sigma_8$ values. For comparison, in each panel, the gray shaded region shows the LF predicted using the CLF posterior obtained for the $(\Omega_\rmm = 0.30, \, \sigma_8 = 0.82)$ cosmology. Except for a tiny deviation of the LF posteriors in the highest luminosity bin for the CLFs inferred for cosmologies with low $\sigma_8$, the LF posteriors are perfectly consistent with each other in a statistical sense. Therefore, the CLF parameters inferred for different cosmologies are mutually consistent not only in terms of their one-dimensional distributions but also in terms of how they map to the predicted $n_{\rm gal}(L)$ distributions.

Next we combine the posteriors for all 9 cosmologies into a single posterior, $P(\calG)$, which represents the posterior for the CLF irrespective of (or `marginalized over') cosmology. This $P(\calG)$ acts as the prior for the galaxy-halo connection during the next step in which we use the galaxy LF to constrain the cosmological parameters. In particular, we use equation~(\ref{evidence_cosmo_new}) to constrain the posterior of the cosmological parameters, $P(\calC|\bD) = P(\Omega_\rmm, \sigma_8|\bD_{\rm LF})$ with $\bD_{\rm LF}$ the mock LF data and with $N_{\rm MC}$ the total number of chain elements in the combined prior $P(\calG)$.

The resulting posterior is shown in Fig.~\ref{fig:cosmo_mock}, with the shaded regions indicating the $84$ and $95$ percent confidence intervals. The solid square at $(\Omega_\rmm,\sigma_8) = (0.305,0.825)$ shows the true input cosmology used for the SMDPL simulation on which the mock is based. For comparison, the gray dotted lines indicate the $3 \times 3$ grid of $(\Omega_\rmm,\sigma_8)$ used to constrain the galaxy-halo connection with \Basilisc, while the red dotted and dashed curves indicate contours of constant $S_8$, as indicated. Note that the cosmological inference is in excellent agreement with the input cosmology, indicating that our methodology yields unbiased constraints on $\Omega_\rmm$ and $\sigma_8$. 

The constraints display a significant degeneracy in the direction of constant $\sigma_8 \Omega_\rmm^2$. This differs significantly from the direction of constant $S_8 \propto \sigma_8 \Omega_\rmm^{0.5}$, which reflects the typical direction of degeneracy that arises when using cluster abundances to constrain cosmology \citep[e.g.,][]{Mantz.etal.2015, Planck.16.ClusterCosmo, Bocquet.etal.2019, Abbott.etal.2020_CC}. This difference is due to the fact that our method is sensitive to the halo mass function over the entire mass range $M_\rmh \gta 10^{12} \Msunh$, while cluster abundance measurements typically only probe the high-mass end ($M_\rmh \gta 10^{14} \Msunh$).

\begin{figure}
\centering
\includegraphics[width=0.5\textwidth]{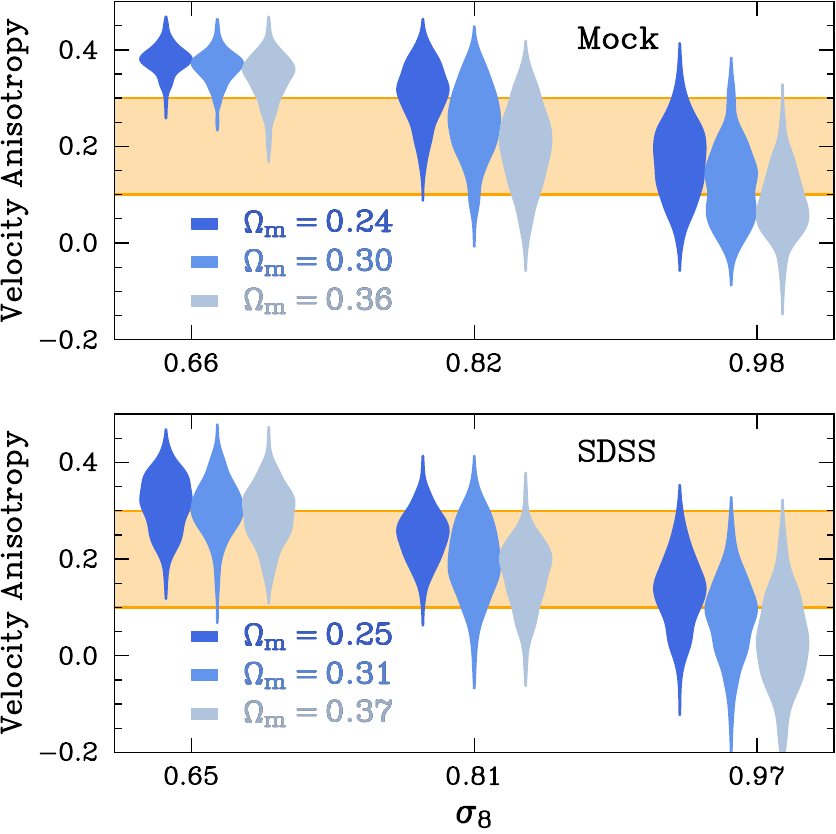}
\caption{Violin plots showing the posterior distributions of the inferred satellite velocity anisotropy, $\beta$, for the different cosmologies of the $3 \times 3$ grid of $(\Omega_\rmm, \sigma_8)$-parameter space used in our analysis. Results are shown for both the mock data (top panel) and the SDSS DR7 data (bottom panel). The primary trend is with respect to $\sigma_8$, shown along the horizontal axis, while different shades of blue corrrespond to different values of $\Omega_\rmm$, as indicated. The orange band in each panel envelopes the range of velocity anisotropies found for subhalos in the SMDPL simulation at the typical halo-centric radii probed by \Basilisc. Note that the posterior constraints on $\beta$ are in excellent agreement with the latter for the \Planc cosmology $(\Omega_\rmm=0.31, \sigma_8=0.81)$.}
\label{fig:Anisotropy_cosmology}
\end{figure}

It is interesting to point out that, unlike all CLF parameters, the inferred velocity anisotropy parameter, $\beta$, shows a weak but significant dependence on cosmology. The top panel of Fig.~\ref{fig:Anisotropy_cosmology} shows the posterior constraints of $\beta$ for different combinations of $\Omega_\rmm$ and $\sigma_8$. As is apparent, the anisotropy is inferred to be larger (more radial orbits) for smaller $\sigma_8$ and with a very weak dependence on $\Omega_\rmm$. Since $\beta$ has no impact on the predicted galaxy LF, this cosmology dependence of the inferred satellite velocity anisotropy is of no significance in our cosmological analysis. It is encouraging, though, that for the input cosmology ($\Omega_\rmm = 0.30$, $\sigma_8 = 0.82$) of the simulation used to construct the mock data, the constraints on $\beta$ are in excellent agreement with the typical velocity anisotropy of dark matter subhalos in cosmological simulations, shown as the orange band. Since, by construction, the mock satellite galaxies have the same orbital anisotropy as the dark matter subhalos\footnote{Mock satellites are positioned inside subhalos, and therefore share their phase-space coordinates.}, this indicates that \Basilisk yields unbiased constraints on the orbital anisotropy of the population of satellite galaxies (see also \papII).
%

\section{Results: Cosmological Constraints}
\label{sec:results}

\begin{figure}
\centering
\includegraphics[width=0.5\textwidth]{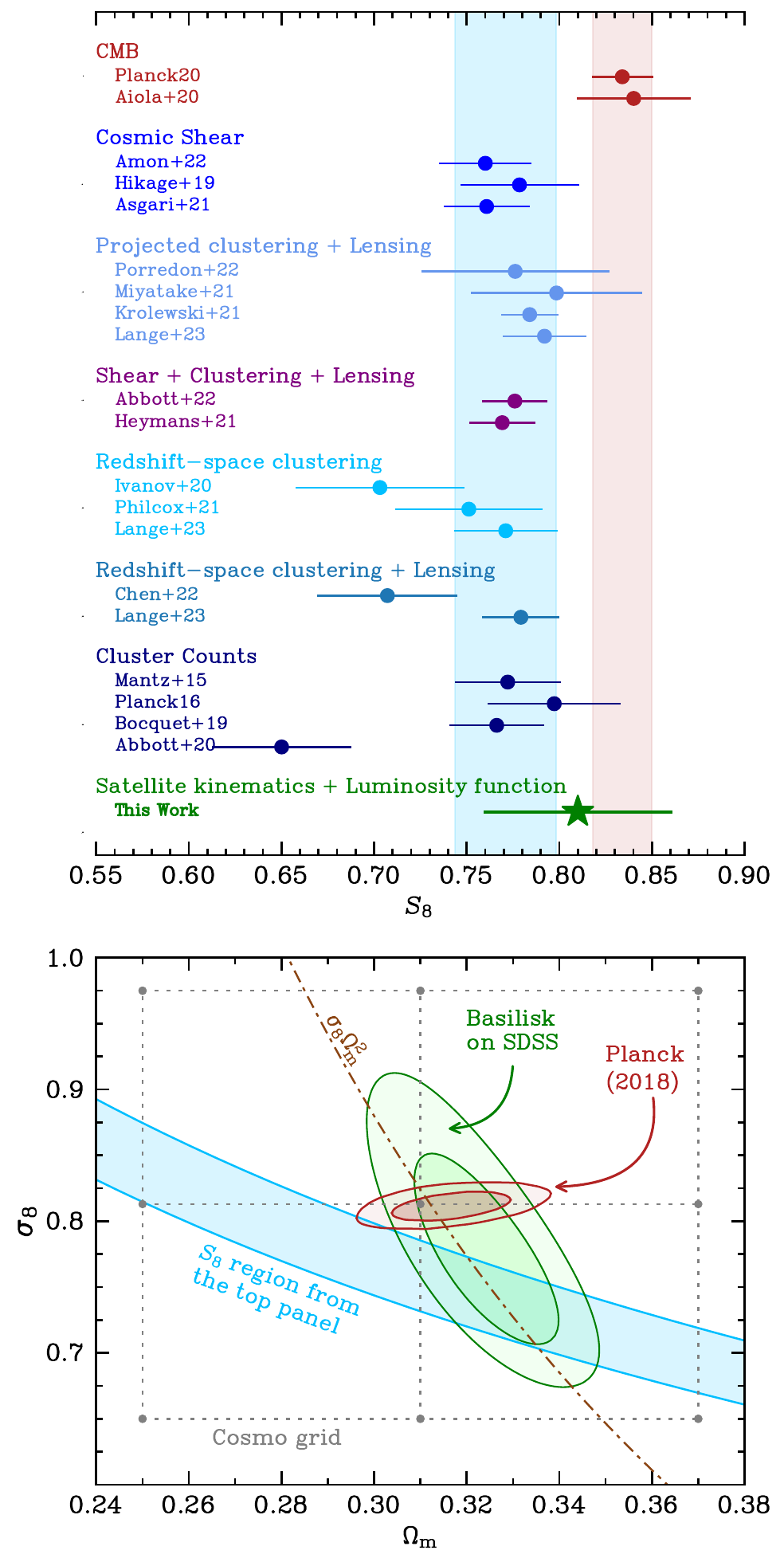}
\caption{Cosmology constraints obtained here using the SDSS-DR7 analysis with \Basilisk compared to the inference from \Planck and various low-redshift probes. {\it Top panel:} the inference in terms of $S_8 = \sigma_8 \sqrt{\Omega_\rmm/0.3}$. The points in different shades of blue/violet show a wide range of low-$z$ analyses, adapted from Fig.~12 of \citet{Lange.etal.23} (errorbars reflect the corresponding $1 \sigma$ confidence intervals). The blue shaded region corresponds to the $1\sigma$ confidence interval of the redshift-space clustering analysis by \citet{Lange.etal.23} and is used as a characteristic constraint for the low-$z$ probes. For comparison, the red shaded region marks the $1 \sigma$ confidence interval from the \Planc CMB data \citep[TT+TE+EE+lowE from][]{Planck.18}.  \Basilisc's inference is shown by the green star at the bottom of the top panel, showing perfect agreement with \Planck. {\it Bottom panel:} the inference in the $(\Omega_\rmm, \, \sigma_8)$-plane. The blue shaded region corresponds to the range of $S_8$ values indicated by the blue band in the top panel, which reflects the weak but systematic $S_8$ tension with respect to the \Planc CMB constraints shown in red. However, \Basilisc's constraints are perfectly consistent with \Planc.}
\label{fig:Cosmo_Constraint}
\end{figure}

We now apply the method described and validated above to the same SDSS data as used in \papII. More specifically, the data derives from the \texttt{bright0} sample of the New York University Value-Added Galaxy Catalog \citep[VAGC;][]{Blanton.etal.05}, constructed using the Seventh Data Release of the SDSS \citep[SDSS DR7;][]{Abazajian.etal.09}. This sample includes $\sim 570,000$ galaxies with a limiting Petrosian magnitude of $m_r < 17.6$. Using the selection criteria outlined in Sections~\ref{sec:selection} and~\ref{sec:datasample}, we obtain the data shown in Fig.~\ref{fig:data}, which we use to constrain the galaxy-halo connection. Similar to the approach used when analyzing the mock data, we first run \Basilisk on a $3\times 3$ grid of $(\Omega_\rmm,\sigma_8)$. Note, though, that this time we apply the baryonic corrections described in Section~\ref{sec:method_baryon} using the fiducial feedback model based on the \EAGLE simulation. Specifically, we multiply the line-of-sight velocity dispersions predicted by \Basilisk with the mass-dependent ratio $\overline{\sigma}_{\rm hydro}/\overline{\sigma}_{\rm DMO}$ shown in the bottom-right panel of Fig.~\ref{fig:baryon_model} (see Appendix~\ref{App:AppA} for details).

For each set of cosmological parameters we first obtain the best fit $\{\gamma, \calR\}$. In each case, the best-fit satellite radial profile resembles a cuspy NFW-like distribution with $\gamma \simeq 1$, with a scale factor ratio $\calR$ ranging between 2 and 2.5, with only a very weak dependence on cosmology. Next, for each cosmology we keep $\gamma$ and $\calR$ fixed at their best-fit values, and run \Basilisk to constrain the full posterior distribution for the 17 free parameters that characterize the galaxy-halo connection, the orbital anisotropy, and the interloper abundance. As in the case of the mock analysis, we find the inference about the galaxy-halo connection to be virtually independent of cosmology, with all 9 CLF posteriors being consistent with each other within their $1\sigma$ extent. Similar to the results shown in Fig.~\ref{fig:mock_param_compare} for the mock data, $\gamma_2$, $\sigma_{13}$, and $\sigma_\rmp$ reveal a weak trend of slightly higher values in the case of a cosmology with higher $\sigma_8$ (not shown), but the differences are small compared to the posterior uncertainties on these parameters for a given cosmology. 

For the final step, we use the combined posterior, $P(\calG)$, to marginalise the likelihood of the SDSS LF data over the uncertainties in the galaxy-halo connection.  The green contours in the bottom panel of Fig.~\ref{fig:Cosmo_Constraint} show the resulting 68 and 95 percent confidence intervals in the  $(\Omega_\rmm,\, \sigma_8)$-plane, while the green star in the top panel indicates the best-fit value for $S_8$, with the errorbar indicating the corresponding $68\%$ confidence interval. For comparison, we also indicate the \Planc TT+TE+EE+lowE results \citep{Planck.18}. In the $S_8$ comparison (top panel), in addition to CMB-based results \citep{Planck.18, Aiola.etal.20}, we also include a collection of low-$z$ results from the literature, adapted from Fig.~12 of \citet{Lange.etal.23}. These constraints are based on a range of complementary techniques and observables applied to a wide variety of data sets, such as cosmic shear \citep{Amon.etal.22, Hikage.etal.2019, Asgari.etal.2021}, projected clustering and lensing \citep{Porredon.etal.2022, Miyatake.etal.2022b, Krolewski.etal.2021, Lange.etal.23}, $3 \times 2$ point analyses \citep{Abbott.etal.2022, Heymans.etal.2021}, redshift-space distortion (RSD) \citep{Ivanov.etal.2020, Philcox.etal.2022, Lange.etal.23}, joint analyses of RSD and lensing \citep{Chen.etal.2022, Lange.etal.23}, and galaxy cluster counts \citep{Mantz.etal.2015, Planck.16.ClusterCosmo, Bocquet.etal.2019, Abbott.etal.2020_CC}. These are representative of the typical constraints inferred from low-$z$ data, revealing a clear tension with the \Planc results. Our \Basilisc-based constraints, on the other hand, are in good agreement with \Planck, with no indication of an $S_8$ tension. Note that the errorbar on $S_8$ inferred from our analysis is somewhat larger compared to that of most other studies. This is partially a consequence of the fact that our results more tightly constrain the combination $\sigma_8 \Omega^2_\rmm$, rather than $\sigma_8 \Omega_\rmm^{0.5}$ (see bottom panel).
\begin{figure}
\includegraphics[width=0.49\textwidth, right]{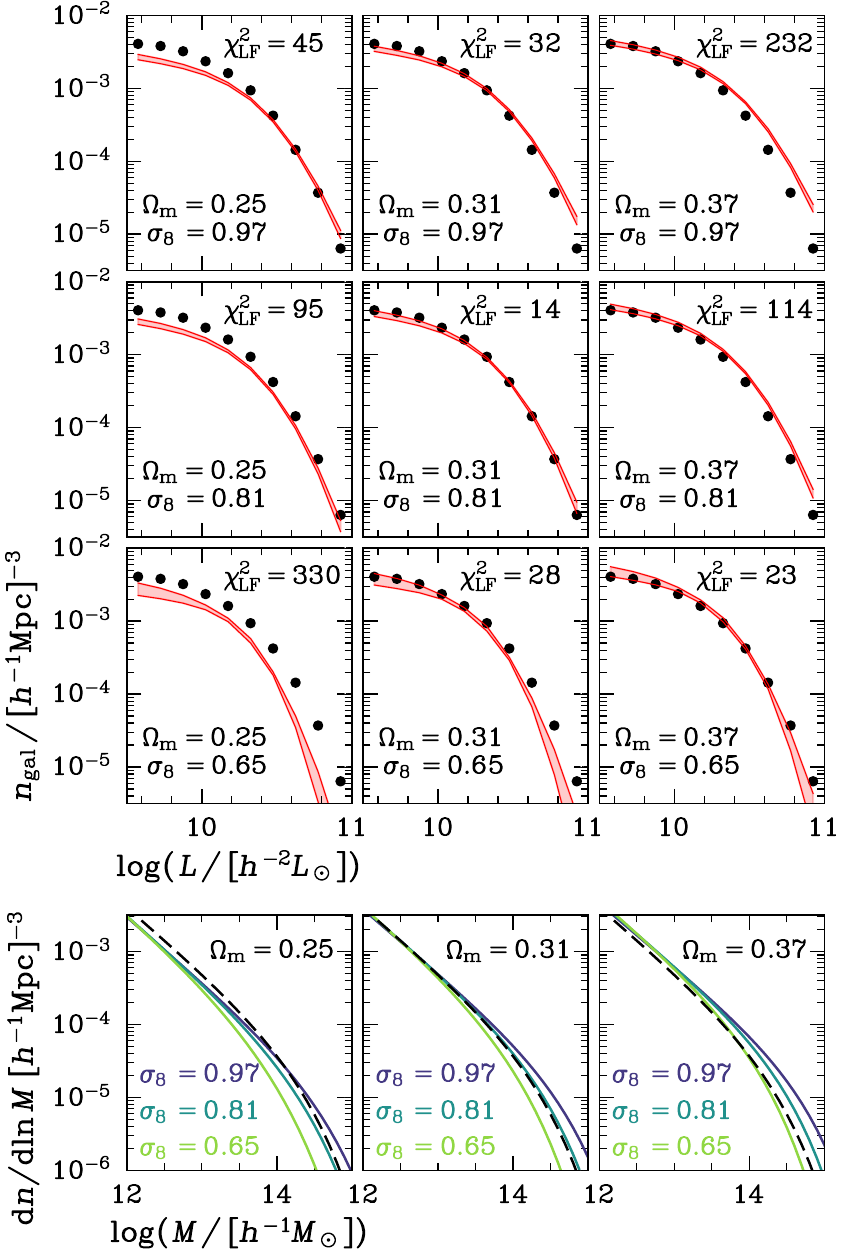}
\caption{The top $3\times 3$ panels compare the SDSS luminosity function (black solid dots) to the predicted luminosity function (red shaded bands, indicating the 68\% confidence intervals), for 9 different cosmologies as indicated. The $\chi^2$ value (equation~[\ref{chi2LF}]) for the best-fit model prediction is indicated in the top-right corner of each panel. The model prediction is in good agreement with the SDSS data for the cosmology with $\Omega_\rmm = 0.31$ and $\sigma_8 = 0.81$ that is consistent with \Planck. For other cosmologies, the fits to the data are significantly worse, especially for the low-$\Omega_\rmm$, low$-\sigma_8$ (bottom-left) and high-$\Omega_\rmm$, high-$\sigma_8$ (top-right) cases. The set of 3 panels in the bottom row shows the \citet{Tinker.etal.08} halo mass functions for the same $9$ cosmologies, making it evident that the predicted luminosity function acts as a proxy for these halo mass functions. The dashed curve in these of these panels shows the mass function for the \Planc best-fit cosmology ($\Omega_\rmm = 0.3158$, $\sigma_8=0.8120$).}
\label{fig:LF_compare}
\end{figure}

Fig~\ref{fig:LF_compare} compares the observed SDSS LF (black dots) to the LF predicted based on \Basilisc's inferred constraints on the galaxy-halo connection (red shaded region, indicating the 68 percent confidence interval). The results are shown for the same $3 \times 3$ cosmologies as used in the analysis of the satellite kinematics and abundance, as indicated. For completeness, we also indicate the $\chi^2$ value (equation~[\ref{chi2LF}]) of the best-fit luminosity function, for each cosmology, in each of the respective panels. As is evident, for $\Omega_\rmm = 0.31$ and $\sigma_8 = 0.81$ (middle of the $3\times 3$ panels), which are the values that are consistent with \Planck, the predicted LF is in nearly perfect agreement with the data. For the other cosmologies, the predicted LF either over- or underestimates the observed abundance of the bright and/or faint galaxies. For comparison, the three separate panels at the bottom of Fig.~\ref{fig:LF_compare} show the $z=0.1$ \citet{Tinker.etal.08} halo mass functions (HMF) for these 9 cosmologies. Here, the black dashed lines show the HMF for the best-fit \Planc cosmology, and is shown for comparison. Note how the HMFs reveal exactly the same trends as the predicted LFs shown in the $3\times3$ top panels. This essentially conveys the crux of our methodology: \Basilisc's precise and accurate galaxy-halo connection inference enables us to use the LF as a proxy for the HMF over a wide range of halo mass.

The bottom panel of Fig.~\ref{fig:Anisotropy_cosmology} shows the inferred posterior distributions for the orbital anisotropy parameter $\beta$ for each of the $3\times 3$ cosmologies used to probe the galaxy-halo connection. As in the case of the mock data, the inferred value of $\beta$ decreases with $\sigma_8$ and with a very weak dependence on $\Omega_\rmm$. Interestingly, the inferred anisotropy parameter is in excellent agreement with the typical orbital anisotropy of dark matter subhalos (indicated by the orange band) for the \Planc cosmology. We consider this to be an independent support of the \Planc cosmology. After all, according to the $\Lambda$CDM paradigm of structure formation, satellite galaxies reside in dark matter subhalos, and thus should have their orbital characteristics.  

For completeness, Table~1 lists the halo-occupation parameters used in our cosmological inference, that is, we list the ranges of the {\it combined} posterior, which is simply the sum of the posteriors for the $3 \times 3$ cosmologies used. 
\begin{table}
\renewcommand{\arraystretch}{1.2}
\caption{Summary of all results: the best-fit values and $1\sigma$ confidence intervals for each of the cosmological parameters (see \S\ref{sec:results}) as well as the the galaxy-halo connection parameters (see \S\ref{sec:ghc} for details), as inferred by \Basilisk from the SDSS DR7 data.}
\smallskip
\begin{tabular}{l c c c}
\hline
  $\,$        & parameter & best-fit & $1\sigma$ interval \\ 
\hline 
Cosmology     & $\Omega_\rmm$ & 0.324 & [0.312, 0.336] \\  
              & $\sigma_8$    & 0.775 & [0.717, 0.842] \\
              & $S_8$         & 0.81 & [0.76, 0.86] \\
              & $\sigma_8 (\Omega_\rmm/0.3)^2$ & 0.91 & [0.86, 0.95]  \smallskip \\ 
\cdashline{1-4} 
Central CLF   & $\log M_1$    & 10.88 & [10.74, 11.02] \\  
              & $\log L_0$    & 9.67 & [9.56, 9.78] \\
              & $\gamma_1$    &  4.09 & [3.36, 4.81]\\
              & $\gamma_2$    & 0.29 & [0.25, 0.33] \\
              & $\sigma_{13}$ & 0.21 & [0.20, 0.22] \\
              & $\sigma_\rmP$ & -0.03 & [-0.04, -0.02] \smallskip \\
\cdashline{1-4}
Satellite CLF & $\alpha_{13}$ & -1.60 & [-1.66, -1.54] \\  
              & $\alpha_\rmP$ &  0.06 & [-0.09, 0.21] \\
              & $\Delta_{13}$ & 0.15 & [0.12, 0.18] \\
              & $\Delta_\rmP$ &  -0.27 & [-0.38, -0.17] \\
              & $b_0$         & -1.57 & [-1.74, -1.40] \\
              & $b_1$         &  1.22 & [1.06, 1.39] \\
              & $b_2$         & -0.07 & [-0.08, -0.05] \smallskip \\
\cdashline{1-4} 
Velocity anisotropy     & $\beta$          & 0.20  & [0.12, 0.27]
\smallskip
\\
\hline
\end{tabular}
\label{table:SDSS_params}
\end{table}

\section{Discussion}
\label{sec:discussion}

In the following subsections, we characterise the effect of the dominant sources of uncertainty in our modeling. This is crucial to evaluate the robustness of our results and point to systematics that will require careful mitigation when dealing with larger datasets from upcoming and future missions.

\subsection{Uncertainties related to baryonic effects}
\label{sec:subsec:varying_baryon}

\begin{figure*}
\centering
\includegraphics[width=\textwidth]{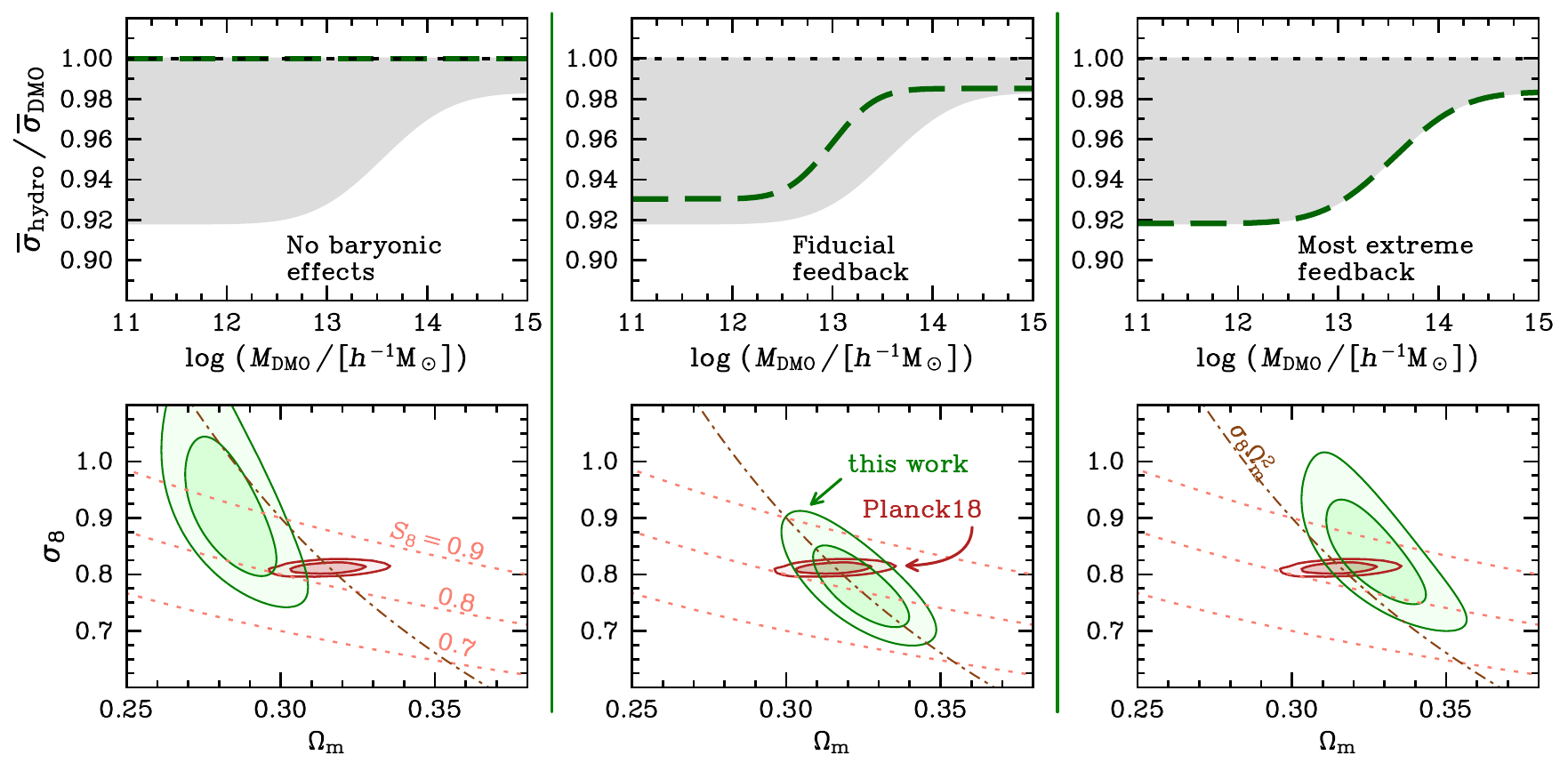}
\caption{Characterizing the effects of baryons on the cosmology inference. The panels on the top show the $M_{\rm hydro}/M_{\rm DMO}$ with a dashed line, and shows the whole range of models tested with a gray shaded region. The ones on the bottom show the corresponding inferred cosmological parameters, and \Planck's inference for comparison. The middle column corresponds to the fiducial feedback model, as obtained by \citet{Baggen.etal.25}. The left column represents the dark matter-only assumption in \texttt{Basilisk}'s modeling, while the column on the right shows the most extreme feedback model tested in this work. All other models fall between the most extreme and the no-feedback scenarios. For all intermediate models, and for the fiducial feedback model used in this work, the inferred cosmology is in perfect agreement with \Planck, while the no-feedback model shows a very weak hint of tension.}
\label{fig:varying_baryon}
\end{figure*}
\begin{figure}
\centering
\includegraphics[width=0.48\textwidth]{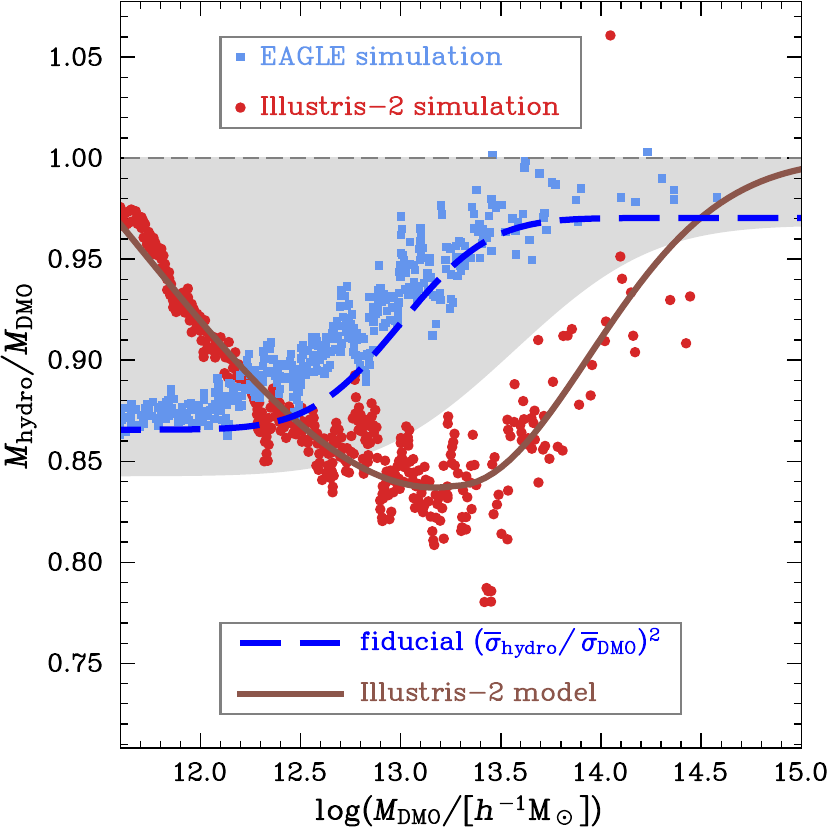}
\caption{The data points, taken from \citet{Beltz-Mohrmann.etal.2020}, show the ratio of virial masses in hydrodynamic simulations to that of the matched halos in dark matter-only (DMO) runs, as a function of $M_{\rm DMO}$, for the \EAGLE (blue squares) and the Illustris-2 (red circles) galaxy formation models. The ratio is mostly below unity due to the mass loss from within the virial radius due to AGN and supernova feedback. However, the mass-dependence of this ratio is very different between the two galaxy formation simulations. The blue dashed line corresponds to $(\overline{\sigma}_{\rm hydro}/\overline{\sigma}_{\rm DMO})^2$ as a function of $M_{\rm DMO}$ based on the detailed model by \citet{Baggen.etal.25} for the baryonic effects on satellite velocity dispersion in the \EAGLE simulation. The fact that it closely follows the blue squares, implies that simply using this mass ratio is enough to capture most of the effects inferred using the more sophisticated modeling. Validated by this agreement, we use a simple model for this mass ratio in Illustris-2 (brown solid line) and modify \Basilisk accordingly to incorporate baryonic effects mimicing that in Illustris-2.}
\label{fig:illustris2_model}
\end{figure}

\begin{figure}
\centering
\includegraphics[width=0.46\textwidth]{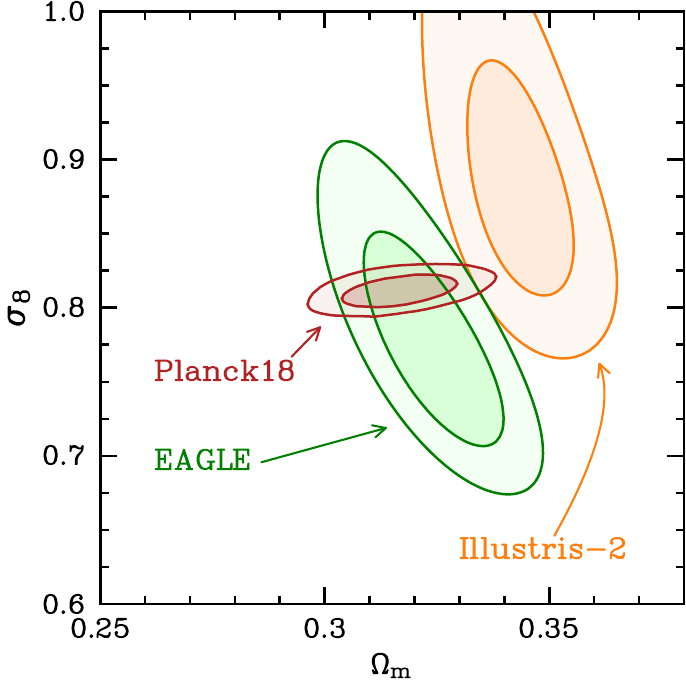}\caption{Demonstrating the power of \Basilisk to constrain subgrid prescription of baryonic physics in hydrodynamic simulations. Upon modeling the altered mass distribution in Illustris-2, compared to DMO runs, \Basilisk yields particularly high values of $S_8$, inconsistent with \Planck. The difference with respect to \EAGLE-based baryonic modeling is statistically significant, and by assuming \Planc inference to be the truth, one can already rule out Illustris-2 feedback prescription using \Basilisk on SDSS data.}
\label{fig:constraining_feedback}
\end{figure}

In the analysis of the SDSS data described in Section~\ref{sec:results} we account for the way baryonic processes related to galaxy formation impact the gravitational potential of the dark matter halo in which the satellites orbit. In particular, we multiply the line-of-sight velocity dispersion of satellite galaxies predicted by \Basilisc, under the assumption that the halo is made entirely of dark matter following an NFW profile, with a halo-mass dependent ratio $\overline{\sigma}_{\rm hydro}/\overline{\sigma}_{\rm DMO}$ computed using the analytical model of B25. As discussed in Section~\ref{sec:method_baryon}, to good approximation this ratio is given by equation~(\ref{eqn:feject}),  which is only a function of $f_{\rm eject}$, the fraction of baryonic mass originally associated with the halo that, due to various feedback processes, has been ejected outside of the halo virial radius. For our fiducial model, the halo-mass dependence of $f_{\rm eject}$ is calibrated against the \EAGLE simulation, which is depicted as the solid line in the top right panel of Fig.~\ref{fig:baryon_model}. However, since the feedback processes responsible for the ejection of baryonic matter in the simulations are not properly resolved, but based on uncertain subgrid models, there are large uncertainties associated with $f_{\rm eject}(M)$, and thus with the `baryonic correction' applied in \Basilisk.

In order to examine how this uncertainty affects our cosmological inference, we proceed as follows. The mass dependence of $\overline{\sigma}_{\rm hydro}/\overline{\sigma}_{\rm DMO}$ measured for the \EAGLE simulation by B25, shown in the lower right panel of  Fig.~\ref{fig:baryon_model}, can be approximated as
\begin{equation}\label{sigma_ratio_fit}
 \dfrac{\overline{\sigma}_{\rm hydro}}{\overline{\sigma}_{\rm DMO}} = A_{\rmb 1} +\dfrac{A_{\rmb 2}-A_{\rmb 1}}{2} \Bigg[ 1 + {\rm erf}\bigg(\dfrac{\log M - \log M_\rmb}{\sigma_\rmb}\bigg)\Bigg]\,,
\end{equation}
which transitions from $\overline{\sigma}_{\rm hydro}/\overline{\sigma}_{\rm DMO} = A_{\rm b1}$ at $M \ll M_\rmb$ to $\overline{\sigma}_{\rm hydro}/\overline{\sigma}_{\rm DMO} = A_{\rm b2}$ at $M \gg M_\rmb$, with $\sigma_\rmb$ controlling the steepness of the transition. For our fiducial model, calibrated against the \EAGLE simulation, $A_{\rmb 1} = 0.93$, $A_{\rmb 1} = 0.985$, $\sigma_\rmb = 0.494$ and $\log M_\rmb = 13 \Msunh$.  We now repeat the entire analysis using different combinations of $A_\rmb$, $\sigma_\rmb$, and $\log M_\rmb$ covering the ranges $A_{\rmb 1} \in [1.0,0.92]$, $A_{\rmb 2} \in [1.0,0.98]$, $\sigma_\rmb \in [0.4,0.8]$, and $\log M_\rmb \in [12.0, 13.7]$. This corresponds to models in which $f_{\rm eject}$ in $10^{12} \Msunh$ halos ranges from $0.0$ (no feedback) to $0.77$ (maximum feedback). The maximum feedback scenario corresponds to the case where {\it all baryons} have been ejected from the low-mass halos, except for the stars that make up the central galaxy (i.e., $M_{\rm gas} = M_{\rm ICL} = 0$ in equation~[\ref{eqn:feject}]). In this limit, $f_{\rm eject} = 1- M_{\rm cen}/(f_{\rm bar} M_{\rm DMO}) \approx 0.77$, where we assume that $M_{\rm cen}$ follows the standard stellar mass -- halo mass relation of \citet{Moster.etal.10}. Following B25, we assume that $f_{\rm eject}$ gradually tends to zero in this extreme feedback model as the halo mass approaches $10^{15} \Msunh$. For comparison, our fiducial model has $f_{\rm eject} = 0.66$ at the low-mass end with a sharper decline in $f_{\rm eject}$ as a function of halo mass, reaching $f_{\rm eject}\approx 0$ at $10^{14} \Msunh$ (see Fig.~\ref{fig:baryon_model}). 

The upper panels of Fig.~\ref{fig:varying_baryon} show three different models for $\overline{\sigma}_{\rm hydro}/\overline{\sigma}_{\rm DMO}$, indicated by the green-dashed lines. From left to right, these correspond to the minimal (no) feedback model, our fiducial model, and the extreme feedback model.  We emphasize that these models include the effect of the adiabatic response of the halo to both the formation of the central galaxy and the ejection of gas, if any (see B25 for details). The gray-shaded regions outline the full range of variation of $\overline{\sigma}_{\rm hydro}/\overline{\sigma}_{\rm DMO}$ probed by these models. The bottom panels show the corresponding cosmological inference compared to that of \Planc. As is apparent, the detailed corrections for baryonic effects have a non-negligible impact. Generally, stronger feedback implies a larger best-fit value for $\Omega_\rmm$, while inference for $\sigma_8$ reveals a nonmonotonic dependence on feedback strength.
Most importantly, though, even for these extreme models considered here, the constraints on $\Omega_\rmm$ and $\sigma_8$ always remain consistent with \Planc $< 2 \sigma$. Hence, we conclude that our main result, that our method to constrain cosmological parameters reveals no $S_8$ tension, is robust to uncertainties arising from poorly constrained baryonic feedback processes.

However, we emphasize that this assertion is based on the notion that the gray-shaded region in the top panels of Fig.~\ref{fig:varying_baryon} is representative of the uncertainties in $f_{\rm eject}(M_\rmh)$. At the low mass end ($M_\rmh \lta 10^{13} \Msunh$), this covers the entire range of gas mass fraction $f_{\rm gas}=0$ to $f_{\rm gas}=1$ (corresponding to $f_{\rm eject} = 0.77$ to $0$, respectively), which is clearly a conservative choice. However, at the massive end ($M_\rmh \gta 10^{14.5}\Msunh$) we have assumed that, in agreement with the \EAGLE simulation, $f_{\rm eject} \to 0$. Although there is observational evidence to support that massive halos indeed hang on to their entire baryonic budget \citep[e.g.,][]{Vikhlinin.etal.2006, Sun.etal.2009, Pratt.etal.2009, Lin.etal.2012, Lovisari.etal.2015}, we acknowledge that there are also examples of cosmological hydrodynamical simulations that make very different predictions. An example is the Illustris-2 simulation \citep[][]{Genel.etal.2014, Vogelsberger.etal.14, Nelson.etal.2015, Pillepich.etal.2015}, which incorporates a subgrid model for AGN feedback that results in the expulsion of large fractions of baryonic material even at $10^{14} \Msunh$. Fig.~\ref{fig:illustris2_model} compares the mass ratios $M_{\rm hydro}/M_{\rm DMO}$ as a function of $M_{\rm DMO}$ in the \EAGLE simulation to those in Illustris-2. The gray shaded region is the same as that shown in the top panels of Fig.~\ref{fig:varying_baryon}, where we have converted between $\sigma_{\rm hydro}/\sigma_{\rm DMO}$ and $M_{\rm hydro}/M_{\rm DMO}$ using equation~(\ref{barcorr}) and the fact that $M_{\rm hydro} = (1 - f_{\rm bar}) M_{\rm DMO} + f_{\rm bar} f_{\rm eject} M_{\rm DMO}$. As is apparent, the $f_{\rm eject}(M_\rmh)$ in Illustris-2 is very different from that in \EAGLE, and the massive halos in the former simulation clearly fall outside of the gray region. Although it is generally accepted that the AGN feedback model adopted in the Illustris-2 simulation is inconsistent with observations \citep[e.g.,][]{Giodini.etal.2009, Lovisari.etal.2015}, as pointed out in \citet{Weinberger.etal.2017} and \citet{Pillepich.etal.2018a}, it is important to remain open-minded about the halo mass dependence of $f_{\rm eject}$ and its potential impact on cosmological inference. Moreover, the original Illustris simulation has recently gained prominence due to its better agreement with the data on the kinetic Sunyaev Zel'dovich effect \citep[see][]{Hadzhiyska.etal.2024, McCarthy.etal.2024}. As a result, it is both interesting and consequential to test how a baryonic model based on the original Illustris simulation translates to \Basilisc's cosmology constraints using satellite kinematics.

Alternatively, one can use the fact that the cosmological inference depends on $f_{\rm eject}(M_\rmh)$ to constrain galaxy formation physics. In particular, one can invert the problem and seek to constrain feedback models for a given cosmology. For example, by assuming the \Planc cosmology as the truth, one can constrain subgrid prescriptions of feedback processes and rule out those that yield results that are irreconcilable with the data. As an illustration of the potential power of such an approach, we repeat our analysis of the abundance and kinematics of satellite galaxies in the SDSS using a baryonic feedback model in agreement with the Illustris-2 simulation. In particular, we change the halo mass dependence of $\sigma_{\rm hydro}/\sigma_{\rm DMO}$ which is consistent with the \EAGLE simulation with one that is consistent with Illustris-2. We do so using the relation between $M_{\rm hydro}$ and $M_{\rm DMO}$ depicted by the thick maroon line in Fig.~\ref{fig:illustris2_model}, and converting it to the corresponding $\sigma_{\rm hydro}/\sigma_{\rm DMO}$ using equation~(\ref{barcorr}). The resulting cosmological constraints are indicated by the orange contours in Fig.~\ref{fig:constraining_feedback}, which is in clear tension with \Planck. Interestingly, this tension is in the opposite direction compared to the $S_8$ tension arising from most low-$z$ probes; i.e., adopting the Illustris-2 feedback model yields values for the $S_8$ parameter that are too high compared to \Planck. Hence, taking the \Planc cosmology for granted, we can claim that satellite kinematics data disfavor the subgrid treatment of AGN feedback in Illustris-2.
\begin{figure}
\centering
\includegraphics[width=0.49\textwidth]{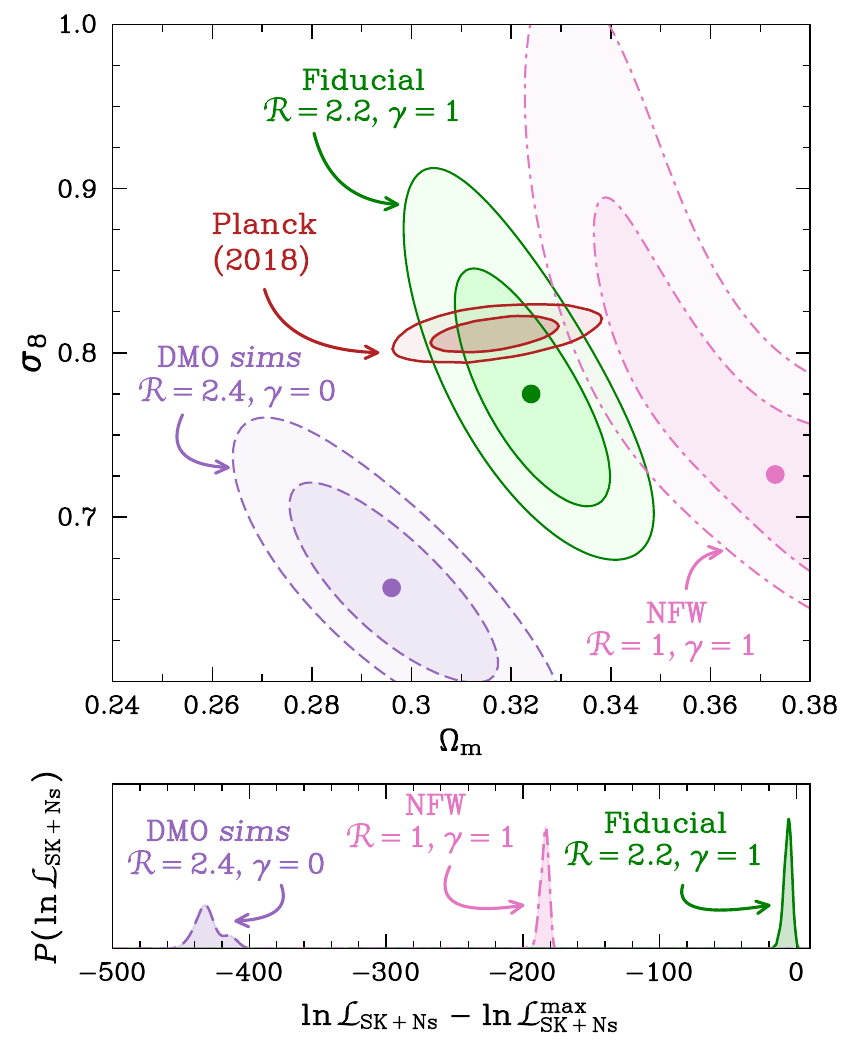}
\caption{
\textit{Top panel:} The impact of the assumed satellite radial profile, $n_{\rm sat}(r|M)$, on the cosmological inference. The green contours correspond to the fiducial model, for which $n_{\rm sat}(r|M)$ is self-consistently inferred from the data. The pink and purple contours show the inference obtained assuming that the radial profile of satellites follows that of the dark matter and that of subhalos in the SMDPL simulation, respectively.
\textit{Bottom panel:} The distributions of $\ln \calL_{\rm SK+Ns} \equiv \ln \calL_{\rm SK} + \ln \calL_{\rm Ns}$ corresponding to the best-fit cosmologies (solid dots) in the top panel (matched in color). Evidently, assuming that $n_{\rm sat}(r|M)$ follows that of the dark matter or that of dark matter subhalos in numerical simulations is clearly disfavoured by the data.
}

\label{fig:calR_cosmology}
\end{figure}

\subsection{Satellite radial profile uncertainty}
\label{App:calR_cosmology}

The cosmological constraints inferred above are obtained by marginalizing over the galaxy-halo connection, which in turn is obtained for the best-fit satellite radial profile at each of the cosmology grid points. For all $(\Omega_\rmm, \sigma_8)$ consistent with our cosmology inference we find that the satellite radial profiles for the SDSS data are consistent with a cuspy inner slope ($\gamma=1$) and with a scale factor that is roughly $\calR = 2.2$ times larger than that of dark matter in each corresponding halo. First fitting for the radial distribution in a cosmology-agnostic way ensures that our results for both the $n_{\rm sat}(r)$ profile and the cosmology are self-consistent.

This approach is starkly different from what is usually done in cosmological analyses. A priori assumption is often made about the radial distribution of satellite galaxies, which acts as a hidden prior in the cosmological inference, be it using real-space or redshift-space galaxy clustering, or weak gravitational lensing. Traditionally, cosmological analyses based on halo occupation modeling make one of two assumptions. Either it is assumed that satellites are unbiased tracers of the underlying dark matter, that is, they follow the same NFW radial profile as the host halo itself, or one proceeds by placing the satellites in the subhalos of a dark matter-only simulation, in which case the radial profile is inherently assumed to be that of the subhalos. However, we emphasize that there are also examples of studies that adopted a more flexible approach, assuming that satellites follow a NFW profile but with a concentration parameter that is different from that of the dark matter \citep[][]{Cacciato.etal.13, Zhai.etal.2023a, Zhai.etal.2023b, Lange.etal.23}.

In this subsection, we evaluate the effect of those two traditional assumptions on our cosmological inference. To that end, we rerun our entire analysis pipeline with either of the two common assumptions: $(\mathcal{R},\gamma)=(1,1)$, which is the NFW satellite profile mimicking the underlying dark matter, and $(\mathcal{R},\gamma)=(2.4,0)$, which is the best-fit radial profile of subhalos in the SMDPL simulation over the observed range of radii (see \papI). The corresponding cosmology inferences are shown in the top panel of Fig.~\ref{fig:calR_cosmology}. For the NFW assumption, the constraints are shown by the pink contours, which are in $\sim 2 \sigma$ tension with \Planck. Interestingly, the direction of the disagreement is opposite to that of the usual growth tension in the literature. For the subhalo-like distribution of satellites, with a cored and extended radial profile, the cosmology constraints are shown with the purple contours in the same panel. It has a significantly lower $\Omega_{\rmm}$ and $\sigma_8$ compared to \Planc and is highly discrepant (at $\sim 5 \sigma$). Thus, if we had naively assumed that the satellites follow the subhalos in the SMDPL simulation (found by \texttt{Rockstar}) we would have inferred a $S_8$ tension with a $\sim 5 \sigma$ confidence.

Interestingly, the different choices for the $n_{\rm sat}(r)$ models yield equally good fits to the luminosity function. Hence, in terms of $\chi^2_{\rm LF}$ we cannot distinguish between these models. However, the different choices on the radial profiles do make a difference in the satellite kinematics$+$abundance fits by \Basilisc. For each radial profile and for its corresponding best-fit cosmology, we rerun \Basilisk and use the combined likelihood $\calL_{\rm SK+Ns} \equiv \calL_{\rm SK}+\calL_{\rm Ns}$ as a goodness-of-fit estimator to see how well each model reproduces the phase-space distribution of satellites. Note that each model has identical parametrization with the same number of free parameters. Therefore, contrasting the likelihoods against each other is a like-for-like comparison. To compare the results for the different $n_{\rm sat}(r)$ in an absolute sense, the bottom panel of Fig.~\ref{fig:calR_cosmology}) shows the distributions of the combined log likelihood $(\ln \calL_{\rm SK+Ns})$ for the posteriors corresponding to the best-fit cosmologies for each of the three different radial profiles (solid green, pink, and purple dots in the top panel). For example, for the $(\calR=2.4, \gamma=0)$ choice, we estimate the $\ln \calL_{\rm SK+Ns}$ distribution (in the steady-state phase of the MCMC, far from its burn-in) for the corresponding best-fit cosmology $(\Omega_\rmm = 0.296, \sigma_8 = 0.657)$. That distribution of $\ln \calL_{\rm SK+Ns}$, which is cosmology-insensitive over the corresponding 95\% confidence interval in the $(\Omega_\rmm,\sigma_8)$ plane, is shown by a purple dashed curve in the bottom panel. The pink dash-dotted curve in the bottom panel is the similar distribution for the NFW radial profile assumption, and the green solid curve for the fiducial self-consistent radial profile modeling. Each of these likelihood distributions is plotted relative to $\ln \calL_{\rm SK+Ns}^{\rm max}$ of the maximum likelihood model corresponding to the fiducial self-consistent choice of the satellite radial profile.

Evidently in terms of $\mathcal{L}_{\rm SK+Ns}$, the fiducial model is significantly more favored compared to the two alternatives. This procedure, extended over all possible variations of the radial profile, enables us to first obtain the best-fit $n_{\rm sat}(r)$, and then use it in the cosmology analysis, to obtain the unbiased results. Without this self-consistent modeling, the cosmology results can become significantly biased. The whole exercise demonstrates the dire need to correctly account for this major source of systematic uncertainty in analyses of small-scale or nonlinear-scale data coming from ongoing and future galaxy surveys.

\section{Summary and Conclusion}
\label{sec:summary}

We have developed a novel method to revisit the $S_8$ tension using the abundance and kinematics of satellite galaxies extracted from the SDSS DR7 along with the total galaxy luminosity function. The method makes use of the recently developed Bayesian hierarchical tool, \Basilisc, which forward models the line-of-sight velocities of central-satellite pairs in their raw form, that is, without resorting to any summary statistic, to constrain the conditional luminosity function (CLF) with exquisite precision and accuracy \citep{Mitra.etal.24}. These constraints on the galaxy-halo connnection are subsequently used to make predictions for the galaxy LF, for different cosmologies. A comparison with the observed LF then allows for constraints on $\Omega_\rmm$ and $\sigma_8$, which are the two main parameters that control the halo mass function from which the galaxy LF is computed.  Conceptually, this is similar to estimating dynamical masses of galaxy clusters and then inferring a cosmological constraint from the cluster mass function or cluster counts above some threshold mass \citep[as in][]{Pratt.etal.2019, Abdullah.etal.2020, Ferragamo.etal.2021}.  However, our method extends this principle to much lower halo masses.

We tested and validated our methodology against realistic mock data, created by populating dark matter halos in the SMDPL simulation \citep{Klypin.etal.16} with mock galaxies. This was used to create a mock SDSS-like redshift survey that properly accounts for redshift space distortions, the SDSS flux limit and angular footprint, redshift errors, and spectroscopic incompleteness (including fiber collisions). Importantly, the construction of the mock data did not make any a priori assumptions regarding the orbital velocity anisotropy, dynamical equilibrium, or the radial or azimuthal profile of satellite galaxies; rather, satellites were simply assigned the phase-space coordinates of dark matter subhalos in the simulation volume. Applying \Basilisk to these mock data yields tight and unbiased constraints on the galaxy-dark matter connection (see Section~\ref{sec:mock_test} and \papII). Most importantly, we demonstrated an unbiased recovery of the true input cosmology of the SMDPL simulation used to create the mock data. The cosmological constraints reveal a significant degeneracy along the direction $\sigma_8  \Omega_\rmm^2 = {\rm constant}$, which differs significantly from the locus along which $S_8 = \sigma_8  \Omega_\rmm^{0.5}$ is fixed. This is primarily due to the fact that our method probes the halo mass function down to lower masses than most other methods.

Applying the same method to data extracted from the SDSS-DR7, we obtain $\Omega_\rmm = 0.324 \pm 0.012$, $\sigma_8 = 0.775 \pm 0.063$, and $S_8 = 0.81 \pm 0.05$, which are in excellent agreement with the \Planc results ($S_8= 0.834 \pm 0.016$). Our constraints are degenerate along constant $\sigma_8 \Omega_\rmm^2$, unlike most low-$z$ analyses. As a result, our strongest cosmological constraint is that $\sigma_8 (\Omega_\rmm/0.3)^2 = 0.91 \pm 0.05$, again in excellent agreement with \Planck. Hence, contrary to many other studies, we find no indication for an $S_8$ tension! Three key aspects of our methodology are worth highlighting in this context:
\begin{itemize}[leftmargin=0.27truecm, labelwidth=0.2truecm]
    
\item {\bf Halo assembly bias:} Several studies have pointed out that halo assembly bias is a key issue in clustering and lensing analyses \citep[][]{Zentner.etal.14, Zentner.etal.19}, specifically in the context of the $S_8$ tension \citep[][]{Lange.etal.19c, Yuan.etal.22, Chaves-Montero.etal.23, Contreras.etal.2023}. Our choice of observables makes our methodology immune to the effects of halo assembly bias because it probes the individual halo potential wells and is therefore insensitive to the clustering of halos.
    
\item {\bf Halo Occupation Model:} Most cosmological studies use HOD models to characterize the galaxy-halo connection, which has been shown to be too restrictive to fully account for the potential complexity in the data \citep{Beltz-Mohrmann.etal.2020}. Instead, we used a far more flexible model based on the CLF, which characterizes the full luminosity dependence of the galaxy-halo connection. In particular, we use an extended 13-parameter CLF model developed in \papIII, where it is shown that the extension with respect to the more commonly used 10-parameter model is (i) warranted by the SDSS data, in that it drastically improves the quality of the fit, and (ii) sufficient, in that adding additional freedom does not cause any further improvement.  Having a sufficiently flexible model for the galaxy-dark matter connection is crucial, as an overly restrictive or inflexible model can lead to systematic bias in the inferred cosmology \citep[e.g.,][]{Szewciw.etal.2022, Contreras.etal.2023}.
    
\item {\bf Baryonic effects:} Several studies have shown that accounting for baryonic effects, especially those arising from various feedback processes related to galaxy formation, can significantly alleviate $S_8$ tension \citep[e.g.,][]{Lange.etal.19c, Chisari.etal.2019, Amodeo.etal.2021, McCarthy.etal.2024}. We use a detailed and flexible model, developed by \citet{Baggen.etal.25}, to account for the way in which baryonic processes modify the gravitational potential in which the satellites orbit, which accounts for the ejection of baryonic matter and the impact that the central galaxy has on shaping the overall mass distribution of the host halo.

\end{itemize}

An interesting by-product of our analysis is a constraint on the average orbital anisotropy of satellite galaxies, $\beta$. Our inferred value for $\beta$ reveals a weak dependence on $\Omega_\rmm$ and, mainly, $\sigma_8$. For the \Planc cosmology, the best-fit value of $\beta$ is in excellent agreement with that of subhalos in dark-matter only simulations, while models with higher (lower) values of $\sigma_8$ yield best-fit values for $\beta$ that are smaller (larger). Since, according to the $\Lambda$CDM paradigm, satellites reside in dark matter subhalos, and should thus have similar orbital ansiotropy, we interpret these results as an independent confirmation of the \Planc cosmology.

The largest uncertainty in our analysis is the treatment of baryonic effects, in particular the fraction of baryons that have been ejected from dark matter halos due to feedback processes. Our fiducial model for $f_{\rm eject}(M)$ is taken from B25 and is calibrated against the \EAGLE simulation. Unfortunately, different hydrodynamical simulations of galaxy formation predict ejected mass fractions that can differ substantially. We have repeated our analysis for two fairly extreme models; a minimal feedback model, with $f_{\rm eject}(M)=0$, and an extreme feedback model for which $f_{\rm eject}$ slowly transitions from $0.77$ at $M < 10^{12} \Msunh$ to zero at $M > 10^{15} \Msunh$. Interestingly, both yield constraints on $S_8$ that are somewhat {\it larger} than what we infer using the fiducial model ($S_8 = 0.89 \pm 0.09$ and $0.87 \pm 0.07$, respectively), while remaining consistent with \Planck's $S_8$ to better than $1\sigma$. We also used a model for $f_{\rm eject}(M)$ that is calibrated to the Illustris-2 simulation, which is known to have a particularly high feedback efficiency (i.e., large $f_{\rm eject}$), especially at the massive end. Using this $f_{\rm eject}(M)$ produces cosmological constraints that are in tension with \Planck. However, interestingly, it implies a value for $S_8$ that is {\it larger} than \Planck, which is opposite to the typical $S_8$ tension. Hence, we conclude that although our detailed results are subject to non-negligible uncertainties in $f_{\rm eject}(M)$, our conclusion that the cosmological inference is consistent with \Planck, without any indication of an $S_8$ tension, appears robust.

We have demonstrated that the cosmological inference is quite sensitive to the assumed radial distribution of satellite galaxies. Although in our analysis this radial distribution is self-consistently inferred from the data, this is not the case in many cosmological analyses (based on clustering and/or lensing) in the literature. Often, the radial profile of satellites is simply assumed to follow an NFW profile \citep[][]{Zheng.2004, Tinker.2007, vdBosch.etal.13}, exactly mirroring that of dark matter particles. Another commonly used approach is to populate dark matter halos in dark-matter-only $N$-body simulations with mock galaxies and to extract summary statistics from the resulting distribution of mock galaxies for comparison with data. Typically, satellite galaxies are placed on dark matter subhalos \citep[][]{Conroy.etal.06, Conroy.Wechsler.09, Behroozi.etal.13c, Reddick.etal.13}, which are known to have a radial profile that is anti-biased with respect to the dark matter. However, subhalos in numerical simulations are subject to artificial disruption\citep[][]{vdBosch.etal.18a, vdBosch.etal.18b, Errani.Penarrubia.20}, and therefore their radial profile in simulations is likely to be systematically biased \citep[e.g.,][]{Campbell.etal.18, Green.etal.21}. We have demonstrated that, using our method, either assumption (satellites trace dark matter or satellites trace subhalos in simulations) yields a cosmological inference that is significantly biased compared to what one infers using a self-consistently inferred satellite profile. In our analysis of the kinematics and abundance of satellite galaxies in SDSS-DR7, we infer that satellite galaxies follow a radial profile that is consistent with an NFW profile, but with a scale radius that is $~\sim 2.2$ times larger than that of dark matter. Importantly, we have demonstrated that this self-consistently inferred profile is clearly preferred by the data over either of the alternatives mentioned above. This underscores the importance of correctly modeling the radial profile of satellite galaxies in analyses that probe small, nonlinear scales. This is especially acute in this era of new forthcoming surveys in which systematic errors can easily overshadow statistical uncertainties in the data.

In the near future, the Dark Energy Spectroscopic Instrument \citep[DESI;][]{Desi.22} Bright Galaxy Survey \citep[BGS;][]{DESI.BGS.23} will provide a high completeness spectroscopic sample that is close to two magnitudes deeper than the SDSS DR7 main galaxy sample used here. Consequently, the BGS \texttt{Bright} sample will have approximately $20 \times$ the number of galaxies. It should be fairly straightforward to apply \Basilisk to the BGS data, with minimal changes to the methodology\footnote{A potential concern, though, is how best to account for incompleteness due to fiber collisions, which will require a somewhat different approach than for the SDSS data.}. This will allow for constraints on the galaxy-halo connection down to fainter galaxies (and thus lower mass halos) and/or to split the sample by secondary galaxy properties (i.e., size, color, star formation rate), which will allow for a higher-dimensional characterization of the galaxy-halo connection. In addition, the vast increase in data volume is also likely to significantly tighten the confidence intervals on the cosmological parameters, compared to what we have achieved here. Alternatively, if one has confidence in the cosmological parameters from \Planc and other probes, one can also invert the problem and use the methodology presented here to put tight constraints on $f_{\rm eject}(M)$, and thus on the efficiency of feedback processes associated with galaxy formation.

In conclusion, using a combination of galaxy abundances and satellite kinematics, extracted from the SDSS-DR7, we obtain constraints on $\Omega_\rmm$ and $\sigma_8$ that are in perfect agreement with \Planck's inference from the CMB. As such, this study joins a growing pool of analyzes that, contrary to many others, do not find $S_8$ tension between low-$z$ probes of large-scale structure and the constraints of CMB fluctuations \citep{Amon.Efstathiou.2022, Arico.etal.2023, Farren.etal.2024, Bocquet.etal.2024, Chen.etal.2024}. Using our novel and complementary technique, it remains to be seen whether the conclusion of no $S_8$ tension remains true as the constraints shrink with future data. As the sensitivity to baryonic modeling will be more severe, we plan to develop better techniques to marginalize over baryonic uncertainty in future analyses. However, joint analyses with other probes such as clustering and weak lensing, and specifically the probes sensitive to the baryonic physics, such as X-ray, thermal-SZ, kinematic-SZ, etc., will help with robust inference on the cosmological parameters while simultaneously constraining the subgrid physics in cosmological hydrodynamical simulations.

\section*{Acknowledgments}

We are grateful to Benedikt Diemer, Joel Primack, Aldo Rodr\'iguez-Puebla, Joop Schaye, Ravi Sheth, Jeremy Tinker and Walter Jaffe for engaging discussions. This work has been supported by the National Aeronautics and Space Administration through Grant No. 19-ATP19-0059 issued as part of the Astrophysics Theory Program. FvdB has received additional support from the National Science Foundation (NSF) through grants AST-2307280 and AST-2407063. This research was supported in part by grant NSF PHY-2309135 to the Kavli Institute for Theoretical Physics (KITP). This work used, primarily for plotting purposes, the following python packages: \texttt{Matplotlib} \citep{Matplotlib_Hunter2007}, \texttt{SciPy} \citep{SciPy_Virtanen2020}, \texttt{NumPy} \citep{numpy_vanderWalt2011}, and \texttt{PyGTC} \citep{PyGTC_Bocquet2016}.


\bibliographystyle{mnras}
\bibliography{references_vdb}

@ARTICLE{Desi.22,
       author = {{DESI Collaboration} and {Abareshi}, B. and {Aguilar}, J. and {Ahlen}, S. and {Alam}, Shadab and {Alexander}, David M. and {Alfarsy}, R. and {Allen}, L. and {Allende Prieto}, C. and {Alves}, O. and {Ameel}, J. and {Armengaud}, E. and {Asorey}, J. and {Aviles}, Alejandro and {Bailey}, S. and {Balaguera-Antol{\'\i}nez}, A. and {Ballester}, O. and {Baltay}, C. and {Bault}, A. and {Beltran}, S.~F. and {Benavides}, B. and {BenZvi}, S. and {Berti}, A. and {Besuner}, R. and {Beutler}, Florian and {Bianchi}, D. and {Blake}, C. and {Blanc}, P. and {Blum}, R. and {Bolton}, A. and {Bose}, S. and {Bramall}, D. and {Brieden}, S. and {Brodzeller}, A. and {Brooks}, D. and {Brownewell}, C. and {Buckley-Geer}, E. and {Cahn}, R.~N. and {Cai}, Z. and {Canning}, R. and {Capasso}, R. and {Carnero Rosell}, A. and {Carton}, P. and {Casas}, R. and {Castander}, F.~J. and {Cervantes-Cota}, J.~L. and {Chabanier}, S. and {Chaussidon}, E. and {Chuang}, C. and {Circosta}, C. and {Cole}, S. and {Cooper}, A.~P. and {da Costa}, L. and {Cousinou}, M. -C. and {Cuceu}, A. and {Davis}, T.~M. and {Dawson}, K. and {de la Cruz-Noriega}, R. and {de la Macorra}, A. and {de Mattia}, A. and {Della Costa}, J. and {Demmer}, P. and {Derwent}, M. and {Dey}, A. and {Dey}, B. and {Dhungana}, G. and {Ding}, Z. and {Dobson}, C. and {Doel}, P. and {Donald-McCann}, J. and {Donaldson}, J. and {Douglass}, K. and {Duan}, Y. and {Dunlop}, P. and {Edelstein}, J. and {Eftekharzadeh}, S. and {Eisenstein}, D.~J. and {Enriquez-Vargas}, M. and {Escoffier}, S. and {Evatt}, M. and {Fagrelius}, P. and {Fan}, X. and {Fanning}, K. and {Fawcett}, V.~A. and {Ferraro}, S. and {Ereza}, J. and {Flaugher}, B. and {Font-Ribera}, A. and {Forero-Romero}, J.~E. and {Frenk}, C.~S. and {Fromenteau}, S. and {G{\"a}nsicke}, B.~T. and {Garcia-Quintero}, C. and {Garrison}, L. and {Gazta{\~n}aga}, E. and {Gerardi}, F. and {Gil-Mar{\'\i}n}, H. and {Gontcho A Gontcho}, S. and {Gonzalez-Morales}, Alma X. and {Gonzalez-de-Rivera}, G. and {Gonzalez-Perez}, V. and {Gordon}, C. and {Graur}, O. and {Green}, D. and {Grove}, C. and {Gruen}, D. and {Gutierrez}, G. and {Guy}, J. and {Hahn}, C. and {Harris}, S. and {Herrera}, D. and {Herrera-Alcantar}, Hiram K. and {Honscheid}, K. and {Howlett}, C. and {Huterer}, D. and {Ir{\v{s}}i{\v{c}}}, V. and {Ishak}, M. and {Jelinsky}, P. and {Jiang}, L. and {Jimenez}, J. and {Jing}, Y.~P. and {Joyce}, R. and {Jullo}, E. and {Juneau}, S. and {Kara{\c{c}}ayl{\i}}, N.~G. and {Karamanis}, M. and {Karcher}, A. and {Karim}, T. and {Kehoe}, R. and {Kent}, S. and {Kirkby}, D. and {Kisner}, T. and {Kitaura}, F. and {Koposov}, S.~E. and {Kov{\'a}cs}, A. and {Kremin}, A. and {Krolewski}, Alex and {L'Huillier}, B. and {Lahav}, O. and {Lambert}, A. and {Lamman}, C. and {Lan}, Ting-Wen and {Landriau}, M. and {Lane}, S. and {Lang}, D. and {Lange}, J.~U. and {Lasker}, J. and {Le Guillou}, L. and {Leauthaud}, A. and {Le Van Suu}, A. and {Levi}, Michael E. and {Li}, T.~S. and {Magneville}, C. and {Manera}, M. and {Manser}, Christopher J. and {Marshall}, B. and {Martini}, Paul and {McCollam}, W. and {McDonald}, P. and {Meisner}, Aaron M. and {Mena-Fern{\'a}ndez}, J. and {Meneses-Rizo}, J. and {Mezcua}, M. and {Miller}, T. and {Miquel}, R. and {Montero-Camacho}, P. and {Moon}, J. and {Moustakas}, J. and {Mueller}, E. and {Mu{\~n}oz-Guti{\'e}rrez}, Andrea and {Myers}, Adam D. and {Nadathur}, S. and {Najita}, J. and {Napolitano}, L. and {Neilsen}, E. and {Newman}, Jeffrey A. and {Nie}, J.~D. and {Ning}, Y. and {Niz}, G. and {Norberg}, P. and {Noriega}, Hern{\'a}n E. and {O'Brien}, T. and {Obuljen}, A. and {Palanque-Delabrouille}, N. and {Palmese}, A. and {Zhiwei}, P. and {Pappalardo}, D. and {PENG}, X. and {Percival}, W.~J. and {Perruchot}, S. and {Pogge}, R. and {Poppett}, C. and {Porredon}, A. and {Prada}, F. and {Prochaska}, J. and {Pucha}, R. and {P{\'e}rez-Fern{\'a}ndez}, A. and {P{\'e}rez-R{\`a}fols}, I. and {Rabinowitz}, D. and {Raichoor}, A.},
        title = "{Overview of the Instrumentation for the Dark Energy Spectroscopic Instrument}",
      journal = {\aj},
         year = 2022,
        month = nov,
       volume = {164},
       number = {5},
          eid = {207},
        pages = {207},
          doi = {10.3847/1538-3881/ac882b},
archivePrefix = {arXiv},
       eprint = {2205.10939},
 primaryClass = {astro-ph.IM},
       adsurl = {https://ui.adsabs.harvard.edu/abs/2022AJ....164..207D}
}

@ARTICLE{DESI.BGS.23,
       author = {{Hahn}, ChangHoon and {Wilson}, Michael J. and {Ruiz-Macias}, Omar and {Cole}, Shaun and {Weinberg}, David H. and {Moustakas}, John and {Kremin}, Anthony and {Tinker}, Jeremy L. and {Smith}, Alex and {Wechsler}, Risa H. and {Ahlen}, Steven and {Alam}, Shadab and {Bailey}, Stephen and {Brooks}, David and {Cooper}, Andrew P. and {Davis}, Tamara M. and {Dawson}, Kyle and {Dey}, Arjun and {Dey}, Biprateep and {Eftekharzadeh}, Sarah and {Eisenstein}, Daniel J. and {Fanning}, Kevin and {Forero-Romero}, Jaime E. and {Frenk}, Carlos S. and {Gazta{\~n}aga}, Enrique and {A Gontcho}, Satya Gontcho and {Guy}, Julien and {Honscheid}, Klaus and {Ishak}, Mustapha and {Juneau}, St{\'e}phanie and {Kehoe}, Robert and {Kisner}, Theodore and {Lan}, Ting-Wen and {Landriau}, Martin and {Le Guillou}, Laurent and {Levi}, Michael E. and {Magneville}, Christophe and {Martini}, Paul and {Meisner}, Aaron and {Myers}, Adam D. and {Nie}, Jundan and {Norberg}, Peder and {Palanque-Delabrouille}, Nathalie and {Percival}, Will J. and {Poppett}, Claire and {Prada}, Francisco and {Raichoor}, Anand and {Ross}, Ashley J. and {Gaines}, Sasha and {Saulder}, Christoph and {Schlafly}, Eddie and {Schlegel}, David and {Sierra-Porta}, David and {Tarle}, Gregory and {Weaver}, Benjamin A. and {Y{\`e}che}, Christophe and {Zarrouk}, Pauline and {Zhou}, Rongpu and {Zhou}, Zhimin and {Zou}, Hu},
        title = "{The DESI Bright Galaxy Survey: Final Target Selection, Design, and Validation}",
      journal = {\aj},
         year = 2023,
        month = jun,
       volume = {165},
       number = {6},
          eid = {253},
        pages = {253},
          doi = {10.3847/1538-3881/accff8},
archivePrefix = {arXiv},
       eprint = {2208.08512},
 primaryClass = {astro-ph.CO},
       adsurl = {https://ui.adsabs.harvard.edu/abs/2023AJ....165..253H}
}

@ARTICLE{Abazajian.etal.09,
   author = {{Abazajian}, K.~N. and {Adelman-McCarthy}, J.~K. and {Ag{\"u}eros}, M.~A. and 
	{Allam}, S.~S. and {Allende Prieto}, C. and {An}, D. and {Anderson}, K.~S.~J. and 
	{Anderson}, S.~F. and {Annis}, J. and {Bahcall}, N.~A. and et al.},
    title = "{The Seventh Data Release of the Sloan Digital Sky Survey}",
  journal = {\apjs},
archivePrefix = "arXiv",
   eprint = {0812.0649},
     year = 2009,
    month = jun,
   volume = 182,
    pages = {543-558},
      doi = {10.1088/0067-0049/182/2/543},
   adsurl = {http://adsabs.harvard.edu/abs/2009ApJS..182..543A}
}

@ARTICLE{Abbott.etal.2020_CC,
       author = {{Abbott}, T.~M.~C. and {Aguena}, M. and {Alarcon}, A. and {Allam}, S. and {Allen}, S. and {Annis}, J. and {Avila}, S. and {Bacon}, D. and {Bechtol}, K. and {Bermeo}, A. and {Bernstein}, G.~M. and {Bertin}, E. and {Bhargava}, S. and {Bocquet}, S. and {Brooks}, D. and {Brout}, D. and {Buckley-Geer}, E. and {Burke}, D.~L. and {Carnero Rosell}, A. and {Carrasco Kind}, M. and {Carretero}, J. and {Castander}, F.~J. and {Cawthon}, R. and {Chang}, C. and {Chen}, X. and {Choi}, A. and {Costanzi}, M. and {Crocce}, M. and {da Costa}, L.~N. and {Davis}, T.~M. and {De Vicente}, J. and {DeRose}, J. and {Desai}, S. and {Diehl}, H.~T. and {Dietrich}, J.~P. and {Dodelson}, S. and {Doel}, P. and {Drlica-Wagner}, A. and {Eckert}, K. and {Eifler}, T.~F. and {Elvin-Poole}, J. and {Estrada}, J. and {Everett}, S. and {Evrard}, A.~E. and {Farahi}, A. and {Ferrero}, I. and {Flaugher}, B. and {Fosalba}, P. and {Frieman}, J. and {Garc{\'\i}a-Bellido}, J. and {Gatti}, M. and {Gaztanaga}, E. and {Gerdes}, D.~W. and {Giannantonio}, T. and {Giles}, P. and {Grandis}, S. and {Gruen}, D. and {Gruendl}, R.~A. and {Gschwend}, J. and {Gutierrez}, G. and {Hartley}, W.~G. and {Hinton}, S.~R. and {Hollowood}, D.~L. and {Honscheid}, K. and {Hoyle}, B. and {Huterer}, D. and {James}, D.~J. and {Jarvis}, M. and {Jeltema}, T. and {Johnson}, M.~W.~G. and {Johnson}, M.~D. and {Kent}, S. and {Krause}, E. and {Kron}, R. and {Kuehn}, K. and {Kuropatkin}, N. and {Lahav}, O. and {Li}, T.~S. and {Lidman}, C. and {Lima}, M. and {Lin}, H. and {MacCrann}, N. and {Maia}, M.~A.~G. and {Mantz}, A. and {Marshall}, J.~L. and {Martini}, P. and {Mayers}, J. and {Melchior}, P. and {Mena-Fern{\'a}ndez}, J. and {Menanteau}, F. and {Miquel}, R. and {Mohr}, J.~J. and {Nichol}, R.~C. and {Nord}, B. and {Ogando}, R.~L.~C. and {Palmese}, A. and {Paz-Chinch{\'o}n}, F. and {Plazas}, A.~A. and {Prat}, J. and {Rau}, M.~M. and {Romer}, A.~K. and {Roodman}, A. and {Rooney}, P. and {Rozo}, E. and {Rykoff}, E.~S. and {Sako}, M. and {Samuroff}, S. and {S{\'a}nchez}, C. and {Sanchez}, E. and {Saro}, A. and {Scarpine}, V. and {Schubnell}, M. and {Scolnic}, D. and {Serrano}, S. and {Sevilla-Noarbe}, I. and {Sheldon}, E. and {Smith}, J. Allyn. and {Smith}, M. and {Suchyta}, E. and {Swanson}, M.~E.~C. and {Tarle}, G. and {Thomas}, D. and {To}, C. and {Troxel}, M.~A. and {Tucker}, D.~L. and {Varga}, T.~N. and {von der Linden}, A. and {Walker}, A.~R. and {Wechsler}, R.~H. and {Weller}, J. and {Wilkinson}, R.~D. and {Wu}, H. and {Yanny}, B. and {Zhang}, Y. and {Zhang}, Z. and {Zuntz}, J. and {DES Collaboration}},
        title = "{Dark Energy Survey Year 1 Results: Cosmological constraints from cluster abundances and weak lensing}",
      journal = {\prd},
         year = 2020,
        month = jul,
       volume = {102},
       number = {2},
          eid = {023509},
        pages = {023509},
          doi = {10.1103/PhysRevD.102.023509},
archivePrefix = {arXiv},
       eprint = {2002.11124},
 primaryClass = {astro-ph.CO},
       adsurl = {https://ui.adsabs.harvard.edu/abs/2020PhRvD.102b3509A}
}

@ARTICLE{Abbott.etal.2022,
       author = {{Abbott}, T.~M.~C. and {Aguena}, M. and {Alarcon}, A. and {Allam}, S. and {Alves}, O. and {Amon}, A. and {Andrade-Oliveira}, F. and {Annis}, J. and {Avila}, S. and {Bacon}, D. and {Baxter}, E. and {Bechtol}, K. and {Becker}, M.~R. and {Bernstein}, G.~M. and {Bhargava}, S. and {Birrer}, S. and {Blazek}, J. and {Brandao-Souza}, A. and {Bridle}, S.~L. and {Brooks}, D. and {Buckley-Geer}, E. and {Burke}, D.~L. and {Camacho}, H. and {Campos}, A. and {Carnero Rosell}, A. and {Carrasco Kind}, M. and {Carretero}, J. and {Castander}, F.~J. and {Cawthon}, R. and {Chang}, C. and {Chen}, A. and {Chen}, R. and {Choi}, A. and {Conselice}, C. and {Cordero}, J. and {Costanzi}, M. and {Crocce}, M. and {da Costa}, L.~N. and {da Silva Pereira}, M.~E. and {Davis}, C. and {Davis}, T.~M. and {De Vicente}, J. and {DeRose}, J. and {Desai}, S. and {Di Valentino}, E. and {Diehl}, H.~T. and {Dietrich}, J.~P. and {Dodelson}, S. and {Doel}, P. and {Doux}, C. and {Drlica-Wagner}, A. and {Eckert}, K. and {Eifler}, T.~F. and {Elsner}, F. and {Elvin-Poole}, J. and {Everett}, S. and {Evrard}, A.~E. and {Fang}, X. and {Farahi}, A. and {Fernandez}, E. and {Ferrero}, I. and {Fert{\'e}}, A. and {Fosalba}, P. and {Friedrich}, O. and {Frieman}, J. and {Garc{\'\i}a-Bellido}, J. and {Gatti}, M. and {Gaztanaga}, E. and {Gerdes}, D.~W. and {Giannantonio}, T. and {Giannini}, G. and {Gruen}, D. and {Gruendl}, R.~A. and {Gschwend}, J. and {Gutierrez}, G. and {Harrison}, I. and {Hartley}, W.~G. and {Herner}, K. and {Hinton}, S.~R. and {Hollowood}, D.~L. and {Honscheid}, K. and {Hoyle}, B. and {Huff}, E.~M. and {Huterer}, D. and {Jain}, B. and {James}, D.~J. and {Jarvis}, M. and {Jeffrey}, N. and {Jeltema}, T. and {Kovacs}, A. and {Krause}, E. and {Kron}, R. and {Kuehn}, K. and {Kuropatkin}, N. and {Lahav}, O. and {Leget}, P. -F. and {Lemos}, P. and {Liddle}, A.~R. and {Lidman}, C. and {Lima}, M. and {Lin}, H. and {MacCrann}, N. and {Maia}, M.~A.~G. and {Marshall}, J.~L. and {Martini}, P. and {McCullough}, J. and {Melchior}, P. and {Mena-Fern{\'a}ndez}, J. and {Menanteau}, F. and {Miquel}, R. and {Mohr}, J.~J. and {Morgan}, R. and {Muir}, J. and {Myles}, J. and {Nadathur}, S. and {Navarro-Alsina}, A. and {Nichol}, R.~C. and {Ogando}, R.~L.~C. and {Omori}, Y. and {Palmese}, A. and {Pandey}, S. and {Park}, Y. and {Paz-Chinch{\'o}n}, F. and {Petravick}, D. and {Pieres}, A. and {Plazas Malag{\'o}n}, A.~A. and {Porredon}, A. and {Prat}, J. and {Raveri}, M. and {Rodriguez-Monroy}, M. and {Rollins}, R.~P. and {Romer}, A.~K. and {Roodman}, A. and {Rosenfeld}, R. and {Ross}, A.~J. and {Rykoff}, E.~S. and {Samuroff}, S. and {S{\'a}nchez}, C. and {Sanchez}, E. and {Sanchez}, J. and {Sanchez Cid}, D. and {Scarpine}, V. and {Schubnell}, M. and {Scolnic}, D. and {Secco}, L.~F. and {Serrano}, S. and {Sevilla-Noarbe}, I. and {Sheldon}, E. and {Shin}, T. and {Smith}, M. and {Soares-Santos}, M. and {Suchyta}, E. and {Swanson}, M.~E.~C. and {Tabbutt}, M. and {Tarle}, G. and {Thomas}, D. and {To}, C. and {Troja}, A. and {Troxel}, M.~A. and {Tucker}, D.~L. and {Tutusaus}, I. and {Varga}, T.~N. and {Walker}, A.~R. and {Weaverdyck}, N. and {Wechsler}, R. and {Weller}, J. and {Yanny}, B. and {Yin}, B. and {Zhang}, Y. and {Zuntz}, J. and {DES Collaboration}},
        title = "{Dark Energy Survey Year 3 results: Cosmological constraints from galaxy clustering and weak lensing}",
      journal = {\prd},
         year = 2022,
        month = jan,
       volume = {105},
       number = {2},
          eid = {023520},
        pages = {023520},
          doi = {10.1103/PhysRevD.105.023520},
archivePrefix = {arXiv},
       eprint = {2105.13549},
 primaryClass = {astro-ph.CO},
       adsurl = {https://ui.adsabs.harvard.edu/abs/2022PhRvD.105b3520A}
}

@ARTICLE{Abdalla.etal.2022,
       author = {{Abdalla}, Elcio and {Abell{\'a}n}, Guillermo Franco and {Aboubrahim}, Amin and {Agnello}, Adriano and {Akarsu}, {\"O}zg{\"u}r and {Akrami}, Yashar and {Alestas}, George and {Aloni}, Daniel and {Amendola}, Luca and {Anchordoqui}, Luis A. and {Anderson}, Richard I. and {Arendse}, Nikki and {Asgari}, Marika and {Ballardini}, Mario and {Barger}, Vernon and {Basilakos}, Spyros and {Batista}, Ronaldo C. and {Battistelli}, Elia S. and {Battye}, Richard and {Benetti}, Micol and {Benisty}, David and {Berlin}, Asher and {de Bernardis}, Paolo and {Berti}, Emanuele and {Bidenko}, Bohdan and {Birrer}, Simon and {Blakeslee}, John P. and {Boddy}, Kimberly K. and {Bom}, Clecio R. and {Bonilla}, Alexander and {Borghi}, Nicola and {Bouchet}, Fran{\c{c}}ois R. and {Braglia}, Matteo and {Buchert}, Thomas and {Buckley-Geer}, Elizabeth and {Calabrese}, Erminia and {Caldwell}, Robert R. and {Camarena}, David and {Capozziello}, Salvatore and {Casertano}, Stefano and {Chen}, Geoff C. -F. and {Chluba}, Jens and {Chen}, Angela and {Chen}, Hsin-Yu and {Chudaykin}, Anton and {Cicoli}, Michele and {Copi}, Craig J. and {Courbin}, Fred and {Cyr-Racine}, Francis-Yan and {Czerny}, Bo{\.z}ena and {Dainotti}, Maria and {D'Amico}, Guido and {Davis}, Anne-Christine and {de Cruz P{\'e}rez}, Javier and {de Haro}, Jaume and {Delabrouille}, Jacques and {Denton}, Peter B. and {Dhawan}, Suhail and {Dienes}, Keith R. and {Di Valentino}, Eleonora and {Du}, Pu and {Eckert}, Dominique and {Escamilla-Rivera}, Celia and {Fert{\'e}}, Agn{\`e}s and {Finelli}, Fabio and {Fosalba}, Pablo and {Freedman}, Wendy L. and {Frusciante}, Noemi and {Gazta{\~n}aga}, Enrique and {Giar{\`e}}, William and {Giusarma}, Elena and {G{\'o}mez-Valent}, Adri{\`a} and {Handley}, Will and {Harrison}, Ian and {Hart}, Luke and {Hazra}, Dhiraj Kumar and {Heavens}, Alan and {Heinesen}, Asta and {Hildebrandt}, Hendrik and {Hill}, J. Colin and {Hogg}, Natalie B. and {Holz}, Daniel E. and {Hooper}, Deanna C. and {Hosseininejad}, Nikoo and {Huterer}, Dragan and {Ishak}, Mustapha and {Ivanov}, Mikhail M. and {Jaffe}, Andrew H. and {Jang}, In Sung and {Jedamzik}, Karsten and {Jimenez}, Raul and {Joseph}, Melissa and {Joudaki}, Shahab and {Kamionkowski}, Marc and {Karwal}, Tanvi and {Kazantzidis}, Lavrentios and {Keeley}, Ryan E. and {Klasen}, Michael and {Komatsu}, Eiichiro and {Koopmans}, L{\'e}on V.~E. and {Kumar}, Suresh and {Lamagna}, Luca and {Lazkoz}, Ruth and {Lee}, Chung-Chi and {Lesgourgues}, Julien and {Levi Said}, Jackson and {Lewis}, Tiffany R. and {L'Huillier}, Benjamin and {Lucca}, Matteo and {Maartens}, Roy and {Macri}, Lucas M. and {Marfatia}, Danny and {Marra}, Valerio and {Martins}, Carlos J.~A.~P. and {Masi}, Silvia and {Matarrese}, Sabino and {Mazumdar}, Arindam and {Melchiorri}, Alessandro and {Mena}, Olga and {Mersini-Houghton}, Laura and {Mertens}, James and {Milakovi{\'c}}, Dinko and {Minami}, Yuto and {Miranda}, Vivian and {Moreno-Pulido}, Cristian and {Moresco}, Michele and {Mota}, David F. and {Mottola}, Emil and {Mozzon}, Simone and {Muir}, Jessica and {Mukherjee}, Ankan and {Mukherjee}, Suvodip and {Naselsky}, Pavel and {Nath}, Pran and {Nesseris}, Savvas and {Niedermann}, Florian and {Notari}, Alessio and {Nunes}, Rafael C. and {{\'O} Colg{\'a}in}, Eoin and {Owens}, Kayla A. and {{\"O}z{\"u}lker}, Emre and {Pace}, Francesco and {Paliathanasis}, Andronikos and {Palmese}, Antonella and {Pan}, Supriya and {Paoletti}, Daniela and {Perez Bergliaffa}, Santiago E. and {Perivolaropoulos}, Leandros and {Pesce}, Dominic W. and {Pettorino}, Valeria and {Philcox}, Oliver H.~E. and {Pogosian}, Levon and {Poulin}, Vivian and {Poulot}, Gaspard and {Raveri}, Marco and {Reid}, Mark J. and {Renzi}, Fabrizio and {Riess}, Adam G. and {Sabla}, Vivian I. and {Salucci}, Paolo and {Salzano}, Vincenzo and {Saridakis}, Emmanuel N. and {Sathyaprakash}, Bangalore S. and {Schmaltz}, Martin and {Sch{\"o}neberg}, Nils and {Scolnic}, Dan and {Sen}, Anjan A. and {Sehgal}, Neelima and {Shafieloo}, Arman and {Sheikh-Jabbari}, M.~M. and {Silk}, Joseph and {Silvestri}, Alessandra and {Skara}, Foteini and {Sloth}, Martin S. and {Soares-Santos}, Marcelle and {Sol{\`a} Peracaula}, Joan and {Songsheng}, Yu-Yang and {Soriano}, Jorge F. and {Staicova}, Denitsa and {Starkman}, Glenn D. and {Szapudi}, Istv{\'a}n and {Teixeira}, Elsa M. and {Thomas}, Brooks and {Treu}, Tommaso and {Trott}, Emery and {van de Bruck}, Carsten and {Vazquez}, J. Alberto and {Verde}, Licia and {Visinelli}, Luca and {Wang}, Deng and {Wang}, Jian-Min and {Wang}, Shao-Jiang and {Watkins}, Richard and {Watson}, Scott and {Webb}, John K. and {Weiner}, Neal and {Weltman}, Amanda and {Witte}, Samuel J. and {Wojtak}, Rados{\l}aw and {Yadav}, Anil Kumar and {Yang}, Weiqiang and {Zhao}, Gong-Bo and {Zumalac{\'a}rregui}, Miguel},
        title = "{Cosmology intertwined: A review of the particle physics, astrophysics, and cosmology associated with the cosmological tensions and anomalies}",
      journal = {Journal of High Energy Astrophysics},
         year = 2022,
        month = jun,
       volume = {34},
        pages = {49-211},
          doi = {10.1016/j.jheap.2022.04.002},
archivePrefix = {arXiv},
       eprint = {2203.06142},
 primaryClass = {astro-ph.CO},
       adsurl = {https://ui.adsabs.harvard.edu/abs/2022JHEAp..34...49A}
}

@ARTICLE{Abdullah.etal.2020,
       author = {{Abdullah}, Mohamed H. and {Klypin}, Anatoly and {Wilson}, Gillian},
        title = "{Cosmological Constraints on {\ensuremath{\Omega}}$_{m}$ and {\ensuremath{\sigma}}$_{8}$ from Cluster Abundances Using the GalWCat19 Optical-spectroscopic SDSS Catalog}",
      journal = {\apj},
         year = 2020,
        month = oct,
       volume = {901},
       number = {2},
          eid = {90},
        pages = {90},
          doi = {10.3847/1538-4357/aba619},
archivePrefix = {arXiv},
       eprint = {2002.11907},
 primaryClass = {astro-ph.CO},
       adsurl = {https://ui.adsabs.harvard.edu/abs/2020ApJ...901...90A}
}

@ARTICLE{Aiola.etal.20,
       author = {{Aiola}, Simone and {Calabrese}, Erminia and {Maurin}, Lo{\"\i}c and {Naess}, Sigurd and {Schmitt}, Benjamin L. and {Abitbol}, Maximilian H. and {Addison}, Graeme E. and {Ade}, Peter A.~R. and {Alonso}, David and {Amiri}, Mandana and {Amodeo}, Stefania and {Angile}, Elio and {Austermann}, Jason E. and {Baildon}, Taylor and {Battaglia}, Nick and {Beall}, James A. and {Bean}, Rachel and {Becker}, Daniel T. and {Bond}, J. Richard and {Bruno}, Sarah Marie and {Calafut}, Victoria and {Campusano}, Luis E. and {Carrero}, Felipe and {Chesmore}, Grace E. and {Cho}, Hsiao-mei and {Choi}, Steve K. and {Clark}, Susan E. and {Cothard}, Nicholas F. and {Crichton}, Devin and {Crowley}, Kevin T. and {Darwish}, Omar and {Datta}, Rahul and {Denison}, Edward V. and {Devlin}, Mark J. and {Duell}, Cody J. and {Duff}, Shannon M. and {Duivenvoorden}, Adriaan J. and {Dunkley}, Jo and {D{\"u}nner}, Rolando and {Essinger-Hileman}, Thomas and {Fankhanel}, Max and {Ferraro}, Simone and {Fox}, Anna E. and {Fuzia}, Brittany and {Gallardo}, Patricio A. and {Gluscevic}, Vera and {Golec}, Joseph E. and {Grace}, Emily and {Gralla}, Megan and {Guan}, Yilun and {Hall}, Kirsten and {Halpern}, Mark and {Han}, Dongwon and {Hargrave}, Peter and {Hasselfield}, Matthew and {Helton}, Jakob M. and {Henderson}, Shawn and {Hensley}, Brandon and {Hill}, J. Colin and {Hilton}, Gene C. and {Hilton}, Matt and {Hincks}, Adam D. and {Hlo{\v{z}}ek}, Ren{\'e}e and {Ho}, Shuay-Pwu Patty and {Hubmayr}, Johannes and {Huffenberger}, Kevin M. and {Hughes}, John P. and {Infante}, Leopoldo and {Irwin}, Kent and {Jackson}, Rebecca and {Klein}, Jeff and {Knowles}, Kenda and {Koopman}, Brian and {Kosowsky}, Arthur and {Lakey}, Vincent and {Li}, Dale and {Li}, Yaqiong and {Li}, Zack and {Lokken}, Martine and {Louis}, Thibaut and {Lungu}, Marius and {MacInnis}, Amanda and {Madhavacheril}, Mathew and {Maldonado}, Felipe and {Mallaby-Kay}, Maya and {Marsden}, Danica and {McMahon}, Jeff and {Menanteau}, Felipe and {Moodley}, Kavilan and {Morton}, Tim and {Namikawa}, Toshiya and {Nati}, Federico and {Newburgh}, Laura and {Nibarger}, John P. and {Nicola}, Andrina and {Niemack}, Michael D. and {Nolta}, Michael R. and {Orlowski-Sherer}, John and {Page}, Lyman A. and {Pappas}, Christine G. and {Partridge}, Bruce and {Phakathi}, Phumlani and {Pisano}, Giampaolo and {Prince}, Heather and {Puddu}, Roberto and {Qu}, Frank J. and {Rivera}, Jesus and {Robertson}, Naomi and {Rojas}, Felipe and {Salatino}, Maria and {Schaan}, Emmanuel and {Schillaci}, Alessandro and {Sehgal}, Neelima and {Sherwin}, Blake D. and {Sierra}, Carlos and {Sievers}, Jon and {Sifon}, Cristobal and {Sikhosana}, Precious and {Simon}, Sara and {Spergel}, David N. and {Staggs}, Suzanne T. and {Stevens}, Jason and {Storer}, Emilie and {Sunder}, Dhaneshwar D. and {Switzer}, Eric R. and {Thorne}, Ben and {Thornton}, Robert and {Trac}, Hy and {Treu}, Jesse and {Tucker}, Carole and {Vale}, Leila R. and {Van Engelen}, Alexander and {Van Lanen}, Jeff and {Vavagiakis}, Eve M. and {Wagoner}, Kasey and {Wang}, Yuhan and {Ward}, Jonathan T. and {Wollack}, Edward J. and {Xu}, Zhilei and {Zago}, Fernando and {Zhu}, Ningfeng},
        title = "{The Atacama Cosmology Telescope: DR4 maps and cosmological parameters}",
      journal = {\jcap},
         year = 2020,
        month = dec,
       volume = {2020},
       number = {12},
          eid = {047},
        pages = {047},
          doi = {10.1088/1475-7516/2020/12/047},
archivePrefix = {arXiv},
       eprint = {2007.07288},
 primaryClass = {astro-ph.CO},
       adsurl = {https://ui.adsabs.harvard.edu/abs/2020JCAP...12..047A}
}

@ARTICLE{Akarsu.etal.2021,
       author = {{Akarsu}, {\"O}zg{\"u}r and {Kumar}, Suresh and {{\"O}z{\"u}lker}, Emre and {Vazquez}, J. Alberto},
        title = "{Relaxing cosmological tensions with a sign switching cosmological constant}",
      journal = {\prd},
         year = 2021,
        month = dec,
       volume = {104},
       number = {12},
          eid = {123512},
        pages = {123512},
          doi = {10.1103/PhysRevD.104.123512},
archivePrefix = {arXiv},
       eprint = {2108.09239},
 primaryClass = {astro-ph.CO},
       adsurl = {https://ui.adsabs.harvard.edu/abs/2021PhRvD.104l3512A}
}

@ARTICLE{Allali.etal.2021,
       author = {{Allali}, Itamar J. and {Hertzberg}, Mark P. and {Rompineve}, Fabrizio},
        title = "{Dark sector to restore cosmological concordance}",
      journal = {\prd},
         year = 2021,
        month = oct,
       volume = {104},
       number = {8},
          eid = {L081303},
        pages = {L081303},
          doi = {10.1103/PhysRevD.104.L081303},
archivePrefix = {arXiv},
       eprint = {2104.12798},
 primaryClass = {astro-ph.CO},
       adsurl = {https://ui.adsabs.harvard.edu/abs/2021PhRvD.104h1303A}
}

@ARTICLE{Amodeo.etal.2021,
       author = {{Amodeo}, Stefania and {Battaglia}, Nicholas and {Schaan}, Emmanuel and {Ferraro}, Simone and {Moser}, Emily and {Aiola}, Simone and {Austermann}, Jason E. and {Beall}, James A. and {Bean}, Rachel and {Becker}, Daniel T. and {Bond}, Richard J. and {Calabrese}, Erminia and {Calafut}, Victoria and {Choi}, Steve K. and {Denison}, Edward V. and {Devlin}, Mark and {Duff}, Shannon M. and {Duivenvoorden}, Adriaan J. and {Dunkley}, Jo and {D{\"u}nner}, Rolando and {Gallardo}, Patricio A. and {Hall}, Kirsten R. and {Han}, Dongwon and {Hill}, J. Colin and {Hilton}, Gene C. and {Hilton}, Matt and {Hlo{\v{z}}ek}, Ren{\'e}e and {Hubmayr}, Johannes and {Huffenberger}, Kevin M. and {Hughes}, John P. and {Koopman}, Brian J. and {MacInnis}, Amanda and {McMahon}, Jeff and {Madhavacheril}, Mathew S. and {Moodley}, Kavilan and {Mroczkowski}, Tony and {Naess}, Sigurd and {Nati}, Federico and {Newburgh}, Laura B. and {Niemack}, Michael D. and {Page}, Lyman A. and {Partridge}, Bruce and {Schillaci}, Alessandro and {Sehgal}, Neelima and {Sif{\'o}n}, Crist{\'o}bal and {Spergel}, David N. and {Staggs}, Suzanne and {Storer}, Emilie R. and {Ullom}, Joel N. and {Vale}, Leila R. and {van Engelen}, Alexander and {Van Lanen}, Jeff and {Vavagiakis}, Eve M. and {Wollack}, Edward J. and {Xu}, Zhilei},
        title = "{Atacama Cosmology Telescope: Modeling the gas thermodynamics in BOSS CMASS galaxies from kinematic and thermal Sunyaev-Zel'dovich measurements}",
      journal = {\prd},
         year = 2021,
        month = mar,
       volume = {103},
       number = {6},
          eid = {063514},
        pages = {063514},
          doi = {10.1103/PhysRevD.103.063514},
archivePrefix = {arXiv},
       eprint = {2009.05558},
 primaryClass = {astro-ph.CO},
       adsurl = {https://ui.adsabs.harvard.edu/abs/2021PhRvD.103f3514A}
}

@ARTICLE{Amon.Efstathiou.2022,
       author = {{Amon}, Alexandra and {Efstathiou}, George},
        title = "{A non-linear solution to the S$_{8}$ tension?}",
      journal = {\mnras},
         year = 2022,
        month = nov,
       volume = {516},
       number = {4},
        pages = {5355-5366},
          doi = {10.1093/mnras/stac2429},
archivePrefix = {arXiv},
       eprint = {2206.11794},
 primaryClass = {astro-ph.CO},
       adsurl = {https://ui.adsabs.harvard.edu/abs/2022MNRAS.516.5355A}
}

@ARTICLE{Amon.etal.22,
       author = {{Amon}, A. and {Gruen}, D. and {Troxel}, M.~A. and {MacCrann}, N. and {Dodelson}, S. and {Choi}, A. and {Doux}, C. and {Secco}, L.~F. and {Samuroff}, S. and {Krause}, E. and {Cordero}, J. and {Myles}, J. and {DeRose}, J. and {Wechsler}, R.~H. and {Gatti}, M. and {Navarro-Alsina}, A. and {Bernstein}, G.~M. and {Jain}, B. and {Blazek}, J. and {Alarcon}, A. and {Fert{\'e}}, A. and {Lemos}, P. and {Raveri}, M. and {Campos}, A. and {Prat}, J. and {S{\'a}nchez}, C. and {Jarvis}, M. and {Alves}, O. and {Andrade-Oliveira}, F. and {Baxter}, E. and {Bechtol}, K. and {Becker}, M.~R. and {Bridle}, S.~L. and {Camacho}, H. and {Carnero Rosell}, A. and {Carrasco Kind}, M. and {Cawthon}, R. and {Chang}, C. and {Chen}, R. and {Chintalapati}, P. and {Crocce}, M. and {Davis}, C. and {Diehl}, H.~T. and {Drlica-Wagner}, A. and {Eckert}, K. and {Eifler}, T.~F. and {Elvin-Poole}, J. and {Everett}, S. and {Fang}, X. and {Fosalba}, P. and {Friedrich}, O. and {Gaztanaga}, E. and {Giannini}, G. and {Gruendl}, R.~A. and {Harrison}, I. and {Hartley}, W.~G. and {Herner}, K. and {Huang}, H. and {Huff}, E.~M. and {Huterer}, D. and {Kuropatkin}, N. and {Leget}, P. and {Liddle}, A.~R. and {McCullough}, J. and {Muir}, J. and {Pandey}, S. and {Park}, Y. and {Porredon}, A. and {Refregier}, A. and {Rollins}, R.~P. and {Roodman}, A. and {Rosenfeld}, R. and {Ross}, A.~J. and {Rykoff}, E.~S. and {Sanchez}, J. and {Sevilla-Noarbe}, I. and {Sheldon}, E. and {Shin}, T. and {Troja}, A. and {Tutusaus}, I. and {Tutusaus}, I. and {Varga}, T.~N. and {Weaverdyck}, N. and {Yanny}, B. and {Yin}, B. and {Zhang}, Y. and {Zuntz}, J. and {Aguena}, M. and {Allam}, S. and {Annis}, J. and {Bacon}, D. and {Bertin}, E. and {Bhargava}, S. and {Brooks}, D. and {Buckley-Geer}, E. and {Burke}, D.~L. and {Carretero}, J. and {Costanzi}, M. and {da Costa}, L.~N. and {Pereira}, M.~E.~S. and {De Vicente}, J. and {Desai}, S. and {Dietrich}, J.~P. and {Doel}, P. and {Ferrero}, I. and {Flaugher}, B. and {Frieman}, J. and {Garc{\'\i}a-Bellido}, J. and {Gaztanaga}, E. and {Gerdes}, D.~W. and {Giannantonio}, T. and {Gschwend}, J. and {Gutierrez}, G. and {Hinton}, S.~R. and {Hollowood}, D.~L. and {Honscheid}, K. and {Hoyle}, B. and {James}, D.~J. and {Kron}, R. and {Kuehn}, K. and {Lahav}, O. and {Lima}, M. and {Lin}, H. and {Maia}, M.~A.~G. and {Marshall}, J.~L. and {Martini}, P. and {Melchior}, P. and {Menanteau}, F. and {Miquel}, R. and {Mohr}, J.~J. and {Morgan}, R. and {Ogando}, R.~L.~C. and {Palmese}, A. and {Paz-Chinch{\'o}n}, F. and {Petravick}, D. and {Pieres}, A. and {Romer}, A.~K. and {Sanchez}, E. and {Scarpine}, V. and {Schubnell}, M. and {Serrano}, S. and {Smith}, M. and {Soares-Santos}, M. and {Tarle}, G. and {Thomas}, D. and {To}, C. and {Weller}, J. and {DES Collaboration}},
        title = "{Dark Energy Survey Year 3 results: Cosmology from cosmic shear and robustness to data calibration}",
      journal = {\prd},
         year = 2022,
        month = jan,
       volume = {105},
       number = {2},
          eid = {023514},
        pages = {023514},
          doi = {10.1103/PhysRevD.105.023514},
archivePrefix = {arXiv},
       eprint = {2105.13543},
 primaryClass = {astro-ph.CO},
       adsurl = {https://ui.adsabs.harvard.edu/abs/2022PhRvD.105b3514A}
}

@ARTICLE{Amon.etal.23,
       author = {{Amon}, A. and {Robertson}, N.~C. and {Miyatake}, H. and {Heymans}, C. and {White}, M. and {DeRose}, J. and {Yuan}, S. and {Wechsler}, R.~H. and {Varga}, T.~N. and {Bocquet}, S. and {Dvornik}, A. and {More}, S. and {Ross}, A.~J. and {Hoekstra}, H. and {Alarcon}, A. and {Asgari}, M. and {Blazek}, J. and {Campos}, A. and {Chen}, R. and {Choi}, A. and {Crocce}, M. and {Diehl}, H.~T. and {Doux}, C. and {Eckert}, K. and {Elvin-Poole}, J. and {Everett}, S. and {Fert{\'e}}, A. and {Gatti}, M. and {Giannini}, G. and {Gruen}, D. and {Gruendl}, R.~A. and {Hartley}, W.~G. and {Herner}, K. and {Hildebrandt}, H. and {Huang}, S. and {Huff}, E.~M. and {Joachimi}, B. and {Lee}, S. and {MacCrann}, N. and {Myles}, J. and {Navarro-Alsina}, A. and {Nishimichi}, T. and {Prat}, J. and {Secco}, L.~F. and {Sevilla-Noarbe}, I. and {Sheldon}, E. and {Shin}, T. and {Tr{\"o}ster}, T. and {Troxel}, M.~A. and {Tutusaus}, I. and {Wright}, A.~H. and {Yin}, B. and {Aguena}, M. and {Allam}, S. and {Annis}, J. and {Bacon}, D. and {Bilicki}, M. and {Brooks}, D. and {Burke}, D.~L. and {Carnero Rosell}, A. and {Carretero}, J. and {Castander}, F.~J. and {Cawthon}, R. and {Costanzi}, M. and {da Costa}, L.~N. and {Pereira}, M.~E.~S. and {de Jong}, J. and {De Vicente}, J. and {Desai}, S. and {Dietrich}, J.~P. and {Doel}, P. and {Ferrero}, I. and {Frieman}, J. and {Garc{\'\i}a-Bellido}, J. and {Gerdes}, D.~W. and {Gschwend}, J. and {Gutierrez}, G. and {Hinton}, S.~R. and {Hollowood}, D.~L. and {Honscheid}, K. and {Huterer}, D. and {Kannawadi}, A. and {Kuehn}, K. and {Kuropatkin}, N. and {Lahav}, O. and {Lima}, M. and {Maia}, M.~A.~G. and {Marshall}, J.~L. and {Menanteau}, F. and {Miquel}, R. and {Mohr}, J.~J. and {Morgan}, R. and {Muir}, J. and {Paz-Chinch{\'o}n}, F. and {Pieres}, A. and {Plazas Malag{\'o}n}, A.~A. and {Porredon}, A. and {Rodriguez-Monroy}, M. and {Roodman}, A. and {Sanchez}, E. and {Serrano}, S. and {Shan}, H. and {Suchyta}, E. and {Swanson}, M.~E.~C. and {Tarle}, G. and {Thomas}, D. and {To}, C. and {Zhang}, Y.},
        title = "{Consistent lensing and clustering in a low-S$_{8}$ Universe with BOSS, DES Year 3, HSC Year 1, and KiDS-1000}",
      journal = {\mnras},
         year = 2023,
        month = jan,
       volume = {518},
       number = {1},
        pages = {477-503},
          doi = {10.1093/mnras/stac2938},
archivePrefix = {arXiv},
       eprint = {2202.07440},
 primaryClass = {astro-ph.CO},
       adsurl = {https://ui.adsabs.harvard.edu/abs/2023MNRAS.518..477A}
}

@ARTICLE{Archidiacono.etal.2019,
       author = {{Archidiacono}, Maria and {Hooper}, Deanna C. and {Murgia}, Riccardo and {Bohr}, Sebastian and {Lesgourgues}, Julien and {Viel}, Matteo},
        title = "{Constraining Dark Matter-Dark Radiation interactions with CMB, BAO, and Lyman-{\ensuremath{\alpha}}}",
      journal = {\jcap},
         year = 2019,
        month = oct,
       volume = {2019},
       number = {10},
          eid = {055},
        pages = {055},
          doi = {10.1088/1475-7516/2019/10/055},
archivePrefix = {arXiv},
       eprint = {1907.01496},
 primaryClass = {astro-ph.CO},
       adsurl = {https://ui.adsabs.harvard.edu/abs/2019JCAP...10..055A}
}

@ARTICLE{Arico.etal.2021,
       author = {{Aric{\`o}}, Giovanni and {Angulo}, Raul E. and {Hern{\'a}ndez-Monteagudo}, Carlos and {Contreras}, Sergio and {Zennaro}, Matteo},
        title = "{Simultaneous modelling of matter power spectrum and bispectrum in the presence of baryons}",
      journal = {\mnras},
         year = 2021,
        month = may,
       volume = {503},
       number = {3},
        pages = {3596-3609},
          doi = {10.1093/mnras/stab699},
archivePrefix = {arXiv},
       eprint = {2009.14225},
 primaryClass = {astro-ph.CO},
       adsurl = {https://ui.adsabs.harvard.edu/abs/2021MNRAS.503.3596A}
}

@ARTICLE{Arico.etal.2023,
       author = {{Aric{\`o}}, Giovanni and {Angulo}, Raul E. and {Zennaro}, Matteo and {Contreras}, Sergio and {Chen}, Angela and {Hern{\'a}ndez-Monteagudo}, Carlos},
        title = "{DES Y3 cosmic shear down to small scales: Constraints on cosmology and baryons}",
      journal = {\aap},
         year = 2023,
        month = oct,
       volume = {678},
          eid = {A109},
        pages = {A109},
          doi = {10.1051/0004-6361/202346539},
archivePrefix = {arXiv},
       eprint = {2303.05537},
 primaryClass = {astro-ph.CO},
       adsurl = {https://ui.adsabs.harvard.edu/abs/2023A&A...678A.109A}
}

@ARTICLE{Asgari.etal.2021,
       author = {{Asgari}, Marika and {Lin}, Chieh-An and {Joachimi}, Benjamin and {Giblin}, Benjamin and {Heymans}, Catherine and {Hildebrandt}, Hendrik and {Kannawadi}, Arun and {St{\"o}lzner}, Benjamin and {Tr{\"o}ster}, Tilman and {van den Busch}, Jan Luca and {Wright}, Angus H. and {Bilicki}, Maciej and {Blake}, Chris and {de Jong}, Jelte and {Dvornik}, Andrej and {Erben}, Thomas and {Getman}, Fedor and {Hoekstra}, Henk and {K{\"o}hlinger}, Fabian and {Kuijken}, Konrad and {Miller}, Lance and {Radovich}, Mario and {Schneider}, Peter and {Shan}, HuanYuan and {Valentijn}, Edwin},
        title = "{KiDS-1000 cosmology: Cosmic shear constraints and comparison between two point statistics}",
      journal = {\aap},
         year = 2021,
        month = jan,
       volume = {645},
          eid = {A104},
        pages = {A104},
          doi = {10.1051/0004-6361/202039070},
archivePrefix = {arXiv},
       eprint = {2007.15633},
 primaryClass = {astro-ph.CO},
       adsurl = {https://ui.adsabs.harvard.edu/abs/2021A&A...645A.104A}
}

@ARTICLE{Ayromlou.etal.2023,
       author = {{Ayromlou}, Mohammadreza and {Nelson}, Dylan and {Pillepich}, Annalisa},
        title = "{Feedback reshapes the baryon distribution within haloes, in halo outskirts, and beyond: the closure radius from dwarfs to massive clusters}",
      journal = {\mnras},
         year = 2023,
        month = oct,
       volume = {524},
       number = {4},
        pages = {5391-5410},
          doi = {10.1093/mnras/stad2046},
archivePrefix = {arXiv},
       eprint = {2211.07659},
 primaryClass = {astro-ph.GA},
       adsurl = {https://ui.adsabs.harvard.edu/abs/2023MNRAS.524.5391A}
}

@ARTICLE{Baggen.etal.25,
       author = {{Baggen}, Josephine F.~W. and {van den Bosch}, Frank C. and {Mitra}, Kaustav},
        title = "{The Impact of Baryonic Effects on the Dynamical Masses Inferred Using Satellite Kinematics}",
      journal = {arXiv e-prints},
         year = 2025,
        month = sep,
          eid = {arXiv:2509.01690},
        pages = {arXiv:2509.01690},
          doi = {10.48550/arXiv.2509.01690},
archivePrefix = {arXiv},
       eprint = {2509.01690},
 primaryClass = {astro-ph.GA},
       adsurl = {https://ui.adsabs.harvard.edu/abs/2025arXiv250901690B}
}

@ARTICLE{Behroozi.etal.13c,
   author = {{Behroozi}, P.~S. and {Wechsler}, R.~H. and {Conroy}, C.},
    title = "{The Average Star Formation Histories of Galaxies in Dark Matter Halos from z = 0-8}",
  journal = {\apj},
archivePrefix = "arXiv",
   eprint = {1207.6105},
 primaryClass = "astro-ph.CO",
     year = 2013,
    month = jun,
   volume = 770,
      eid = {57},
    pages = {57},
      doi = {10.1088/0004-637X/770/1/57},
   adsurl = {http://adsabs.harvard.edu/abs/2013ApJ...770...57B}
}

@ARTICLE{Beltz-Mohrmann.etal.2020,
       author = {{Beltz-Mohrmann}, Gillian D. and {Berlind}, Andreas A. and {Szewciw}, Adam O.},
        title = "{Testing the accuracy of halo occupation distribution modelling using hydrodynamic simulations}",
      journal = {\mnras},
         year = 2020,
        month = feb,
       volume = {491},
       number = {4},
        pages = {5771-5788},
          doi = {10.1093/mnras/stz3442},
archivePrefix = {arXiv},
       eprint = {1908.11448},
 primaryClass = {astro-ph.CO},
       adsurl = {https://ui.adsabs.harvard.edu/abs/2020MNRAS.491.5771B}
}

@ARTICLE{Beltz-Mohrmann.Berlind.2021,
       author = {{Beltz-Mohrmann}, Gillian D. and {Berlind}, Andreas A.},
        title = "{The Impact of Baryonic Physics on the Abundance, Clustering, and Concentration of Halos}",
      journal = {\apj},
         year = 2021,
        month = nov,
       volume = {921},
       number = {2},
          eid = {112},
        pages = {112},
          doi = {10.3847/1538-4357/ac1e27},
archivePrefix = {arXiv},
       eprint = {2103.05076},
 primaryClass = {astro-ph.CO},
       adsurl = {https://ui.adsabs.harvard.edu/abs/2021ApJ...921..112B}
}

@ARTICLE{Berlind.etal.03,
   author = {{Berlind}, A.~A. and {Weinberg}, D.~H. and {Benson}, A.~J. and 
	{Baugh}, C.~M. and {Cole}, S. and {Dav{\'e}}, R. and {Frenk}, C.~S. and 
	{Jenkins}, A. and {Katz}, N. and {Lacey}, C.~G.},
    title = "{The Halo Occupation Distribution and the Physics of Galaxy Formation}",
  journal = {\apj},
   eprint = {astro-ph/0212357},
     year = 2003,
    month = aug,
   volume = 593,
    pages = {1-25},
      doi = {10.1086/376517},
   adsurl = {http://adsabs.harvard.edu/abs/2003ApJ...593....1B}
}

@ARTICLE{Binney.Mamon.82,
       author = {{Binney}, J. and {Mamon}, G.~A.},
        title = "{M/L and velocity anisotropy from observations of spherical galaxies, or must M87 have a massive black hole ?}",
      journal = {\mnras},
         year = 1982,
        month = jul,
       volume = {200},
        pages = {361-375},
          doi = {10.1093/mnras/200.2.361},
       adsurl = {https://ui.adsabs.harvard.edu/abs/1982MNRAS.200..361B}
}

@ARTICLE{Blanton.etal.05,
   author = {{Blanton}, M.~R. and {Schlegel}, D.~J. and {Strauss}, M.~A. and 
	{Brinkmann}, J. and {Finkbeiner}, D. and {Fukugita}, M. and 
	{Gunn}, J.~E. and {Hogg}, D.~W. and {Ivezi{\'c}}, {\v Z}. and 
	{Knapp}, G.~R. and {Lupton}, R.~H. and {Munn}, J.~A. and {Schneider}, D.~P. and 
	{Tegmark}, M. and {Zehavi}, I.},
    title = "{New York University Value-Added Galaxy Catalog: A Galaxy Catalog Based on New Public Surveys}",
  journal = {\aj},
   eprint = {astro-ph/0410166},
     year = 2005,
    month = jun,
   volume = 129,
    pages = {2562-2578},
      doi = {10.1086/429803},
   adsurl = {http://adsabs.harvard.edu/abs/2005AJ....129.2562B}
}

@ARTICLE{Blanton.Berlind.07,
       author = {{Blanton}, Michael R. and {Berlind}, Andreas A.},
        title = "{What Aspects of Galaxy Environment Matter?}",
      journal = {\apj},
         year = 2007,
        month = aug,
       volume = {664},
       number = {2},
        pages = {791-803},
          doi = {10.1086/512478},
archivePrefix = {arXiv},
       eprint = {astro-ph/0608353},
 primaryClass = {astro-ph},
       adsurl = {https://ui.adsabs.harvard.edu/abs/2007ApJ...664..791B}
}

@ARTICLE{Blumenthal.etal.86,
       author = {{Blumenthal}, G.~R. and {Faber}, S.~M. and {Flores}, R. and {Primack}, J.~R.},
        title = "{Contraction of Dark Matter Galactic Halos Due to Baryonic Infall}",
      journal = {\apj},
         year = 1986,
        month = feb,
       volume = {301},
        pages = {27},
          doi = {10.1086/163867},
       adsurl = {https://ui.adsabs.harvard.edu/abs/1986ApJ...301...27B}
}

@ARTICLE{Bocquet.etal.2019,
       author = {{Bocquet}, S. and {Dietrich}, J.~P. and {Schrabback}, T. and {Bleem}, L.~E. and {Klein}, M. and {Allen}, S.~W. and {Applegate}, D.~E. and {Ashby}, M.~L.~N. and {Bautz}, M. and {Bayliss}, M. and {Benson}, B.~A. and {Brodwin}, M. and {Bulbul}, E. and {Canning}, R.~E.~A. and {Capasso}, R. and {Carlstrom}, J.~E. and {Chang}, C.~L. and {Chiu}, I. and {Cho}, H. -M. and {Clocchiatti}, A. and {Crawford}, T.~M. and {Crites}, A.~T. and {de Haan}, T. and {Desai}, S. and {Dobbs}, M.~A. and {Foley}, R.~J. and {Forman}, W.~R. and {Garmire}, G.~P. and {George}, E.~M. and {Gladders}, M.~D. and {Gonzalez}, A.~H. and {Grandis}, S. and {Gupta}, N. and {Halverson}, N.~W. and {Hlavacek-Larrondo}, J. and {Hoekstra}, H. and {Holder}, G.~P. and {Holzapfel}, W.~L. and {Hou}, Z. and {Hrubes}, J.~D. and {Huang}, N. and {Jones}, C. and {Khullar}, G. and {Knox}, L. and {Kraft}, R. and {Lee}, A.~T. and {von der Linden}, A. and {Luong-Van}, D. and {Mantz}, A. and {Marrone}, D.~P. and {McDonald}, M. and {McMahon}, J.~J. and {Meyer}, S.~S. and {Mocanu}, L.~M. and {Mohr}, J.~J. and {Morris}, R.~G. and {Padin}, S. and {Patil}, S. and {Pryke}, C. and {Rapetti}, D. and {Reichardt}, C.~L. and {Rest}, A. and {Ruhl}, J.~E. and {Saliwanchik}, B.~R. and {Saro}, A. and {Sayre}, J.~T. and {Schaffer}, K.~K. and {Shirokoff}, E. and {Stalder}, B. and {Stanford}, S.~A. and {Staniszewski}, Z. and {Stark}, A.~A. and {Story}, K.~T. and {Strazzullo}, V. and {Stubbs}, C.~W. and {Vanderlinde}, K. and {Vieira}, J.~D. and {Vikhlinin}, A. and {Williamson}, R. and {Zenteno}, A.},
        title = "{Cluster Cosmology Constraints from the 2500 deg$^{2}$ SPT-SZ Survey: Inclusion of Weak Gravitational Lensing Data from Magellan and the Hubble Space Telescope}",
      journal = {\apj},
         year = 2019,
        month = jun,
       volume = {878},
       number = {1},
          eid = {55},
        pages = {55},
          doi = {10.3847/1538-4357/ab1f10},
archivePrefix = {arXiv},
       eprint = {1812.01679},
 primaryClass = {astro-ph.CO},
       adsurl = {https://ui.adsabs.harvard.edu/abs/2019ApJ...878...55B}
}

@ARTICLE{Bocquet.etal.2024,
       author = {{Bocquet}, S. and {Grandis}, S. and {Bleem}, L.~E. and {Klein}, M. and {Mohr}, J.~J. and {Schrabback}, T. and {Abbott}, T.~M.~C. and {Ade}, P.~A.~R. and {Aguena}, M. and {Alarcon}, A. and {Allam}, S. and {Allen}, S.~W. and {Alves}, O. and {Amon}, A. and {Anderson}, A.~J. and {Annis}, J. and {Ansarinejad}, B. and {Austermann}, J.~E. and {Avila}, S. and {Bacon}, D. and {Bayliss}, M. and {Beall}, J.~A. and {Bechtol}, K. and {Becker}, M.~R. and {Bender}, A.~N. and {Benson}, B.~A. and {Bernstein}, G.~M. and {Bhargava}, S. and {Bianchini}, F. and {Brodwin}, M. and {Brooks}, D. and {Bryant}, L. and {Campos}, A. and {Canning}, R.~E.~A. and {Carlstrom}, J.~E. and {Carnero Rosell}, A. and {Carrasco Kind}, M. and {Carretero}, J. and {Castander}, F.~J. and {Cawthon}, R. and {Chang}, C.~L. and {Chang}, C. and {Chaubal}, P. and {Chen}, R. and {Chiang}, H.~C. and {Choi}, A. and {Chou}, T. -L. and {Citron}, R. and {Corbett Moran}, C. and {Cordero}, J. and {Costanzi}, M. and {Crawford}, T.~M. and {Crites}, A.~T. and {da Costa}, L.~N. and {Pereira}, M.~E.~S. and {Davis}, C. and {Davis}, T.~M. and {DeRose}, J. and {Desai}, S. and {de Haan}, T. and {Diehl}, H.~T. and {Dobbs}, M.~A. and {Dodelson}, S. and {Doux}, C. and {Drlica-Wagner}, A. and {Eckert}, K. and {Elvin-Poole}, J. and {Everett}, S. and {Everett}, W. and {Ferrero}, I. and {Fert{\'e}}, A. and {Flores}, A.~M. and {Frieman}, J. and {Gallicchio}, J. and {Garc{\'\i}a-Bellido}, J. and {Gatti}, M. and {George}, E.~M. and {Giannini}, G. and {Gladders}, M.~D. and {Gruen}, D. and {Gruendl}, R.~A. and {Gupta}, N. and {Gutierrez}, G. and {Halverson}, N.~W. and {Harrison}, I. and {Hartley}, W.~G. and {Herner}, K. and {Hinton}, S.~R. and {Holder}, G.~P. and {Hollowood}, D.~L. and {Holzapfel}, W.~L. and {Honscheid}, K. and {Hrubes}, J.~D. and {Huang}, N. and {Hubmayr}, J. and {Huff}, E.~M. and {Huterer}, D. and {Irwin}, K.~D. and {James}, D.~J. and {Jarvis}, M. and {Khullar}, G. and {Kim}, K. and {Knox}, L. and {Kraft}, R. and {Krause}, E. and {Kuehn}, K. and {Kuropatkin}, N. and {K{\'e}ruzor{\'e}}, F. and {Lahav}, O. and {Lee}, A.~T. and {Leget}, P. -F. and {Li}, D. and {Lin}, H. and {Lowitz}, A. and {MacCrann}, N. and {Mahler}, G. and {Mantz}, A. and {Marshall}, J.~L. and {McCullough}, J. and {McDonald}, M. and {McMahon}, J.~J. and {Mena-Fern{\'a}ndez}, J. and {Menanteau}, F. and {Meyer}, S.~S. and {Miquel}, R. and {Montgomery}, J. and {Myles}, J. and {Natoli}, T. and {Navarro-Alsina}, A. and {Nibarger}, J.~P. and {Noble}, G.~I. and {Novosad}, V. and {Ogando}, R.~L.~C. and {Omori}, Y. and {Padin}, S. and {Pandey}, S. and {Paschos}, P. and {Patil}, S. and {Pieres}, A. and {Plazas Malag{\'o}n}, A.~A. and {Porredon}, A. and {Prat}, J. and {Pryke}, C. and {Raveri}, M. and {Reichardt}, C.~L. and {Roberson}, J. and {Rollins}, R.~P. and {Romero}, C. and {Roodman}, A. and {Ruhl}, J.~E. and {Rykoff}, E.~S. and {Saliwanchik}, B.~R. and {Salvati}, L. and {S{\'a}nchez}, C. and {Sanchez}, E. and {Sanchez Cid}, D. and {Saro}, A. and {Schaffer}, K.~K. and {Secco}, L.~F. and {Sevilla-Noarbe}, I. and {Sharon}, K. and {Sheldon}, E. and {Shin}, T. and {Sievers}, C. and {Smecher}, G. and {Smith}, M. and {Somboonpanyakul}, T. and {Sommer}, M. and {Stalder}, B. and {Stark}, A.~A. and {Stephen}, J. and {Strazzullo}, V. and {Suchyta}, E. and {Tarle}, G. and {To}, C. and {Troxel}, M.~A. and {Tucker}, C. and {Tutusaus}, I. and {Varga}, T.~N. and {Veach}, T. and {Vieira}, J.~D. and {Vikhlinin}, A. and {von der Linden}, A. and {Wang}, G. and {Weaverdyck}, N. and {Weller}, J. and {Whitehorn}, N. and {Wu}, W.~L.~K. and {Yanny}, B. and {Yefremenko}, V. and {Yin}, B. and {Young}, M. and {Zebrowski}, J.~A. and {Zhang}, Y. and {Zohren}, H. and {Zuntz}, J. and {(SPT} and {DES Collaborations)}},
        title = "{SPT clusters with DES and HST weak lensing. II. Cosmological constraints from the abundance of massive halos}",
      journal = {\prd},
         year = 2024,
        month = oct,
       volume = {110},
       number = {8},
          eid = {083510},
        pages = {083510},
          doi = {10.1103/PhysRevD.110.083510},
archivePrefix = {arXiv},
       eprint = {2401.02075},
 primaryClass = {astro-ph.CO},
       adsurl = {https://ui.adsabs.harvard.edu/abs/2024PhRvD.110h3510B}
}

@ARTICLE{Bridle.King.07,
       author = {{Bridle}, Sarah and {King}, Lindsay},
        title = "{Dark energy constraints from cosmic shear power spectra: impact of intrinsic alignments on photometric redshift requirements}",
      journal = {New Journal of Physics},
         year = 2007,
        month = dec,
       volume = {9},
       number = {12},
        pages = {444},
          doi = {10.1088/1367-2630/9/12/444},
archivePrefix = {arXiv},
       eprint = {0705.0166},
 primaryClass = {astro-ph},
       adsurl = {https://ui.adsabs.harvard.edu/abs/2007NJPh....9..444B}
}

@ARTICLE{Bryan.Norman.98,
   author = {{Bryan}, G.~L. and {Norman}, M.~L.},
    title = "{Statistical Properties of X-Ray Clusters: Analytic and Numerical Comparisons}",
  journal = {\apj},
   eprint = {astro-ph/9710107},
     year = 1998,
    month = mar,
   volume = 495,
    pages = {80-99},
      doi = {10.1086/305262},
   adsurl = {http://adsabs.harvard.edu/abs/1998ApJ...495...80B}
}

@ARTICLE{Cacciato.etal.09,
   author = {{Cacciato}, M. and {van den Bosch}, F.~C. and {More}, S. and 
	{Li}, R. and {Mo}, H.~J. and {Yang}, X.},
    title = "{Galaxy clustering and galaxy-galaxy lensing: a promising union to constrain cosmological parameters}",
  journal = {\mnras},
archivePrefix = "arXiv",
   eprint = {0807.4932},
     year = 2009,
    month = apr,
   volume = 394,
    pages = {929-946},
      doi = {10.1111/j.1365-2966.2008.14362.x},
   adsurl = {http://adsabs.harvard.edu/abs/2009MNRAS.394..929C}
}

@ARTICLE{Cacciato.etal.13,
   author = {{Cacciato}, M. and {van den Bosch}, F.~C. and {More}, S. and 
	{Mo}, H. and {Yang}, X.},
    title = "{Cosmological constraints from a combination of galaxy clustering and lensing - III. Application to SDSS data}",
  journal = {\mnras},
archivePrefix = "arXiv",
   eprint = {1207.0503},
     year = 2013,
    month = apr,
   volume = 430,
    pages = {767-786},
      doi = {10.1093/mnras/sts525},
   adsurl = {http://adsabs.harvard.edu/abs/2013MNRAS.430..767C}
}

@ARTICLE{Campbell.etal.18,
       author = {{Campbell}, Duncan and {van den Bosch}, Frank C. and {Padmanabhan}, Nikhil and {Mao}, Yao-Yuan and {Zentner}, Andrew R. and {Lange}, Johannes U. and {Jiang}, Fangzhou and {Villarreal}, Antonia Sierra},
        title = "{The galaxy clustering crisis in abundance matching}",
      journal = {\mnras},
         year = 2018,
        month = jun,
       volume = {477},
       number = {1},
        pages = {359-383},
          doi = {10.1093/mnras/sty495},
archivePrefix = {arXiv},
       eprint = {1705.06347},
 primaryClass = {astro-ph.GA},
       adsurl = {https://ui.adsabs.harvard.edu/abs/2018MNRAS.477..359C}
}

@ARTICLE{Carlberg.etal.97,
   author = {{Carlberg}, R.~G. and {Yee}, H.~K.~C. and {Ellingson}, E. and 
	{Morris}, S.~L. and {Abraham}, R. and {Gravel}, P. and {Pritchet}, C.~J. and 
	{Smecker-Hane}, T. and {Hartwick}, F.~D.~A. and {Hesser}, J.~E. and 
	{Hutchings}, J.~B. and {Oke}, J.~B.},
    title = "{The Average Mass Profile of Galaxy Clusters}",
  journal = {\apjl},
   eprint = {astro-ph/9703107},
     year = 1997,
    month = aug,
   volume = 485,
    pages = {L13-L16},
      doi = {10.1086/310801},
   adsurl = {http://adsabs.harvard.edu/abs/1997ApJ...485L..13C}
}

@ARTICLE{Chaves-Montero.etal.23,
       author = {{Chaves-Montero}, Jon{\'a}s and {Angulo}, Raul E. and {Contreras}, Sergio},
        title = "{The galaxy formation origin of the lensing is low problem}",
      journal = {\mnras},
         year = 2023,
        month = may,
       volume = {521},
       number = {1},
        pages = {937-951},
          doi = {10.1093/mnras/stad243},
archivePrefix = {arXiv},
       eprint = {2211.01744},
 primaryClass = {astro-ph.CO},
       adsurl = {https://ui.adsabs.harvard.edu/abs/2023MNRAS.521..937C}
}

@ARTICLE{Chen.08,
   author = {{Chen}, J.},
    title = "{Color dependence in the spatial distribution of satellite galaxies}",
  journal = {\aap},
archivePrefix = "arXiv",
   eprint = {0711.0989},
     year = 2008,
    month = jun,
   volume = 484,
    pages = {347-354},
      doi = {10.1051/0004-6361:20079018},
   adsurl = {http://adsabs.harvard.edu/abs/2008A%26A...484..347C}
}

@ARTICLE{Chen.etal.2022,
       author = {{Chen}, Shi-Fan and {White}, Martin and {DeRose}, Joseph and {Kokron}, Nickolas},
        title = "{Cosmological analysis of three-dimensional BOSS galaxy clustering and Planck CMB lensing cross correlations via Lagrangian perturbation theory}",
      journal = {\jcap},
         year = 2022,
        month = jul,
       volume = {2022},
       number = {7},
          eid = {041},
        pages = {041},
          doi = {10.1088/1475-7516/2022/07/041},
archivePrefix = {arXiv},
       eprint = {2204.10392},
 primaryClass = {astro-ph.CO},
       adsurl = {https://ui.adsabs.harvard.edu/abs/2022JCAP...07..041C}
}

@ARTICLE{Chen.etal.2024,
       author = {{Chen}, S. and {DeRose}, J. and {Zhou}, R. and {White}, M. and {Ferraro}, S. and {Blake}, C. and {Lange}, J.~U. and {Wechsler}, R.~H. and {Aguilar}, J. and {Ahlen}, S. and {Brooks}, D. and {Claybaugh}, T. and {Dawson}, K. and {de la Macorra}, A. and {Doel}, P. and {Font-Ribera}, A. and {Gazta{\~n}aga}, E. and {Gontcho}, S. Gontcho A and {Gutierrez}, G. and {Honscheid}, K. and {Howlett}, C. and {Kehoe}, R. and {Kirkby}, D. and {Kisner}, T. and {Kremin}, A. and {Landriau}, M. and {Le Guillou}, L. and {Manera}, M. and {Meisner}, A. and {Miquel}, R. and {Newman}, J.~A. and {Niz}, G. and {Palanque-Delabrouille}, N. and {Percival}, W.~J. and {Prada}, F. and {Rossi}, G. and {Sanchez}, E. and {Schlegel}, D. and {Schubnell}, M. and {Sprayberry}, D. and {Tarl{\'e}}, G. and {Weaver}, B.~A.},
        title = "{Not all lensing is low: An analysis of DESI$\times$DES using the Lagrangian Effective Theory of LSS}",
      journal = {arXiv e-prints},
         year = 2024,
        month = jul,
          eid = {arXiv:2407.04795},
        pages = {arXiv:2407.04795},
          doi = {10.48550/arXiv.2407.04795},
archivePrefix = {arXiv},
       eprint = {2407.04795},
 primaryClass = {astro-ph.CO},
       adsurl = {https://ui.adsabs.harvard.edu/abs/2024arXiv240704795C}
}

@ARTICLE{Chisari.etal.2019,
       author = {{Chisari}, Nora Elisa and {Mead}, Alexander J. and {Joudaki}, Shahab and {Ferreira}, Pedro G. and {Schneider}, Aurel and {Mohr}, Joseph and {Tr{\"o}ster}, Tilman and {Alonso}, David and {McCarthy}, Ian G. and {Martin-Alvarez}, Sergio and {Devriendt}, Julien and {Slyz}, Adrianne and {van Daalen}, Marcel P.},
        title = "{Modelling baryonic feedback for survey cosmology}",
      journal = {The Open Journal of Astrophysics},
         year = 2019,
        month = jun,
       volume = {2},
       number = {1},
          eid = {4},
        pages = {4},
          doi = {10.21105/astro.1905.06082},
archivePrefix = {arXiv},
       eprint = {1905.06082},
 primaryClass = {astro-ph.CO},
       adsurl = {https://ui.adsabs.harvard.edu/abs/2019OJAp....2E...4C}
}

@ARTICLE{Chudaykin.etal.2022,
       author = {{Chudaykin}, Anton and {Gorbunov}, Dmitry and {Nedelko}, Nikita},
        title = "{Exploring ${\Lambda}$CDM extensions with SPT-3G and Planck data: 4$\sigma$ evidence for neutrino masses and implications of extended dark energy models for cosmological tensions}",
      journal = {arXiv e-prints},
         year = 2022,
        month = mar,
          eid = {arXiv:2203.03666},
        pages = {arXiv:2203.03666},
          doi = {10.48550/arXiv.2203.03666},
archivePrefix = {arXiv},
       eprint = {2203.03666},
 primaryClass = {astro-ph.CO},
       adsurl = {https://ui.adsabs.harvard.edu/abs/2022arXiv220303666C}
}

@ARTICLE{Conroy.etal.06,
   author = {{Conroy}, C. and {Wechsler}, R.~H. and {Kravtsov}, A.~V.},
    title = "{Modeling Luminosity-dependent Galaxy Clustering through Cosmic Time}",
  journal = {\apj},
   eprint = {arXiv:astro-ph/0512234},
     year = 2006,
    month = aug,
   volume = 647,
    pages = {201-214},
      doi = {10.1086/503602},
   adsurl = {http://adsabs.harvard.edu/abs/2006ApJ...647..201C}
}

@ARTICLE{Conroy.Wechsler.09,
   author = {{Conroy}, C. and {Wechsler}, R.~H.},
    title = "{Connecting Galaxies, Halos, and Star Formation Rates Across Cosmic Time}",
  journal = {\apj},
archivePrefix = "arXiv",
   eprint = {0805.3346},
     year = 2009,
    month = may,
   volume = 696,
    pages = {620-635},
      doi = {10.1088/0004-637X/696/1/620},
   adsurl = {http://adsabs.harvard.edu/abs/2009ApJ...696..620C}
}

@ARTICLE{Contreras.etal.2023,
       author = {{Contreras}, Sergio and {Chaves-Montero}, Jon{\'a}s and {Angulo}, Raul E.},
        title = "{Consistent clustering and lensing of SDSS-III BOSS galaxies with an extended abundance matching formalism}",
      journal = {\mnras},
         year = 2023,
        month = oct,
       volume = {525},
       number = {2},
        pages = {3149-3161},
          doi = {10.1093/mnras/stad2434},
archivePrefix = {arXiv},
       eprint = {2305.09637},
 primaryClass = {astro-ph.CO},
       adsurl = {https://ui.adsabs.harvard.edu/abs/2023MNRAS.525.3149C}
}

@ARTICLE{Cooray.06,
   author = {{Cooray}, A.},
    title = "{Halo model at its best: constraints on conditional luminosity functions from measured galaxy statistics}",
  journal = {\mnras},
   eprint = {astro-ph/0509033},
     year = 2006,
    month = jan,
   volume = 365,
    pages = {842-866},
      doi = {10.1111/j.1365-2966.2005.09747.x},
   adsurl = {http://adsabs.harvard.edu/abs/2006MNRAS.365..842C}
}

@ARTICLE{Costanzi.etal.2021,
       author = {{Costanzi}, M. and {Saro}, A. and {Bocquet}, S. and {Abbott}, T.~M.~C. and {Aguena}, M. and {Allam}, S. and {Amara}, A. and {Annis}, J. and {Avila}, S. and {Bacon}, D. and {Benson}, B.~A. and {Bhargava}, S. and {Brooks}, D. and {Buckley-Geer}, E. and {Burke}, D.~L. and {Carnero Rosell}, A. and {Carrasco Kind}, M. and {Carretero}, J. and {Choi}, A. and {da Costa}, L.~N. and {Pereira}, M.~E.~S. and {De Vicente}, J. and {Desai}, S. and {Diehl}, H.~T. and {Dietrich}, J.~P. and {Doel}, P. and {Eifler}, T.~F. and {Everett}, S. and {Ferrero}, I. and {Fert{\'e}}, A. and {Flaugher}, B. and {Fosalba}, P. and {Frieman}, J. and {Garc{\'\i}a-Bellido}, J. and {Gaztanaga}, E. and {Gerdes}, D.~W. and {Giannantonio}, T. and {Giles}, P. and {Grandis}, S. and {Gruen}, D. and {Gruendl}, R.~A. and {Gupta}, N. and {Gutierrez}, G. and {Hartley}, W.~G. and {Hinton}, S.~R. and {Hollowood}, D.~L. and {Honscheid}, K. and {James}, D.~J. and {Jeltema}, T. and {Krause}, E. and {Kuehn}, K. and {Kuropatkin}, N. and {Lahav}, O. and {Lima}, M. and {MacCrann}, N. and {Maia}, M.~A.~G. and {Marshall}, J.~L. and {Menanteau}, F. and {Miquel}, R. and {Mohr}, J.~J. and {Morgan}, R. and {Myles}, J. and {Ogando}, R.~L.~C. and {Palmese}, A. and {Paz-Chinch{\'o}n}, F. and {Plazas}, A.~A. and {Rapetti}, D. and {Reichardt}, C.~L. and {Romer}, A.~K. and {Roodman}, A. and {Ruppin}, F. and {Salvati}, L. and {Samuroff}, S. and {Sanchez}, E. and {Scarpine}, V. and {Serrano}, S. and {Sevilla-Noarbe}, I. and {Singh}, P. and {Smith}, M. and {Soares-Santos}, M. and {Stark}, A.~A. and {Suchyta}, E. and {Swanson}, M.~E.~C. and {Tarle}, G. and {Thomas}, D. and {To}, C. and {Tucker}, D.~L. and {Varga}, T.~N. and {Wechsler}, R.~H. and {Zhang}, Z. and {DES} and {SPT Collaborations}},
        title = "{Cosmological constraints from DES Y1 cluster abundances and SPT multiwavelength data}",
      journal = {\prd},
         year = 2021,
        month = feb,
       volume = {103},
       number = {4},
          eid = {043522},
        pages = {043522},
          doi = {10.1103/PhysRevD.103.043522},
archivePrefix = {arXiv},
       eprint = {2010.13800},
 primaryClass = {astro-ph.CO},
       adsurl = {https://ui.adsabs.harvard.edu/abs/2021PhRvD.103d3522C}
}

@ARTICLE{Crain.etal.15,
       author = {{Crain}, Robert A. and {Schaye}, Joop and {Bower}, Richard G. and {Furlong}, Michelle and {Schaller}, Matthieu and {Theuns}, Tom and {Dalla Vecchia}, Claudio and {Frenk}, Carlos S. and {McCarthy}, Ian G. and {Helly}, John C. and {Jenkins}, Adrian and {Rosas-Guevara}, Yetli M. and {White}, Simon D.~M. and {Trayford}, James W.},
        title = "{The EAGLE simulations of galaxy formation: calibration of subgrid physics and model variations}",
      journal = {\mnras},
         year = 2015,
        month = jun,
       volume = {450},
       number = {2},
        pages = {1937-1961},
          doi = {10.1093/mnras/stv725},
archivePrefix = {arXiv},
       eprint = {1501.01311},
 primaryClass = {astro-ph.GA},
       adsurl = {https://ui.adsabs.harvard.edu/abs/2015MNRAS.450.1937C}
}

@ARTICLE{Croton.etal.2007,
       author = {{Croton}, Darren J. and {Gao}, Liang and {White}, Simon D.~M.},
        title = "{Halo assembly bias and its effects on galaxy clustering}",
      journal = {\mnras},
         year = 2007,
        month = feb,
       volume = {374},
       number = {4},
        pages = {1303-1309},
          doi = {10.1111/j.1365-2966.2006.11230.x},
archivePrefix = {arXiv},
       eprint = {astro-ph/0605636},
 primaryClass = {astro-ph},
       adsurl = {https://ui.adsabs.harvard.edu/abs/2007MNRAS.374.1303C}
}

@ARTICLE{Dalal.etal.08,
       author = {{Dalal}, Neal and {White}, Martin and {Bond}, J. Richard and {Shirokov}, Alexander},
        title = "{Halo Assembly Bias in Hierarchical Structure Formation}",
      journal = {\apj},
         year = 2008,
        month = nov,
       volume = {687},
       number = {1},
        pages = {12-21},
          doi = {10.1086/591512},
archivePrefix = {arXiv},
       eprint = {0803.3453},
 primaryClass = {astro-ph},
       adsurl = {https://ui.adsabs.harvard.edu/abs/2008ApJ...687...12D}
}

@ARTICLE{Dekel.Silk.86,
       author = {{Dekel}, A. and {Silk}, J.},
        title = "{The Origin of Dwarf Galaxies, Cold Dark Matter, and Biased Galaxy Formation}",
      journal = {\apj},
         year = 1986,
        month = apr,
       volume = {303},
        pages = {39},
          doi = {10.1086/164050},
       adsurl = {https://ui.adsabs.harvard.edu/abs/1986ApJ...303...39D}
}

@ARTICLE{Diemand.etal.04,
   author = {{Diemand}, J. and {Moore}, B. and {Stadel}, J. and {Kazantzidis}, S.
	},
    title = "{Two-body relaxation in cold dark matter simulations}",
  journal = {\mnras},
   eprint = {astro-ph/0304549},
     year = 2004,
    month = mar,
   volume = 348,
    pages = {977-986},
      doi = {10.1111/j.1365-2966.2004.07424.x},
   adsurl = {http://adsabs.harvard.edu/abs/2004MNRAS.348..977D}
}

@ARTICLE{Diemer.Kravtsov.15,
       author = {{Diemer}, Benedikt and {Kravtsov}, Andrey V.},
        title = "{A Universal Model for Halo Concentrations}",
      journal = {\apj},
         year = 2015,
        month = jan,
       volume = {799},
       number = {1},
          eid = {108},
        pages = {108},
          doi = {10.1088/0004-637X/799/1/108},
archivePrefix = {arXiv},
       eprint = {1407.4730},
 primaryClass = {astro-ph.CO},
       adsurl = {https://ui.adsabs.harvard.edu/abs/2015ApJ...799..108D}
}

@ARTICLE{diValentino.etal.2020,
       author = {{Di Valentino}, Eleonora and {Melchiorri}, Alessandro and {Mena}, Olga and {Vagnozzi}, Sunny},
        title = "{Interacting dark energy in the early 2020s: A promising solution to the H$_{0}$ and cosmic shear tensions}",
      journal = {Physics of the Dark Universe},
         year = 2020,
        month = dec,
       volume = {30},
          eid = {100666},
        pages = {100666},
          doi = {10.1016/j.dark.2020.100666},
archivePrefix = {arXiv},
       eprint = {1908.04281},
 primaryClass = {astro-ph.CO},
       adsurl = {https://ui.adsabs.harvard.edu/abs/2020PDU....3000666D}
}

@ARTICLE{Errani.Penarrubia.20,
       author = {{Errani}, Rapha{\"e}l and {Pe{\~n}arrubia}, Jorge},
        title = "{Can tides disrupt cold dark matter subhaloes?}",
      journal = {\mnras},
         year = 2020,
        month = feb,
       volume = {491},
       number = {4},
        pages = {4591-4601},
          doi = {10.1093/mnras/stz3349},
archivePrefix = {arXiv},
       eprint = {1906.01642},
 primaryClass = {astro-ph.GA},
       adsurl = {https://ui.adsabs.harvard.edu/abs/2020MNRAS.491.4591E}
}

@ARTICLE{Faltenbacher.White.10,
       author = {{Faltenbacher}, Andreas and {White}, Simon D.~M.},
        title = "{Assembly Bias and the Dynamical Structure of Dark Matter Halos}",
      journal = {\apj},
         year = 2010,
        month = jan,
       volume = {708},
       number = {1},
        pages = {469-473},
          doi = {10.1088/0004-637X/708/1/469},
archivePrefix = {arXiv},
       eprint = {0909.4302},
 primaryClass = {astro-ph.CO},
       adsurl = {https://ui.adsabs.harvard.edu/abs/2010ApJ...708..469F}
}

@ARTICLE{Farren.etal.2024,
       author = {{Farren}, Gerrit S. and {Krolewski}, Alex and {MacCrann}, Niall and {Ferraro}, Simone and {Abril-Cabezas}, Irene and {An}, Rui and {Atkins}, Zachary and {Battaglia}, Nicholas and {Bond}, J. Richard and {Calabrese}, Erminia and {Choi}, Steve K. and {Darwish}, Omar and {Devlin}, Mark J. and {Duivenvoorden}, Adriaan J. and {Dunkley}, Jo and {Hill}, J. Colin and {Hilton}, Matt and {Huffenberger}, Kevin M. and {Kim}, Joshua and {Louis}, Thibaut and {Madhavacheril}, Mathew S. and {Marques}, Gabriela A. and {McMahon}, Jeff and {Moodley}, Kavilan and {Page}, Lyman A. and {Partridge}, Bruce and {Qu}, Frank J. and {Schaan}, Emmanuel and {Sehgal}, Neelima and {Sherwin}, Blake D. and {Sif{\'o}n}, Crist{\'o}bal and {Staggs}, Suzanne T. and {Van Engelen}, Alexander and {Vargas}, Cristian and {Wenzl}, Lukas and {White}, Martin and {Wollack}, Edward J.},
        title = "{The Atacama Cosmology Telescope: Cosmology from Cross-correlations of unWISE Galaxies and ACT DR6 CMB Lensing}",
      journal = {\apj},
         year = 2024,
        month = may,
       volume = {966},
       number = {2},
          eid = {157},
        pages = {157},
          doi = {10.3847/1538-4357/ad31a5},
archivePrefix = {arXiv},
       eprint = {2309.05659},
 primaryClass = {astro-ph.CO},
       adsurl = {https://ui.adsabs.harvard.edu/abs/2024ApJ...966..157F}
}

@ARTICLE{Ferragamo.etal.2021,
       author = {{Ferragamo}, A. and {Barrena}, R. and {Rubi{\~n}o-Mart{\'\i}n}, J.~A. and {Aguado-Barahona}, A. and {Streblyanska}, A. and {Tramonte}, D. and {G{\'e}nova-Santos}, R.~T. and {Hempel}, A. and {Lietzen}, H.},
        title = "{Velocity dispersion and dynamical mass for 270 galaxy clusters in the Planck PSZ1 catalogue}",
      journal = {\aap},
         year = 2021,
        month = nov,
       volume = {655},
          eid = {A115},
        pages = {A115},
          doi = {10.1051/0004-6361/202140382},
archivePrefix = {arXiv},
       eprint = {2109.04967},
 primaryClass = {astro-ph.CO},
       adsurl = {https://ui.adsabs.harvard.edu/abs/2021A&A...655A.115F}
}

@ARTICLE{Gao.etal.05,
       author = {{Gao}, Liang and {Springel}, Volker and {White}, Simon D.~M.},
        title = "{The age dependence of halo clustering}",
      journal = {\mnras},
         year = 2005,
        month = oct,
       volume = {363},
       number = {1},
        pages = {L66-L70},
          doi = {10.1111/j.1745-3933.2005.00084.x},
archivePrefix = {arXiv},
       eprint = {astro-ph/0506510},
 primaryClass = {astro-ph},
       adsurl = {https://ui.adsabs.harvard.edu/abs/2005MNRAS.363L..66G}
}

@ARTICLE{Gao.White.07,
       author = {{Gao}, Liang and {White}, Simon D.~M.},
        title = "{Assembly bias in the clustering of dark matter haloes}",
      journal = {\mnras},
         year = 2007,
        month = apr,
       volume = {377},
       number = {1},
        pages = {L5-L9},
          doi = {10.1111/j.1745-3933.2007.00292.x},
archivePrefix = {arXiv},
       eprint = {astro-ph/0611921},
 primaryClass = {astro-ph},
       adsurl = {https://ui.adsabs.harvard.edu/abs/2007MNRAS.377L...5G}
}

@ARTICLE{Genel.etal.2014,
       author = {{Genel}, Shy and {Vogelsberger}, Mark and {Springel}, Volker and {Sijacki}, Debora and {Nelson}, Dylan and {Snyder}, Greg and {Rodriguez-Gomez}, Vicente and {Torrey}, Paul and {Hernquist}, Lars},
        title = "{Introducing the Illustris project: the evolution of galaxy populations across cosmic time}",
      journal = {\mnras},
         year = 2014,
        month = nov,
       volume = {445},
       number = {1},
        pages = {175-200},
          doi = {10.1093/mnras/stu1654},
archivePrefix = {arXiv},
       eprint = {1405.3749},
 primaryClass = {astro-ph.CO},
       adsurl = {https://ui.adsabs.harvard.edu/abs/2014MNRAS.445..175G}
}

@ARTICLE{Gil-Marin.etal.2017,
       author = {{Gil-Mar{\'\i}n}, H{\'e}ctor and {Percival}, Will J. and {Verde}, Licia and {Brownstein}, Joel R. and {Chuang}, Chia-Hsun and {Kitaura}, Francisco-Shu and {Rodr{\'\i}guez-Torres}, Sergio A. and {Olmstead}, Matthew D.},
        title = "{The clustering of galaxies in the SDSS-III Baryon Oscillation Spectroscopic Survey: RSD measurement from the power spectrum and bispectrum of the DR12 BOSS galaxies}",
      journal = {\mnras},
         year = 2017,
        month = feb,
       volume = {465},
       number = {2},
        pages = {1757-1788},
          doi = {10.1093/mnras/stw2679},
archivePrefix = {arXiv},
       eprint = {1606.00439},
 primaryClass = {astro-ph.CO},
       adsurl = {https://ui.adsabs.harvard.edu/abs/2017MNRAS.465.1757G}
}

@ARTICLE{Giodini.etal.2009,
       author = {{Giodini}, S. and {Pierini}, D. and {Finoguenov}, A. and {Pratt}, G.~W. and {Boehringer}, H. and {Leauthaud}, A. and {Guzzo}, L. and {Aussel}, H. and {Bolzonella}, M. and {Capak}, P. and {Elvis}, M. and {Hasinger}, G. and {Ilbert}, O. and {Kartaltepe}, J.~S. and {Koekemoer}, A.~M. and {Lilly}, S.~J. and {Massey}, R. and {McCracken}, H.~J. and {Rhodes}, J. and {Salvato}, M. and {Sanders}, D.~B. and {Scoville}, N.~Z. and {Sasaki}, S. and {Smolcic}, V. and {Taniguchi}, Y. and {Thompson}, D. and {COSMOS Collaboration}},
        title = "{Stellar and Total Baryon Mass Fractions in Groups and Clusters Since Redshift 1}",
      journal = {\apj},
         year = 2009,
        month = sep,
       volume = {703},
       number = {1},
        pages = {982-993},
          doi = {10.1088/0004-637X/703/1/982},
archivePrefix = {arXiv},
       eprint = {0904.0448},
 primaryClass = {astro-ph.CO},
       adsurl = {https://ui.adsabs.harvard.edu/abs/2009ApJ...703..982G}
}

@ARTICLE{Goodman.Weare.10,
   author = {{Goodman}, J. and {Weare}, J.},
    title = "{Ensemble samplers with affine invariance}",
  journal = {Communications in Applied Mathematics and Computational Science, Vol.~5, No.~1, p.~65-80, 2010},
     year = 2010,
   volume = 5,
    pages = {65-80},
      doi = {10.2140/camcos.2010.5.65},
   adsurl = {http://adsabs.harvard.edu/abs/2010CAMCS...5...65G}
}

@ARTICLE{Green.etal.21,
       author = {{Green}, Sheridan B. and {van den Bosch}, Frank C. and {Jiang}, Fangzhou},
        title = "{The tidal evolution of dark matter substructure - II. The impact of artificial disruption on subhalo mass functions and radial profiles}",
      journal = {\mnras},
         year = 2021,
        month = may,
       volume = {503},
       number = {3},
        pages = {4075-4091},
          doi = {10.1093/mnras/stab696},
archivePrefix = {arXiv},
       eprint = {2103.01227},
 primaryClass = {astro-ph.GA},
       adsurl = {https://ui.adsabs.harvard.edu/abs/2021MNRAS.503.4075G}
}

@ARTICLE{Guo.etal.12a,
   author = {{Guo}, Q. and {Cole}, S. and {Eke}, V. and {Frenk}, C.},
    title = "{Satellite galaxy number density profiles in the Sloan Digital Sky Survey}",
  journal = {\mnras},
archivePrefix = "arXiv",
   eprint = {1201.1296},
     year = 2012,
    month = nov,
   volume = 427,
    pages = {428-441},
      doi = {10.1111/j.1365-2966.2012.21882.x},
   adsurl = {http://adsabs.harvard.edu/abs/2012MNRAS.427..428G}
}

@ARTICLE{Hadzhiyska.etal.2024,
       author = {{Hadzhiyska}, B. and {Ferraro}, S. and {Ried Guachalla}, B. and {Schaan}, E. and {Aguilar}, J. and {Battaglia}, N. and {Bond}, J.~R. and {Brooks}, D. and {Calabrese}, E. and {Choi}, S.~K. and {Claybaugh}, T. and {Coulton}, W.~R. and {Dawson}, K. and {Devlin}, M. and {Dey}, B. and {Doel}, P. and {Duivenvoorden}, A.~J. and {Dunkley}, J. and {Farren}, G.~S. and {Font-Ribera}, A. and {Forero-Romero}, J.~E. and {Gallardo}, P.~A. and {Gazta{\~n}aga}, E. and {Gontcho Gontcho}, S. and {Gralla}, M. and {Le Guillou}, L. and {Gutierrez}, G. and {Guy}, J. and {Hill}, J.~C. and {Hlo{\v{z}}ek}, R. and {Honscheid}, K. and {Juneau}, S. and {Kisner}, T. and {Kremin}, A. and {Landriau}, M. and {Liu}, R.~H. and {Louis}, T. and {MacCrann}, N. and {de Macorra}, A. and {Madhavacheril}, M. and {Manera}, M. and {Meisner}, A. and {Miquel}, R. and {Moodley}, K. and {Moustakas}, J. and {Mroczkowski}, T. and {Naess}, S. and {Newman}, J. and {Niemack}, M.~D. and {Niz}, G. and {Page}, L. and {Palanque-Delabrouille}, N. and {Partridge}, B. and {Percival}, W.~J. and {Prada}, F. and {Qu}, F.~J. and {Rossi}, G. and {Sanchez}, E. and {Schlegel}, D. and {Schubnell}, M. and {Sehgal}, N. and {Seo}, H. and {Sif{\'o}n}, C. and {Spergel}, D. and {Sprayberry}, D. and {Staggs}, S. and {Tarl{\'e}}, G. and {Vargas}, C. and {Vavagiakis}, E.~M. and {Weaver}, B.~A. and {Wollack}, E.~J. and {Zhou}, R. and {Zou}, H.},
        title = "{Evidence for large baryonic feedback at low and intermediate redshifts from kinematic Sunyaev-Zel'dovich observations with ACT and DESI photometric galaxies}",
      journal = {arXiv e-prints},
         year = 2024,
        month = jul,
          eid = {arXiv:2407.07152},
        pages = {arXiv:2407.07152},
          doi = {10.48550/arXiv.2407.07152},
archivePrefix = {arXiv},
       eprint = {2407.07152},
 primaryClass = {astro-ph.CO},
       adsurl = {https://ui.adsabs.harvard.edu/abs/2024arXiv240707152H}
}

@ARTICLE{Han.etal.16,
   author = {{Han}, J. and {Cole}, S. and {Frenk}, C.~S. and {Jing}, Y.},
    title = "{A unified model for the spatial and mass distribution of subhaloes}",
  journal = {\mnras},
archivePrefix = "arXiv",
   eprint = {1509.02175},
     year = 2016,
    month = apr,
   volume = 457,
    pages = {1208-1223},
      doi = {10.1093/mnras/stv2900},
   adsurl = {http://adsabs.harvard.edu/abs/2016MNRAS.457.1208H}
}

@ARTICLE{Hartlap.etal.07,
       author = {{Hartlap}, J. and {Simon}, P. and {Schneider}, P.},
        title = "{Why your model parameter confidences might be too optimistic. Unbiased estimation of the inverse covariance matrix}",
      journal = {\aap},
         year = 2007,
        month = mar,
       volume = {464},
       number = {1},
        pages = {399-404},
          doi = {10.1051/0004-6361:20066170},
archivePrefix = {arXiv},
       eprint = {astro-ph/0608064},
 primaryClass = {astro-ph},
       adsurl = {https://ui.adsabs.harvard.edu/abs/2007A&A...464..399H}
}

@ARTICLE{Hearin.etal.16,
   author = {{Hearin}, A.~P. and {Zentner}, A.~R. and {van den Bosch}, F.~C. and 
	{Campbell}, D. and {Tollerud}, E.},
    title = "{Introducing decorated HODs: modelling assembly bias in the galaxy-halo connection}",
  journal = {\mnras},
archivePrefix = "arXiv",
   eprint = {1512.03050},
     year = 2016,
    month = aug,
   volume = 460,
    pages = {2552-2570},
      doi = {10.1093/mnras/stw840},
   adsurl = {http://adsabs.harvard.edu/abs/2016MNRAS.460.2552H}
}

@ARTICLE{Heisenberg.etal.2023,
       author = {{Heisenberg}, Lavinia and {Villarrubia-Rojo}, Hector and {Zosso}, Jann},
        title = "{Simultaneously solving the H$_{0}$ and {\ensuremath{\sigma}}$_{8}$ tensions with late dark energy}",
      journal = {Physics of the Dark Universe},
         year = 2023,
        month = feb,
       volume = {39},
          eid = {101163},
        pages = {101163},
          doi = {10.1016/j.dark.2022.101163},
archivePrefix = {arXiv},
       eprint = {2201.11623},
 primaryClass = {astro-ph.CO},
       adsurl = {https://ui.adsabs.harvard.edu/abs/2023PDU....3901163H}
}

@ARTICLE{Heymans.etal.06,
       author = {{Heymans}, Catherine and {Van Waerbeke}, Ludovic and {Bacon}, David and {Berge}, Joel and {Bernstein}, Gary and {Bertin}, Emmanuel and {Bridle}, Sarah and {Brown}, Michael L. and {Clowe}, Douglas and {Dahle}, H{\r{a}}kon and {Erben}, Thomas and {Gray}, Meghan and {Hetterscheidt}, Marco and {Hoekstra}, Henk and {Hudelot}, Patrick and {Jarvis}, Mike and {Kuijken}, Konrad and {Margoniner}, Vera and {Massey}, Richard and {Mellier}, Yannick and {Nakajima}, Reiko and {Refregier}, Alexandre and {Rhodes}, Jason and {Schrabback}, Tim and {Wittman}, David},
        title = "{The Shear Testing Programme - I. Weak lensing analysis of simulated ground-based observations}",
      journal = {\mnras},
         year = 2006,
        month = may,
       volume = {368},
       number = {3},
        pages = {1323-1339},
          doi = {10.1111/j.1365-2966.2006.10198.x},
archivePrefix = {arXiv},
       eprint = {astro-ph/0506112},
 primaryClass = {astro-ph},
       adsurl = {https://ui.adsabs.harvard.edu/abs/2006MNRAS.368.1323H}
}

@ARTICLE{Heymans.etal.2021,
       author = {{Heymans}, Catherine and {Tr{\"o}ster}, Tilman and {Asgari}, Marika and {Blake}, Chris and {Hildebrandt}, Hendrik and {Joachimi}, Benjamin and {Kuijken}, Konrad and {Lin}, Chieh-An and {S{\'a}nchez}, Ariel G. and {van den Busch}, Jan Luca and {Wright}, Angus H. and {Amon}, Alexandra and {Bilicki}, Maciej and {de Jong}, Jelte and {Crocce}, Martin and {Dvornik}, Andrej and {Erben}, Thomas and {Fortuna}, Maria Cristina and {Getman}, Fedor and {Giblin}, Benjamin and {Glazebrook}, Karl and {Hoekstra}, Henk and {Joudaki}, Shahab and {Kannawadi}, Arun and {K{\"o}hlinger}, Fabian and {Lidman}, Chris and {Miller}, Lance and {Napolitano}, Nicola R. and {Parkinson}, David and {Schneider}, Peter and {Shan}, HuanYuan and {Valentijn}, Edwin A. and {Verdoes Kleijn}, Gijs and {Wolf}, Christian},
        title = "{KiDS-1000 Cosmology: Multi-probe weak gravitational lensing and spectroscopic galaxy clustering constraints}",
      journal = {\aap},
         year = 2021,
        month = feb,
       volume = {646},
          eid = {A140},
        pages = {A140},
          doi = {10.1051/0004-6361/202039063},
archivePrefix = {arXiv},
       eprint = {2007.15632},
 primaryClass = {astro-ph.CO},
       adsurl = {https://ui.adsabs.harvard.edu/abs/2021A&A...646A.140H}
}

@ARTICLE{Hikage.etal.2019,
       author = {{Hikage}, Chiaki and {Oguri}, Masamune and {Hamana}, Takashi and {More}, Surhud and {Mandelbaum}, Rachel and {Takada}, Masahiro and {K{\"o}hlinger}, Fabian and {Miyatake}, Hironao and {Nishizawa}, Atsushi J. and {Aihara}, Hiroaki and {Armstrong}, Robert and {Bosch}, James and {Coupon}, Jean and {Ducout}, Anne and {Ho}, Paul and {Hsieh}, Bau-Ching and {Komiyama}, Yutaka and {Lanusse}, Fran{\c{c}}ois and {Leauthaud}, Alexie and {Lupton}, Robert H. and {Medezinski}, Elinor and {Mineo}, Sogo and {Miyama}, Shoken and {Miyazaki}, Satoshi and {Murata}, Ryoma and {Murayama}, Hitoshi and {Shirasaki}, Masato and {Sif{\'o}n}, Crist{\'o}bal and {Simet}, Melanie and {Speagle}, Joshua and {Spergel}, David N. and {Strauss}, Michael A. and {Sugiyama}, Naoshi and {Tanaka}, Masayuki and {Utsumi}, Yousuke and {Wang}, Shiang-Yu and {Yamada}, Yoshihiko},
        title = "{Cosmology from cosmic shear power spectra with Subaru Hyper Suprime-Cam first-year data}",
      journal = {\pasj},
         year = 2019,
        month = apr,
       volume = {71},
       number = {2},
          eid = {43},
        pages = {43},
          doi = {10.1093/pasj/psz010},
archivePrefix = {arXiv},
       eprint = {1809.09148},
 primaryClass = {astro-ph.CO},
       adsurl = {https://ui.adsabs.harvard.edu/abs/2019PASJ...71...43H}
}

@ARTICLE{Hildebrandt.etal.2020,
       author = {{Hildebrandt}, H. and {K{\"o}hlinger}, F. and {van den Busch}, J.~L. and {Joachimi}, B. and {Heymans}, C. and {Kannawadi}, A. and {Wright}, A.~H. and {Asgari}, M. and {Blake}, C. and {Hoekstra}, H. and {Joudaki}, S. and {Kuijken}, K. and {Miller}, L. and {Morrison}, C.~B. and {Tr{\"o}ster}, T. and {Amon}, A. and {Archidiacono}, M. and {Brieden}, S. and {Choi}, A. and {de Jong}, J.~T.~A. and {Erben}, T. and {Giblin}, B. and {Mead}, A. and {Peacock}, J.~A. and {Radovich}, M. and {Schneider}, P. and {Sif{\'o}n}, C. and {Tewes}, M.},
        title = "{KiDS+VIKING-450: Cosmic shear tomography with optical and infrared data}",
      journal = {\aap},
         year = 2020,
        month = jan,
       volume = {633},
          eid = {A69},
        pages = {A69},
          doi = {10.1051/0004-6361/201834878},
archivePrefix = {arXiv},
       eprint = {1812.06076},
 primaryClass = {astro-ph.CO},
       adsurl = {https://ui.adsabs.harvard.edu/abs/2020A&A...633A..69H}
}

@ARTICLE{Hirata.Seljak.04,
       author = {{Hirata}, Christopher M. and {Seljak}, Uro{\v{s}}},
        title = "{Intrinsic alignment-lensing interference as a contaminant of cosmic shear}",
      journal = {\prd},
         year = 2004,
        month = sep,
       volume = {70},
       number = {6},
          eid = {063526},
        pages = {063526},
          doi = {10.1103/PhysRevD.70.063526},
archivePrefix = {arXiv},
       eprint = {astro-ph/0406275},
 primaryClass = {astro-ph},
       adsurl = {https://ui.adsabs.harvard.edu/abs/2004PhRvD..70f3526H}
}

@ARTICLE{Huff.Mandelbaum.17,
       author = {{Huff}, Eric and {Mandelbaum}, Rachel},
        title = "{Metacalibration: Direct Self-Calibration of Biases in Shear Measurement}",
      journal = {arXiv e-prints},
         year = 2017,
        month = feb,
          eid = {arXiv:1702.02600},
        pages = {arXiv:1702.02600},
          doi = {10.48550/arXiv.1702.02600},
archivePrefix = {arXiv},
       eprint = {1702.02600},
 primaryClass = {astro-ph.CO},
       adsurl = {https://ui.adsabs.harvard.edu/abs/2017arXiv170202600H}
}

@ARTICLE{Ivanov.etal.2020,
       author = {{Ivanov}, Mikhail M. and {Simonovi{\'c}}, Marko and {Zaldarriaga}, Matias},
        title = "{Cosmological parameters from the BOSS galaxy power spectrum}",
      journal = {\jcap},
         year = 2020,
        month = may,
       volume = {2020},
       number = {5},
          eid = {042},
        pages = {042},
          doi = {10.1088/1475-7516/2020/05/042},
archivePrefix = {arXiv},
       eprint = {1909.05277},
 primaryClass = {astro-ph.CO},
       adsurl = {https://ui.adsabs.harvard.edu/abs/2020JCAP...05..042I}
}

@ARTICLE{Jiang.vdBosch.17,
   author = {{Jiang}, F. and {van den Bosch}, F.~C.},
    title = "{Statistics of dark matter substructure - III. Halo-to-halo variance}",
  journal = {\mnras},
archivePrefix = "arXiv",
   eprint = {1610.02399},
     year = 2017,
    month = nov,
   volume = 472,
    pages = {657-674},
      doi = {10.1093/mnras/stx1979},
   adsurl = {http://adsabs.harvard.edu/abs/2017MNRAS.472..657J}
}

@ARTICLE{Jing.etal.98,
       author = {{Jing}, Y.~P. and {Mo}, H.~J. and {B{\"o}rner}, G.},
        title = "{Spatial Correlation Function and Pairwise Velocity Dispersion of Galaxies: Cold Dark Matter Models versus the Las Campanas Survey}",
      journal = {\apj},
         year = 1998,
        month = feb,
       volume = {494},
       number = {1},
        pages = {1-12},
          doi = {10.1086/305209},
archivePrefix = {arXiv},
       eprint = {astro-ph/9707106},
 primaryClass = {astro-ph},
       adsurl = {https://ui.adsabs.harvard.edu/abs/1998ApJ...494....1J}
}

@ARTICLE{Joachimi.etal.15,
       author = {{Joachimi}, Benjamin and {Cacciato}, Marcello and {Kitching}, Thomas D. and {Leonard}, Adrienne and {Mandelbaum}, Rachel and {Sch{\"a}fer}, Bj{\"o}rn Malte and {Sif{\'o}n}, Crist{\'o}bal and {Hoekstra}, Henk and {Kiessling}, Alina and {Kirk}, Donnacha and {Rassat}, Anais},
        title = "{Galaxy Alignments: An Overview}",
      journal = {\ssr},
         year = 2015,
        month = nov,
       volume = {193},
       number = {1-4},
        pages = {1-65},
          doi = {10.1007/s11214-015-0177-4},
archivePrefix = {arXiv},
       eprint = {1504.05456},
 primaryClass = {astro-ph.GA},
       adsurl = {https://ui.adsabs.harvard.edu/abs/2015SSRv..193....1J}
}

@ARTICLE{Joudaki.etal.2017,
       author = {{Joudaki}, Shahab and {Blake}, Chris and {Heymans}, Catherine and {Choi}, Ami and {Harnois-Deraps}, Joachim and {Hildebrandt}, Hendrik and {Joachimi}, Benjamin and {Johnson}, Andrew and {Mead}, Alexander and {Parkinson}, David and {Viola}, Massimo and {van Waerbeke}, Ludovic},
        title = "{CFHTLenS revisited: assessing concordance with Planck including astrophysical systematics}",
      journal = {\mnras},
         year = 2017,
        month = feb,
       volume = {465},
       number = {2},
        pages = {2033-2052},
          doi = {10.1093/mnras/stw2665},
archivePrefix = {arXiv},
       eprint = {1601.05786},
 primaryClass = {astro-ph.CO},
       adsurl = {https://ui.adsabs.harvard.edu/abs/2017MNRAS.465.2033J}
}

@ARTICLE{Joudaki.etal.2020,
       author = {{Joudaki}, S. and {Hildebrandt}, H. and {Traykova}, D. and {Chisari}, N.~E. and {Heymans}, C. and {Kannawadi}, A. and {Kuijken}, K. and {Wright}, A.~H. and {Asgari}, M. and {Erben}, T. and {Hoekstra}, H. and {Joachimi}, B. and {Miller}, L. and {Tr{\"o}ster}, T. and {van den Busch}, J.~L.},
        title = "{KiDS+VIKING-450 and DES-Y1 combined: Cosmology with cosmic shear}",
      journal = {\aap},
         year = 2020,
        month = jun,
       volume = {638},
          eid = {L1},
        pages = {L1},
          doi = {10.1051/0004-6361/201936154},
archivePrefix = {arXiv},
       eprint = {1906.09262},
 primaryClass = {astro-ph.CO},
       adsurl = {https://ui.adsabs.harvard.edu/abs/2020A&A...638L...1J}
}

@INCOLLECTION{Kazantzidis.&.Perivolaropoulos.2021,
       author = {{Kazantzidis}, Lavrentios and {Perivolaropoulos}, Leandros},
        title = "{{\ensuremath{\sigma}}$_{8}$ Tension. Is Gravity Getting Weaker at Low z? Observational Evidence and Theoretical Implications}",
    booktitle = {Modified Gravity and Cosmology; An Update by the CANTATA Network},
         year = 2021,
       editor = {{Saridakis}, Emmanuel N. and {Lazkoz}, Ruth and {Salzano}, Vincenzo and {Moniz}, Paulo Vargas and {Capozziello}, Salvatore and {Beltr{\'a}n Jim{\'e}nez}, Jose and {De Laurentis}, Mariafelicia and {Olmo}, Gonzalo J.},
        pages = {507-537},
          doi = {10.1007/978-3-030-83715-0_33},
       adsurl = {https://ui.adsabs.harvard.edu/abs/2021mgca.book..507K},
}

@ARTICLE{Kilbinger.etal.2013,
       author = {{Kilbinger}, Martin and {Fu}, Liping and {Heymans}, Catherine and {Simpson}, Fergus and {Benjamin}, Jonathan and {Erben}, Thomas and {Harnois-D{\'e}raps}, Joachim and {Hoekstra}, Henk and {Hildebrandt}, Hendrik and {Kitching}, Thomas D. and {Mellier}, Yannick and {Miller}, Lance and {Van Waerbeke}, Ludovic and {Benabed}, Karim and {Bonnett}, Christopher and {Coupon}, Jean and {Hudson}, Michael J. and {Kuijken}, Konrad and {Rowe}, Barnaby and {Schrabback}, Tim and {Semboloni}, Elisabetta and {Vafaei}, Sanaz and {Velander}, Malin},
        title = "{CFHTLenS: combined probe cosmological model comparison using 2D weak gravitational lensing}",
      journal = {\mnras},
         year = 2013,
        month = apr,
       volume = {430},
       number = {3},
        pages = {2200-2220},
          doi = {10.1093/mnras/stt041},
archivePrefix = {arXiv},
       eprint = {1212.3338},
 primaryClass = {astro-ph.CO},
       adsurl = {https://ui.adsabs.harvard.edu/abs/2013MNRAS.430.2200K}
}

@ARTICLE{Klypin.etal.16,
   author = {{Klypin}, A. and {Yepes}, G. and {Gottl{\"o}ber}, S. and {Prada}, F. and 
	{He{\ss}}, S.},
    title = "{MultiDark simulations: the story of dark matter halo concentrations and density profiles}",
  journal = {\mnras},
archivePrefix = "arXiv",
   eprint = {1411.4001},
     year = 2016,
    month = apr,
   volume = 457,
    pages = {4340-4359},
      doi = {10.1093/mnras/stw248},
   adsurl = {http://adsabs.harvard.edu/abs/2016MNRAS.457.4340K}
}

@ARTICLE{Komatsu.etal.2011,
       author = {{Komatsu}, E. and {Smith}, K.~M. and {Dunkley}, J. and {Bennett}, C.~L. and {Gold}, B. and {Hinshaw}, G. and {Jarosik}, N. and {Larson}, D. and {Nolta}, M.~R. and {Page}, L. and {Spergel}, D.~N. and {Halpern}, M. and {Hill}, R.~S. and {Kogut}, A. and {Limon}, M. and {Meyer}, S.~S. and {Odegard}, N. and {Tucker}, G.~S. and {Weiland}, J.~L. and {Wollack}, E. and {Wright}, E.~L.},
        title = "{Seven-year Wilkinson Microwave Anisotropy Probe (WMAP) Observations: Cosmological Interpretation}",
      journal = {\apjs},
         year = 2011,
        month = feb,
       volume = {192},
       number = {2},
          eid = {18},
        pages = {18},
          doi = {10.1088/0067-0049/192/2/18},
archivePrefix = {arXiv},
       eprint = {1001.4538},
 primaryClass = {astro-ph.CO},
       adsurl = {https://ui.adsabs.harvard.edu/abs/2011ApJS..192...18K}
}

@ARTICLE{Krolewski.etal.2021,
       author = {{Krolewski}, Alex and {Ferraro}, Simone and {White}, Martin},
        title = "{Cosmological constraints from unWISE and Planck CMB lensing tomography}",
      journal = {\jcap},
         year = 2021,
        month = dec,
       volume = {2021},
       number = {12},
          eid = {028},
        pages = {028},
          doi = {10.1088/1475-7516/2021/12/028},
archivePrefix = {arXiv},
       eprint = {2105.03421},
 primaryClass = {astro-ph.CO},
       adsurl = {https://ui.adsabs.harvard.edu/abs/2021JCAP...12..028K}
}

@ARTICLE{Kumar.etal.2019,
       author = {{Kumar}, Suresh and {Nunes}, Rafael C. and {Yadav}, Santosh Kumar},
        title = "{Dark sector interaction: a remedy of the tensions between CMB and LSS data}",
      journal = {European Physical Journal C},
         year = 2019,
        month = jul,
       volume = {79},
       number = {7},
          eid = {576},
        pages = {576},
          doi = {10.1140/epjc/s10052-019-7087-7},
archivePrefix = {arXiv},
       eprint = {1903.04865},
 primaryClass = {astro-ph.CO},
       adsurl = {https://ui.adsabs.harvard.edu/abs/2019EPJC...79..576K}
}

@ARTICLE{Kunz.etal.2015,
       author = {{Kunz}, Martin and {Nesseris}, Savvas and {Sawicki}, Ignacy},
        title = "{Using dark energy to suppress power at small scales}",
      journal = {\prd},
         year = 2015,
        month = sep,
       volume = {92},
       number = {6},
          eid = {063006},
        pages = {063006},
          doi = {10.1103/PhysRevD.92.063006},
archivePrefix = {arXiv},
       eprint = {1507.01486},
 primaryClass = {astro-ph.CO},
       adsurl = {https://ui.adsabs.harvard.edu/abs/2015PhRvD..92f3006K}
}

@ARTICLE{Lacerna.Padilla.11,
       author = {{Lacerna}, Ivan and {Padilla}, Nelson},
        title = "{The nature of assembly bias - I. Clues from a {\ensuremath{\Lambda}}CDM cosmology}",
      journal = {\mnras},
         year = 2011,
        month = apr,
       volume = {412},
       number = {2},
        pages = {1283-1294},
          doi = {10.1111/j.1365-2966.2010.17988.x},
archivePrefix = {arXiv},
       eprint = {1011.1498},
 primaryClass = {astro-ph.CO},
       adsurl = {https://ui.adsabs.harvard.edu/abs/2011MNRAS.412.1283L}
}

@ARTICLE{Lange.etal.19a,
   author = {{Lange}, J.~U. and {van den Bosch}, F.~C. and {Zentner}, A.~R. and 
	{Wang}, K. and {Villarreal}, A.~S.},
    title = "{Maturing satellite kinematics into a competitive probe of the galaxy-halo connection}",
  journal = {\mnras},
archivePrefix = "arXiv",
   eprint = {1810.10511},
     year = 2019,
    month = feb,
   volume = 482,
    pages = {4824-4845},
      doi = {10.1093/mnras/sty2950},
   adsurl = {http://adsabs.harvard.edu/abs/2019MNRAS.482.4824L}
}

@ARTICLE{Lange.etal.19b,
   author = {{Lange}, J.~U. and {van den Bosch}, F.~C. and {Zentner}, A.~R. and 
	{Wang}, K. and {Villarreal}, A.~S.},
    title = "{Updated results on the galaxy-halo connection from satellite kinematics in SDSS}",
  journal = {\mnras},
archivePrefix = "arXiv",
   eprint = {1811.03596},
     year = 2019,
    month = aug,
   volume = 487,
    pages = {3112-3129},
      doi = {10.1093/mnras/stz1466},
   adsurl = {https://ui.adsabs.harvard.edu/abs/2019MNRAS.487.3112L}
}

@ARTICLE{Lange.etal.19c,
       author = {{Lange}, Johannes U. and {Yang}, Xiaohu and {Guo}, Hong and {Luo}, Wentao and {van den Bosch}, Frank C.},
        title = "{New perspectives on the BOSS small-scale lensing discrepancy for the Planck {\ensuremath{\Lambda}}CDM cosmology}",
      journal = {\mnras},
         year = 2019,
        month = oct,
       volume = {488},
       number = {4},
        pages = {5771-5787},
          doi = {10.1093/mnras/stz2124},
archivePrefix = {arXiv},
       eprint = {1906.08680},
 primaryClass = {astro-ph.CO},
       adsurl = {https://ui.adsabs.harvard.edu/abs/2019MNRAS.488.5771L}
}

@ARTICLE{Lange.etal.19d,
       author = {{Lange}, Johannes U. and {van den Bosch}, Frank C. and {Zentner}, Andrew R. and {Wang}, Kuan and {Hearin}, Andrew P. and {Guo}, Hong},
        title = "{Cosmological Evidence Modelling: a new simulation-based approach to constrain cosmology on non-linear scales}",
      journal = {\mnras},
         year = 2019,
        month = dec,
       volume = {490},
       number = {2},
        pages = {1870-1878},
          doi = {10.1093/mnras/stz2664},
archivePrefix = {arXiv},
       eprint = {1909.03107},
 primaryClass = {astro-ph.CO},
       adsurl = {https://ui.adsabs.harvard.edu/abs/2019MNRAS.490.1870L}
}

@ARTICLE{Lange.etal.21,
       author = {{Lange}, Johannes U. and {Leauthaud}, Alexie and {Singh}, Sukhdeep and {Guo}, Hong and {Zhou}, Rongpu and {Smith}, Tristan L. and {Cyr-Racine}, Francis-Yan},
        title = "{On the halo-mass and radial scale dependence of the lensing is low effect}",
      journal = {\mnras},
         year = 2021,
        month = apr,
       volume = {502},
       number = {2},
        pages = {2074-2086},
          doi = {10.1093/mnras/stab189},
archivePrefix = {arXiv},
       eprint = {2011.02377},
 primaryClass = {astro-ph.CO},
       adsurl = {https://ui.adsabs.harvard.edu/abs/2021MNRAS.502.2074L}
}

@ARTICLE{Lange.etal.23,
       author = {{Lange}, Johannes U. and {Hearin}, Andrew P. and {Leauthaud}, Alexie and {van den Bosch}, Frank C. and {Xhakaj}, Enia and {Guo}, Hong and {Wechsler}, Risa H. and {DeRose}, Joseph},
        title = "{Constraints on S$_{8}$ from a full-scale and full-shape analysis of redshift-space clustering and galaxy-galaxy lensing in BOSS}",
      journal = {\mnras},
         year = 2023,
        month = apr,
       volume = {520},
       number = {4},
        pages = {5373-5393},
          doi = {10.1093/mnras/stad473},
archivePrefix = {arXiv},
       eprint = {2301.08692},
 primaryClass = {astro-ph.CO},
       adsurl = {https://ui.adsabs.harvard.edu/abs/2023MNRAS.520.5373L}
}

@ARTICLE{Leauthaud.etal.17,
   author = {{Leauthaud}, A. and {Saito}, S. and {Hilbert}, S. and {Barreira}, A. and 
	{More}, S. and {White}, M. and {Alam}, S. and {Behroozi}, P. and 
	{Bundy}, K. and {Coupon}, J. and {Erben}, T. and {Heymans}, C. and 
	{Hildebrandt}, H. and {Mandelbaum}, R. and {Miller}, L. and 
	{Moraes}, B. and {Pereira}, M.~E.~S. and {Rodr{\'{\i}}guez-Torres}, S.~A. and 
	{Schmidt}, F. and {Shan}, H.-Y. and {Viel}, M. and {Villaescusa-Navarro}, F.
	},
    title = "{Lensing is low: cosmology, galaxy formation or new physics?}",
  journal = {\mnras},
archivePrefix = "arXiv",
   eprint = {1611.08606},
     year = 2017,
    month = may,
   volume = 467,
    pages = {3024-3047},
      doi = {10.1093/mnras/stx258},
   adsurl = {http://adsabs.harvard.edu/abs/2017MNRAS.467.3024L}
}

@ARTICLE{Lehmann.etal.17,
       author = {{Lehmann}, Benjamin V. and {Mao}, Yao-Yuan and {Becker}, Matthew R. and {Skillman}, Samuel W. and {Wechsler}, Risa H.},
        title = "{The Concentration Dependence of the Galaxy-Halo Connection: Modeling Assembly Bias with Abundance Matching}",
      journal = {\apj},
         year = 2017,
        month = jan,
       volume = {834},
       number = {1},
          eid = {37},
        pages = {37},
          doi = {10.3847/1538-4357/834/1/37},
archivePrefix = {arXiv},
       eprint = {1510.05651},
 primaryClass = {astro-ph.CO},
       adsurl = {https://ui.adsabs.harvard.edu/abs/2017ApJ...834...37L}
}

@ARTICLE{Li.etal.2008,
       author = {{Li}, Yun and {Mo}, H.~J. and {Gao}, L.},
        title = "{On halo formation times and assembly bias}",
      journal = {\mnras},
         year = 2008,
        month = sep,
       volume = {389},
       number = {3},
        pages = {1419-1426},
          doi = {10.1111/j.1365-2966.2008.13667.x},
archivePrefix = {arXiv},
       eprint = {0803.2250},
 primaryClass = {astro-ph},
       adsurl = {https://ui.adsabs.harvard.edu/abs/2008MNRAS.389.1419L}
}

@ARTICLE{Li.etal.2016,
       author = {{Li}, Zhigang and {Jing}, Y.~P. and {Zhang}, Pengjie and {Cheng}, Dalong},
        title = "{Measurement of A Redshift-space Power Spectrum for BOSS Galaxies and the Growth Rate at Redshift 0.57}",
      journal = {\apj},
         year = 2016,
        month = dec,
       volume = {833},
       number = {2},
          eid = {287},
        pages = {287},
          doi = {10.3847/1538-4357/833/2/287},
archivePrefix = {arXiv},
       eprint = {1609.03697},
 primaryClass = {astro-ph.CO},
       adsurl = {https://ui.adsabs.harvard.edu/abs/2016ApJ...833..287L}
}

@ARTICLE{Lin.etal.04,
   author = {{Lin}, Y.-T. and {Mohr}, J.~J. and {Stanford}, S.~A.},
    title = "{K-Band Properties of Galaxy Clusters and Groups: Luminosity Function, Radial Distribution, and Halo Occupation Number}",
  journal = {\apj},
   eprint = {astro-ph/0402308},
     year = 2004,
    month = aug,
   volume = 610,
    pages = {745-761},
      doi = {10.1086/421714},
   adsurl = {http://adsabs.harvard.edu/abs/2004ApJ...610..745L}
}

@ARTICLE{Lin.etal.2012,
       author = {{Lin}, Yen-Ting and {Stanford}, S. Adam and {Eisenhardt}, Peter R.~M. and {Vikhlinin}, Alexey and {Maughan}, Ben J. and {Kravtsov}, Andrey},
        title = "{Baryon Content of Massive Galaxy Clusters at z = 0-0.6}",
      journal = {\apjl},
         year = 2012,
        month = jan,
       volume = {745},
       number = {1},
          eid = {L3},
        pages = {L3},
          doi = {10.1088/2041-8205/745/1/L3},
archivePrefix = {arXiv},
       eprint = {1112.1705},
 primaryClass = {astro-ph.CO},
       adsurl = {https://ui.adsabs.harvard.edu/abs/2012ApJ...745L...3L}
}

@article{Lokas.02,
  title={Dark matter distribution in dwarf spheroidal galaxies},
  author={{\L}okas, Ewa L},
  journal={Monthly Notices of the Royal Astronomical Society},
  volume={333},
  number={3},
  pages={697--708},
  year={2002},
  publisher={Blackwell Science Ltd Oxford, UK}
}

@article{Lokas.Mamon.03,
  title={Dark matter distribution in the Coma cluster from galaxy kinematics: breaking the mass--anisotropy degeneracy},
  author={{\L}okas, Ewa L and Mamon, Gary A},
  journal={Monthly Notices of the Royal Astronomical Society},
  volume={343},
  number={2},
  pages={401--412},
  year={2003},
  publisher={The Royal Astronomical Society}
}

@ARTICLE{Loureiro.etal.2022,
       author = {{Loureiro}, A. and {Whittaker}, L. and {Spurio Mancini}, A. and {Joachimi}, B. and {Cuceu}, A. and {Asgari}, M. and {St{\"o}lzner}, B. and {Tr{\"o}ster}, T. and {Wright}, A.~H. and {Bilicki}, M. and {Dvornik}, A. and {Giblin}, B. and {Heymans}, C. and {Hildebrandt}, H. and {Shan}, H. and {Amara}, A. and {Auricchio}, N. and {Bodendorf}, C. and {Bonino}, D. and {Branchini}, E. and {Brescia}, M. and {Capobianco}, V. and {Carbone}, C. and {Carretero}, J. and {Castellano}, M. and {Cavuoti}, S. and {Cimatti}, A. and {Cledassou}, R. and {Congedo}, G. and {Conversi}, L. and {Copin}, Y. and {Corcione}, L. and {Cropper}, M. and {Da Silva}, A. and {Douspis}, M. and {Dubath}, F. and {Duncan}, C.~A.~J. and {Dupac}, X. and {Dusini}, S. and {Farrens}, S. and {Ferriol}, S. and {Fosalba}, P. and {Frailis}, M. and {Franceschi}, E. and {Fumana}, M. and {Garilli}, B. and {Gillis}, B. and {Giocoli}, C. and {Grazian}, A. and {Grupp}, F. and {Haugan}, S.~V.~H. and {Holmes}, W. and {Hormuth}, F. and {Jahnke}, K. and {K{\"u}mmel}, M. and {Kermiche}, S. and {Kiessling}, A. and {Kilbinger}, M. and {Kitching}, T. and {Kuijken}, K. and {Kunz}, M. and {Kurki-Suonio}, H. and {Ligori}, S. and {Lilje}, P.~B. and {Lloro}, I. and {Mansutti}, O. and {Marggraf}, O. and {Markovic}, K. and {Marulli}, F. and {Massey}, R. and {Meneghetti}, M. and {Meylan}, G. and {Moresco}, M. and {Morin}, B. and {Moscardini}, L. and {Munari}, E. and {Niemi}, S.~M. and {Padilla}, C. and {Paltani}, S. and {Pasian}, F. and {Pedersen}, K. and {Pettorino}, V. and {Pires}, S. and {Poncet}, M. and {Popa}, L. and {Raison}, F. and {Rhodes}, J. and {Rix}, H. and {Roncarelli}, M. and {Saglia}, R. and {Schneider}, P. and {Secroun}, A. and {Serrano}, S. and {Sirignano}, C. and {Sirri}, G. and {Stanco}, L. and {Starck}, J.~L. and {Tallada-Cresp{\'\i}}, P. and {Taylor}, A.~N. and {Tereno}, I. and {Toledo-Moreo}, R. and {Torradeflot}, F. and {Valentijn}, E.~A. and {Wang}, Y. and {Welikala}, N. and {Weller}, J. and {Zamorani}, G. and {Zoubian}, J. and {Andreon}, S. and {Baldi}, M. and {Camera}, S. and {Farinelli}, R. and {Polenta}, G. and {Tessore}, N.},
        title = "{KiDS and Euclid: Cosmological implications of a pseudo angular power spectrum analysis of KiDS-1000 cosmic shear tomography}",
      journal = {\aap},
         year = 2022,
        month = sep,
       volume = {665},
          eid = {A56},
        pages = {A56},
          doi = {10.1051/0004-6361/202142481},
archivePrefix = {arXiv},
       eprint = {2110.06947},
 primaryClass = {astro-ph.CO},
       adsurl = {https://ui.adsabs.harvard.edu/abs/2022A&A...665A..56L}
}

@ARTICLE{Lovisari.etal.2015,
       author = {{Lovisari}, L. and {Reiprich}, T.~H. and {Schellenberger}, G.},
        title = "{Scaling properties of a complete X-ray selected galaxy group sample}",
      journal = {\aap},
         year = 2015,
        month = jan,
       volume = {573},
          eid = {A118},
        pages = {A118},
          doi = {10.1051/0004-6361/201423954},
archivePrefix = {arXiv},
       eprint = {1409.3845},
 primaryClass = {astro-ph.CO},
       adsurl = {https://ui.adsabs.harvard.edu/abs/2015A&A...573A.118L}
}

@ARTICLE{Lucca.2021,
       author = {{Lucca}, Matteo},
        title = "{Dark energy-dark matter interactions as a solution to the S$_{8}$ tension}",
      journal = {Physics of the Dark Universe},
         year = 2021,
        month = dec,
       volume = {34},
          eid = {100899},
        pages = {100899},
          doi = {10.1016/j.dark.2021.100899},
archivePrefix = {arXiv},
       eprint = {2105.09249},
 primaryClass = {astro-ph.CO},
       adsurl = {https://ui.adsabs.harvard.edu/abs/2021PDU....3400899L}
}

@ARTICLE{Macaulay.etal.2013,
       author = {{Macaulay}, E. and {Wehus}, I.~K. and {Eriksen}, H.~K.},
        title = "{Lower Growth Rate from Recent Redshift Space Distortion Measurements than Expected from Planck}",
      journal = {\prl},
         year = 2013,
        month = oct,
       volume = {111},
       number = {16},
          eid = {161301},
        pages = {161301},
          doi = {10.1103/PhysRevLett.111.161301},
archivePrefix = {arXiv},
       eprint = {1303.6583},
 primaryClass = {astro-ph.CO},
       adsurl = {https://ui.adsabs.harvard.edu/abs/2013PhRvL.111p1301M}
}

@ARTICLE{Mandelbaum.etal.13,
   author = {{Mandelbaum}, R. and {Slosar}, A. and {Baldauf}, T. and {Seljak}, U. and 
	{Hirata}, C.~M. and {Nakajima}, R. and {Reyes}, R. and {Smith}, R.~E.
	},
    title = "{Cosmological parameter constraints from galaxy-galaxy lensing and galaxy clustering with the SDSS DR7}",
  journal = {\mnras},
archivePrefix = "arXiv",
   eprint = {1207.1120},
     year = 2013,
    month = jun,
   volume = 432,
    pages = {1544-1575},
      doi = {10.1093/mnras/stt572},
   adsurl = {http://adsabs.harvard.edu/abs/2013MNRAS.432.1544M}
}

@ARTICLE{Mantz.etal.2015,
       author = {{Mantz}, Adam B. and {von der Linden}, Anja and {Allen}, Steven W. and {Applegate}, Douglas E. and {Kelly}, Patrick L. and {Morris}, R. Glenn and {Rapetti}, David A. and {Schmidt}, Robert W. and {Adhikari}, Saroj and {Allen}, Mark T. and {Burchat}, Patricia R. and {Burke}, David L. and {Cataneo}, Matteo and {Donovan}, David and {Ebeling}, Harald and {Shandera}, Sarah and {Wright}, Adam},
        title = "{Weighing the giants - IV. Cosmology and neutrino mass}",
      journal = {\mnras},
         year = 2015,
        month = jan,
       volume = {446},
       number = {3},
        pages = {2205-2225},
          doi = {10.1093/mnras/stu2096},
archivePrefix = {arXiv},
       eprint = {1407.4516},
 primaryClass = {astro-ph.CO},
       adsurl = {https://ui.adsabs.harvard.edu/abs/2015MNRAS.446.2205M}
}

@ARTICLE{McAlpine.etal.16,
       author = {{McAlpine}, S. and {Helly}, J.~C. and {Schaller}, M. and {Trayford}, J.~W. and {Qu}, Y. and {Furlong}, M. and {Bower}, R.~G. and {Crain}, R.~A. and {Schaye}, J. and {Theuns}, T. and {Dalla Vecchia}, C. and {Frenk}, C.~S. and {McCarthy}, I.~G. and {Jenkins}, A. and {Rosas-Guevara}, Y. and {White}, S.~D.~M. and {Baes}, M. and {Camps}, P. and {Lemson}, G.},
        title = "{The EAGLE simulations of galaxy formation: Public release of halo and galaxy catalogues}",
      journal = {Astronomy and Computing},
         year = 2016,
        month = apr,
       volume = {15},
        pages = {72-89},
          doi = {10.1016/j.ascom.2016.02.004},
archivePrefix = {arXiv},
       eprint = {1510.01320},
 primaryClass = {astro-ph.GA},
       adsurl = {https://ui.adsabs.harvard.edu/abs/2016A&C....15...72M}
}

@ARTICLE{McCarthy.etal.2024,
       author = {{McCarthy}, Ian G. and {Amon}, Alexandra and {Schaye}, Joop and {Schaan}, Emmanuel and {Angulo}, Raul E. and {Salcido}, Jaime and {Schaller}, Matthieu and {Bigwood}, Leah and {Elbers}, Willem and {Kugel}, Roi and {Helly}, John C. and {Forouhar Moreno}, Victor J. and {Frenk}, Carlos S. and {McGibbon}, Robert J. and {Ondaro-Mallea}, Lurdes and {van Daalen}, Marcel P.},
        title = "{FLAMINGO: combining kinetic SZ effect and galaxy-galaxy lensing measurements to gauge the impact of feedback on large-scale structure}",
      journal = {arXiv e-prints},
         year = 2024,
        month = oct,
          eid = {arXiv:2410.19905},
        pages = {arXiv:2410.19905},
          doi = {10.48550/arXiv.2410.19905},
archivePrefix = {arXiv},
       eprint = {2410.19905},
 primaryClass = {astro-ph.CO},
       adsurl = {https://ui.adsabs.harvard.edu/abs/2024arXiv241019905M}
}

@ARTICLE{Mitra.etal.24,
       author = {{Mitra}, Kaustav and {van den Bosch}, Frank C. and {Lange}, Johannes U.},
        title = "{BASILISK II. Improved constraints on the galaxy-halo connection from satellite kinematics in SDSS}",
      journal = {\mnras},
         year = 2024,
        month = sep,
       volume = {533},
       number = {3},
        pages = {3647-3675},
          doi = {10.1093/mnras/stae2030},
archivePrefix = {arXiv},
       eprint = {2409.03105},
 primaryClass = {astro-ph.CO},
       adsurl = {https://ui.adsabs.harvard.edu/abs/2024MNRAS.533.3647M}
}

@ARTICLE{Mitra.etal.25,
       author = {{Mitra}, Kaustav and {van den Bosch}, Frank C.},
        title = "{BASILISK III. Stress-testing the Conditional Luminosity Function model}",
      journal = {\mnras},
         year = 2025,
        month = sep,
       volume = {in prep.},
       number = {},
        pages = {},
}

@article{Miyatake.etal.2022b,
  title = {Cosmological inference from an emulator based halo model. II. Joint analysis of galaxy-galaxy weak lensing and galaxy clustering from HSC-Y1 and SDSS},
  author = {Miyatake, Hironao and Sugiyama, Sunao and Takada, Masahiro and Nishimichi, Takahiro and Shirasaki, Masato and Kobayashi, Yosuke and Mandelbaum, Rachel and More, Surhud and Oguri, Masamune and Osato, Ken and Park, Youngsoo and Takahashi, Ryuichi and Coupon, Jean and Hikage, Chiaki and Hsieh, Bau-Ching and Komiyama, Yutaka and Leauthaud, Alexie and Li, Xiangchong and Luo, Wentao and Lupton, Robert H. and Miyazaki, Satoshi and Murayama, Hitoshi and Nishizawa, Atsushi J. and Price, Paul A. and Simet, Melanie and Speagle, Joshua S. and Strauss, Michael A. and Tanaka, Masayuki and Yoshida, Naoki},
  journal = {Phys. Rev. D},
  volume = {106},
  issue = {8},
  pages = {083520},
  numpages = {39},
  year = {2022},
  month = {Oct},
  publisher = {American Physical Society},
  doi = {10.1103/PhysRevD.106.083520},
  url = {https://link.aps.org/doi/10.1103/PhysRevD.106.083520}
}

@ARTICLE{More.13,
       author = {{More}, Surhud},
        title = "{Cosmological Dependence of the Measurements of Luminosity Function, Projected Clustering and Galaxy-Galaxy Lensing Signal}",
      journal = {\apjl},
         year = 2013,
        month = nov,
       volume = {777},
       number = {2},
          eid = {L26},
        pages = {L26},
          doi = {10.1088/2041-8205/777/2/L26},
archivePrefix = {arXiv},
       eprint = {1309.2943},
 primaryClass = {astro-ph.CO},
       adsurl = {https://ui.adsabs.harvard.edu/abs/2013ApJ...777L..26M}
}

@ARTICLE{More.etal.11,
   author = {{More}, S. and {van den Bosch}, F.~C. and {Cacciato}, M. and 
	{Skibba}, R. and {Mo}, H.~J. and {Yang}, X.},
    title = "{Satellite kinematics - III. Halo masses of central galaxies in SDSS}",
  journal = {\mnras},
archivePrefix = "arXiv",
   eprint = {1003.3203},
     year = 2011,
    month = jan,
   volume = 410,
    pages = {210-226},
      doi = {10.1111/j.1365-2966.2010.17436.x},
   adsurl = {http://adsabs.harvard.edu/abs/2011MNRAS.410..210M}
}

@ARTICLE{More.etal.09b,
   author = {{More}, S. and {van den Bosch}, F.~C. and {Cacciato}, M. and 
	{Mo}, H.~J. and {Yang}, X. and {Li}, R.},
    title = "{Satellite kinematics - II. The halo mass-luminosity relation of central galaxies in SDSS}",
  journal = {\mnras},
archivePrefix = "arXiv",
   eprint = {0807.4532},
     year = 2009,
    month = jan,
   volume = 392,
    pages = {801-816},
      doi = {10.1111/j.1365-2966.2008.14095.x},
   adsurl = {http://adsabs.harvard.edu/abs/2009MNRAS.392..801M}
}

@ARTICLE{Moster.etal.10,
       author = {{Moster}, Benjamin P. and {Somerville}, Rachel S. and {Maulbetsch}, Christian and {van den Bosch}, Frank C. and {Macci{\`o}}, Andrea V. and {Naab}, Thorsten and {Oser}, Ludwig},
        title = "{Constraints on the Relationship between Stellar Mass and Halo Mass at Low and High Redshift}",
      journal = {\apj},
         year = 2010,
        month = feb,
       volume = {710},
       number = {2},
        pages = {903-923},
          doi = {10.1088/0004-637X/710/2/903},
archivePrefix = {arXiv},
       eprint = {0903.4682},
 primaryClass = {astro-ph.CO},
       adsurl = {https://ui.adsabs.harvard.edu/abs/2010ApJ...710..903M}
}

@ARTICLE{Navarro.etal.97,
   author = {{Navarro}, J.~F. and {Frenk}, C.~S. and {White}, S.~D.~M.},
    title = "{A Universal Density Profile from Hierarchical Clustering}",
  journal = {\apj},
   eprint = {astro-ph/9611107},
     year = 1997,
    month = dec,
   volume = 490,
    pages = {493-508},
      doi = {10.1086/304888},
   adsurl = {http://adsabs.harvard.edu/abs/1997ApJ...490..493N}
}

@ARTICLE{Nelson.etal.2015,
       author = {{Nelson}, D. and {Pillepich}, A. and {Genel}, S. and {Vogelsberger}, M. and {Springel}, V. and {Torrey}, P. and {Rodriguez-Gomez}, V. and {Sijacki}, D. and {Snyder}, G.~F. and {Griffen}, B. and {Marinacci}, F. and {Blecha}, L. and {Sales}, L. and {Xu}, D. and {Hernquist}, L.},
        title = "{The illustris simulation: Public data release}",
      journal = {Astronomy and Computing},
         year = 2015,
        month = nov,
       volume = {13},
        pages = {12-37},
          doi = {10.1016/j.ascom.2015.09.003},
archivePrefix = {arXiv},
       eprint = {1504.00362},
 primaryClass = {astro-ph.CO},
       adsurl = {https://ui.adsabs.harvard.edu/abs/2015A&C....13...12N}
}

@ARTICLE{Nesseris.etal.2017,
       author = {{Nesseris}, Savvas and {Pantazis}, George and {Perivolaropoulos}, Leandros},
        title = "{Tension and constraints on modified gravity parametrizations of G$_{eff}$(z ) from growth rate and Planck data}",
      journal = {\prd},
         year = 2017,
        month = jul,
       volume = {96},
       number = {2},
          eid = {023542},
        pages = {023542},
          doi = {10.1103/PhysRevD.96.023542},
archivePrefix = {arXiv},
       eprint = {1703.10538},
 primaryClass = {astro-ph.CO},
       adsurl = {https://ui.adsabs.harvard.edu/abs/2017PhRvD..96b3542N}
}

@ARTICLE{Nguyen.etal.2023,
       author = {{Nguyen}, Nhat-Minh and {Huterer}, Dragan and {Wen}, Yuewei},
        title = "{Evidence for Suppression of Structure Growth in the Concordance Cosmological Model}",
      journal = {\prl},
         year = 2023,
        month = sep,
       volume = {131},
       number = {11},
          eid = {111001},
        pages = {111001},
          doi = {10.1103/PhysRevLett.131.111001},
archivePrefix = {arXiv},
       eprint = {2302.01331},
 primaryClass = {astro-ph.CO},
       adsurl = {https://ui.adsabs.harvard.edu/abs/2023PhRvL.131k1001N}
}

@ARTICLE{Norberg.etal.08,
   author = {{Norberg}, P. and {Frenk}, C.~S. and {Cole}, S.},
    title = "{Massive dark matter haloes around bright isolated galaxies in the 2dFGRS}",
  journal = {\mnras},
archivePrefix = "arXiv",
   eprint = {0710.5473},
     year = 2008,
    month = jan,
   volume = 383,
    pages = {646-662},
      doi = {10.1111/j.1365-2966.2007.12583.x},
   adsurl = {http://adsabs.harvard.edu/abs/2008MNRAS.383..646N}
}

@ARTICLE{Perivolaropoulos.&.Skara.2022,
       author = {{Perivolaropoulos}, L. and {Skara}, F.},
        title = "{Challenges for {\ensuremath{\Lambda}}CDM: An update}",
      journal = {\nar},
         year = 2022,
        month = dec,
       volume = {95},
          eid = {101659},
        pages = {101659},
          doi = {10.1016/j.newar.2022.101659},
archivePrefix = {arXiv},
       eprint = {2105.05208},
 primaryClass = {astro-ph.CO},
       adsurl = {https://ui.adsabs.harvard.edu/abs/2022NewAR..9501659P}
}

@ARTICLE{Philcox.etal.2022,
       author = {{Philcox}, Oliver H.~E. and {Ivanov}, Mikhail M.},
        title = "{BOSS DR12 full-shape cosmology: {\ensuremath{\Lambda}} CDM constraints from the large-scale galaxy power spectrum and bispectrum monopole}",
      journal = {\prd},
         year = 2022,
        month = feb,
       volume = {105},
       number = {4},
          eid = {043517},
        pages = {043517},
          doi = {10.1103/PhysRevD.105.043517},
archivePrefix = {arXiv},
       eprint = {2112.04515},
 primaryClass = {astro-ph.CO},
       adsurl = {https://ui.adsabs.harvard.edu/abs/2022PhRvD.105d3517P}
}

@INPROCEEDINGS{Pillepich.etal.2015,
       author = {{Pillepich}, Annalisa and {Torrey}, Paul and {Snyder}, Gregory and {Nelson}, Dylan and {Genel}, Shy and {Vogelsberger}, Mark and {Sijacki}, Debora and {Springel}, Volker and {Hernquist}, Lars},
        title = "{The Illustris simulation overview: mock observations of galaxies and their stellar haloes}",
    booktitle = {IAU General Assembly},
         year = 2015,
       volume = {29},
        month = aug,
          eid = {2258509},
        pages = {2258509},
       adsurl = {https://ui.adsabs.harvard.edu/abs/2015IAUGA..2258509P}
}

@ARTICLE{Pillepich.etal.2018a,
       author = {{Pillepich}, Annalisa and {Springel}, Volker and {Nelson}, Dylan and {Genel}, Shy and {Naiman}, Jill and {Pakmor}, R{\"u}diger and {Hernquist}, Lars and {Torrey}, Paul and {Vogelsberger}, Mark and {Weinberger}, Rainer and {Marinacci}, Federico},
        title = "{Simulating galaxy formation with the IllustrisTNG model}",
      journal = {\mnras},
         year = 2018,
        month = jan,
       volume = {473},
       number = {3},
        pages = {4077-4106},
          doi = {10.1093/mnras/stx2656},
archivePrefix = {arXiv},
       eprint = {1703.02970},
 primaryClass = {astro-ph.GA},
       adsurl = {https://ui.adsabs.harvard.edu/abs/2018MNRAS.473.4077P}
}

@ARTICLE{Planck.16.ClusterCosmo,
       author = {{Planck Collaboration} and {Ade}, P.~A.~R. and {Aghanim}, N. and {Arnaud}, M. and {Ashdown}, M. and {Aumont}, J. and {Baccigalupi}, C. and {Banday}, A.~J. and {Barreiro}, R.~B. and {Bartlett}, J.~G. and {Bartolo}, N. and {Battaner}, E. and {Battye}, R. and {Benabed}, K. and {Beno{\^\i}t}, A. and {Benoit-L{\'e}vy}, A. and {Bernard}, J. -P. and {Bersanelli}, M. and {Bielewicz}, P. and {Bock}, J.~J. and {Bonaldi}, A. and {Bonavera}, L. and {Bond}, J.~R. and {Borrill}, J. and {Bouchet}, F.~R. and {Bucher}, M. and {Burigana}, C. and {Butler}, R.~C. and {Calabrese}, E. and {Cardoso}, J. -F. and {Catalano}, A. and {Challinor}, A. and {Chamballu}, A. and {Chary}, R. -R. and {Chiang}, H.~C. and {Christensen}, P.~R. and {Church}, S. and {Clements}, D.~L. and {Colombi}, S. and {Colombo}, L.~P.~L. and {Combet}, C. and {Comis}, B. and {Couchot}, F. and {Coulais}, A. and {Crill}, B.~P. and {Curto}, A. and {Cuttaia}, F. and {Danese}, L. and {Davies}, R.~D. and {Davis}, R.~J. and {de Bernardis}, P. and {de Rosa}, A. and {de Zotti}, G. and {Delabrouille}, J. and {D{\'e}sert}, F. -X. and {Diego}, J.~M. and {Dolag}, K. and {Dole}, H. and {Donzelli}, S. and {Dor{\'e}}, O. and {Douspis}, M. and {Ducout}, A. and {Dupac}, X. and {Efstathiou}, G. and {Elsner}, F. and {En{\ss}lin}, T.~A. and {Eriksen}, H.~K. and {Falgarone}, E. and {Fergusson}, J. and {Finelli}, F. and {Forni}, O. and {Frailis}, M. and {Fraisse}, A.~A. and {Franceschi}, E. and {Frejsel}, A. and {Galeotta}, S. and {Galli}, S. and {Ganga}, K. and {Giard}, M. and {Giraud-H{\'e}raud}, Y. and {Gjerl{\o}w}, E. and {Gonz{\'a}lez-Nuevo}, J. and {G{\'o}rski}, K.~M. and {Gratton}, S. and {Gregorio}, A. and {Gruppuso}, A. and {Gudmundsson}, J.~E. and {Hansen}, F.~K. and {Hanson}, D. and {Harrison}, D.~L. and {Henrot-Versill{\'e}}, S. and {Hern{\'a}ndez-Monteagudo}, C. and {Herranz}, D. and {Hildebrandt}, S.~R. and {Hivon}, E. and {Hobson}, M. and {Holmes}, W.~A. and {Hornstrup}, A. and {Hovest}, W. and {Huffenberger}, K.~M. and {Hurier}, G. and {Jaffe}, A.~H. and {Jaffe}, T.~R. and {Jones}, W.~C. and {Juvela}, M. and {Keih{\"a}nen}, E. and {Keskitalo}, R. and {Kisner}, T.~S. and {Kneissl}, R. and {Knoche}, J. and {Kunz}, M. and {Kurki-Suonio}, H. and {Lagache}, G. and {L{\"a}hteenm{\"a}ki}, A. and {Lamarre}, J. -M. and {Lasenby}, A. and {Lattanzi}, M. and {Lawrence}, C.~R. and {Leonardi}, R. and {Lesgourgues}, J. and {Levrier}, F. and {Liguori}, M. and {Lilje}, P.~B. and {Linden-V{\o}rnle}, M. and {L{\'o}pez-Caniego}, M. and {Lubin}, P.~M. and {Mac{\'\i}as-P{\'e}rez}, J.~F. and {Maggio}, G. and {Maino}, D. and {Mandolesi}, N. and {Mangilli}, A. and {Maris}, M. and {Martin}, P.~G. and {Mart{\'\i}nez-Gonz{\'a}lez}, E. and {Masi}, S. and {Matarrese}, S. and {McGehee}, P. and {Meinhold}, P.~R. and {Melchiorri}, A. and {Melin}, J. -B. and {Mendes}, L. and {Mennella}, A. and {Migliaccio}, M. and {Mitra}, S. and {Miville-Desch{\^e}nes}, M. -A. and {Moneti}, A. and {Montier}, L. and {Morgante}, G. and {Mortlock}, D. and {Moss}, A. and {Munshi}, D. and {Murphy}, J.~A. and {Naselsky}, P. and {Nati}, F. and {Natoli}, P. and {Netterfield}, C.~B. and {N{\o}rgaard-Nielsen}, H.~U. and {Noviello}, F. and {Novikov}, D. and {Novikov}, I. and {Oxborrow}, C.~A. and {Paci}, F. and {Pagano}, L. and {Pajot}, F. and {Paoletti}, D. and {Partridge}, B. and {Pasian}, F. and {Patanchon}, G. and {Pearson}, T.~J. and {Perdereau}, O. and {Perotto}, L. and {Perrotta}, F. and {Pettorino}, V. and {Piacentini}, F. and {Piat}, M. and {Pierpaoli}, E. and {Pietrobon}, D. and {Plaszczynski}, S. and {Pointecouteau}, E. and {Polenta}, G. and {Popa}, L. and {Pratt}, G.~W. and {Pr{\'e}zeau}, G. and {Prunet}, S. and {Puget}, J. -L. and {Rachen}, J.~P. and {Rebolo}, R. and {Reinecke}, M. and {Remazeilles}, M. and {Renault}, C. and {Renzi}, A. and {Ristorcelli}, I. and {Rocha}, G. and {Roman}, M. and {Rosset}, C. and {Rossetti}, M. and {Roudier}, G. and {Rubi{\~n}o-Mart{\'\i}n}, J.~A. and {Rusholme}, B. and {Sandri}, M.},
        title = "{Planck 2015 results. XXIV. Cosmology from Sunyaev-Zeldovich cluster counts}",
      journal = {\aap},
         year = 2016,
        month = sep,
       volume = {594},
          eid = {A24},
        pages = {A24},
          doi = {10.1051/0004-6361/201525833},
archivePrefix = {arXiv},
       eprint = {1502.01597},
 primaryClass = {astro-ph.CO},
       adsurl = {https://ui.adsabs.harvard.edu/abs/2016A&A...594A..24P}
}

@article{Porredon.etal.2022,
  title = {Dark Energy Survey Year 3 results: Cosmological constraints from galaxy clustering and galaxy-galaxy lensing using the MagLim lens sample},
  author = {Porredon, A. and Crocce, M. and Elvin-Poole, J. and Cawthon, R. and Giannini, G. and De Vicente, J. and Carnero Rosell, A. and Ferrero, I. and Krause, E. and Fang, X. and Prat, J. and Rodriguez-Monroy, M. and Pandey, S. and Pocino, A. and Castander, F. J. and Choi, A. and Amon, A. and Tutusaus, I. and Dodelson, S. and Sevilla-Noarbe, I. and Fosalba, P. and Gaztanaga, E. and Alarcon, A. and Alves, O. and Andrade-Oliveira, F. and Baxter, E. and Bechtol, K. and Becker, M. R. and Bernstein, G. M. and Blazek, J. and Camacho, H. and Campos, A. and Carrasco Kind, M. and Chintalapati, P. and Cordero, J. and DeRose, J. and Di Valentino, E. and Doux, C. and Eifler, T. F. and Everett, S. and Fert\'e, A. and Friedrich, O. and Gatti, M. and Gruen, D. and Harrison, I. and Hartley, W. G. and Herner, K. and Huff, E. M. and Huterer, D. and Jain, B. and Jarvis, M. and Lee, S. and Lemos, P. and MacCrann, N. and Mena-Fern\'andez, J. and Muir, J. and Myles, J. and Park, Y. and Raveri, M. and Rosenfeld, R. and Ross, A. J. and Rykoff, E. S. and Samuroff, S. and S\'anchez, C. and Sanchez, E. and Sanchez, J. and Sanchez Cid, D. and Scolnic, D. and Secco, L. F. and Sheldon, E. and Troja, A. and Troxel, M. A. and Weaverdyck, N. and Yanny, B. and Zuntz, J. and Abbott, T. M. C. and Aguena, M. and Allam, S. and Annis, J. and Avila, S. and Bacon, D. and Bertin, E. and Bhargava, S. and Brooks, D. and Buckley-Geer, E. and Burke, D. L. and Carretero, J. and Costanzi, M. and da Costa, L. N. and Pereira, M. E. S. and Davis, T. M. and Desai, S. and Diehl, H. T. and Dietrich, J. P. and Doel, P. and Drlica-Wagner, A. and Eckert, K. and Evrard, A. E. and Flaugher, B. and Frieman, J. and Garc\'{\i}a-Bellido, J. and Gerdes, D. W. and Giannantonio, T. and Gruendl, R. A. and Gschwend, J. and Gutierrez, G. and Hinton, S. R. and Hollowood, D. L. and Honscheid, K. and Hoyle, B. and James, D. J. and Kuehn, K. and Kuropatkin, N. and Lahav, O. and Lidman, C. and Lima, M. and Lin, H. and Maia, M. A. G. and Marshall, J. L. and Martini, P. and Melchior, P. and Menanteau, F. and Miquel, R. and Mohr, J. J. and Morgan, R. and Ogando, R. L. C. and Palmese, A. and Paz-Chinch\'on, F. and Petravick, D. and Pieres, A. and Plazas Malag\'on, A. A. and Romer, A. K. and Santiago, B. and Scarpine, V. and Schubnell, M. and Serrano, S. and Smith, M. and Soares-Santos, M. and Suchyta, E. and Tarle, G. and Thomas, D. and To, C. and Varga, T. N. and Weller, J.},
  collaboration = {DES Collaboration},
  journal = {Phys. Rev. D},
  volume = {106},
  issue = {10},
  pages = {103530},
  numpages = {28},
  year = {2022},
  month = {Nov},
  publisher = {American Physical Society},
  doi = {10.1103/PhysRevD.106.103530},
  url = {https://link.aps.org/doi/10.1103/PhysRevD.106.103530}
}


\appendix
\numberwithin{figure}{section}
\numberwithin{table}{section}
\numberwithin{equation}{section}

\section{The satellite kinematics likelihood}
\label{App:AppA}

The data vector for the satellite kinematics (equation~[\ref{dataprim}]) contains the projected phase-space coordinates $\dV$ and $\Rp$ of all secondaries associated with the $N_+$ primaries. We make the reasonable assumption that the data for different primaries are independent. Additionally, for a primary with more than one secondary, we assume that the phase-space coordinates of the secondaries are not correlated with each other. This implies that 
\begin{equation}\label{LikelihoodSK}
\calL_{\rm SK} \equiv \calL(\bD_{\rm SK}|{\btheta}) = \prod\limits_{i=1}^{N_{+}} \, \prod\limits_{j=1}^{\Nsi} P(\dVij, \Rij | \Lci, \zci, \Nsi, \btheta)
\end{equation}
Here, $P(\dV, \Rp | \Lpri, \zpri, \Ns)$ is the probability that a secondary galaxy around a primary at redshift $\zpri$, with a luminosity $\Lpri$, and with a total of $\Ns$ detected secondaries, has projected phase-space coordinates $(\dV, \Rp)$. For true satellites, the probability is computed by assuming that each satellite galaxy is a virialized steady-state tracer of the gravitational potential well in which it orbits. Throughout, we assume dark matter halos to be spherical and to have NFW \citep{Navarro.etal.97} density profiles characterized by the concentration-mass relation of \cite{Diemer.Kravtsov.15} with zero scatter. Hence, host halos are completely specified by their virial mass, $\Mh$, alone\footnote{Throughout this paper, we define virial quantities according to the virial overdensities given by the fitting formula of \cite{Bryan.Norman.98}.}, which implies that we can factor the likelihood as
\begin{equation}\label{Mmarg}
\calL_{\rm SK} = \prod\limits_{i=1}^{N_{+}} \, \int \rmd \Mh \, P(\Mh | \Lci, \zci, \Nsi) \, \prod\limits_{j=1}^{\Nsi} P(\dVij, \Rij|\Mh, \Lci, \zci)
\end{equation}
This equation describes a marginalization over halo mass, which serves as a latent variable for each individual primary, accentuating the hierarchical nature of our inference procedure. Note that the `prior for the halo mass is informed by $\Lpri$, $\zpri$, and $\Ns$ according to the model $\btheta$.

Using the Bayes theorem, we have that
\begin{equation}\label{ProbMass}
P(M|L,z,\Ns) = \frac{P(\Ns|M,L,z) \, P(L|M,z)  \, n(M,z)}{\int \rmd M \, P(\Ns|M,L,z) \, P(L|M,z) \, n(M,z)}
\end{equation}
with $n(M,z)$ the halo mass function at redshift $z$ which we compute using the model of \citet{Tinker.etal.08}.

The number of secondaries, $N_\rms$, associated with a particular primary consists of both satellites (galaxies that belong to the same dark matter host halo as the primary) and interlopers (those that do not). The probability $P(\Ns|M,\Lpri,\zpri)$ is computed under the assumption that both the number of interlopers and the number of satellite galaxies obey Poisson statistics.
This implies that
\begin{equation}\label{PNsec}
P(N_\rms|M,L,z) = \frac{\lambda^{N_{\rms}}_{\rm tot} \, \rme^{-\lambda_{\rm tot}}}{N_\rms!}\,,
\end{equation}
where $\lambda_{\rm tot} = f_{\rm corr} \left[\lambda_{\rm sat} + \lambda_{\rm int}\right]$ is the expectation value for the number of secondaries. Here, $\lambda_{\rm int}(\Lpri,\zpri)$ and $\lambda_{\rm sat}(M,\Lpri,\zpri)$ are the expectation values for the number of interlopers and satellites, respectively, and $f_{\rm corr}$ is a correction factor that accounts for fiber collisions and aperture incompleteness. Detailed expressions for $\lambda_{\rm sat}$, $\lambda_{\rm int}$, and $f_{\rm corr}$ can be found in \papII. 
Briefly, $\lambda_{\rm sat}(M,\Lpri,\zpri)$ is computed by integrating the satellite CLF $\Phi_\rms(L|M)$, $\lambda_{\rm int}(M,\Lpri,\zpri)$ is the expected number of foreground and background galaxies in the conical selection volume, modified by an effective bias term which contains three free nuisance parameters that are constrained simultaneously with all other physical parameters, and $f_{\rm corr}$ is computed using the fiber collision correction scheme of \citet{Lange.etal.19a}.

The function $P(\Lpri|M,\zpri)$ describes the probability that a halo of mass $M$ at redshift $\zpri$ hosts a primary of luminosity $\Lpri$.  If all primaries are true centrals, then $P(\Lpri|M,\zpri) = \Phi_\rmc(\Lpri|M)$. However, in practice, it is unavoidable that some primaries are misidentified satellites, and such impurities need to be accounted for. As detailed in \papII, this is done using a detailed forward model based on the CLF.

The probability $P(\dV, \Rp |\Mh, \Lc, \zc)$ is computed using a detailed model for the line-of-sight kinematics of the secondaries. Since secondaries consist of both true satellites and interlopers, which have distinct phase-space distributions, we write
\begin{equation}\label{PdVall}
P(\dV, \Rp|M,L,z) = f_{\rm int}\, P_{\rm int}(\dV, \Rp|L,z)\, +\, [1 - f_{\rm int}] \, P_{\rm sat}(\dV, \Rp|M,L,z) 
\end{equation}
with the interloper fraction defined as
\begin{equation}\label{fint}
f_{\rm int} = f_{\rm int}(M,L,z) = \frac{\lambda_{\rm int}(L,z)}{\lambda_{\rm tot}(M,L,z)}
\end{equation}
Detailed descriptions of $P_{\rm sat}(\dV, \Rp|M,L,z)$ and $P_{\rm int}(\dV, \Rp|L,z)$ are given in Section~4.2.3 in \papII. Briefly, the line-of-sight distribution of true satellites in a halo of mass $M$ at redshift $z$ at a projected separation $\Rp$ is calculated by modeling $P(\dV|\Rp,M,z)$ as a generalized Gaussian with a projected velocity dispersion, $\sigma_{\rm los}(\Rp|M,L,z)$, and a line-of-sight kurtosis, $\kappa_{\rm los}(\Rp|M,L,z)$. These quantities are computed using the second and fourth order spherical Jeans equations \citep[e.g.,][]{Binney.Mamon.82, Lokas.02, Lokas.Mamon.03}. Whenever we apply baryonic corrections to account for how the process of galaxy formation modifies the potential wells in which the satellites orbit, we multiply $\sigma_{\rm los}(\Rp|M,L,z)$ with the halo mass-dependent correction factor $\overline{\sigma}_{\rm hydro}/\overline{\sigma}_{\rm DMO}$ given by equation~(\ref{barcorr}). Unless stated otherwise, we use the fiducial $\overline{\sigma}_{\rm hydro}/\overline{\sigma}_{\rm DMO}$ shown in the bottom-right panel of Fig.~\ref{fig:baryon_model}, which is based on the \EAGLE simulation (see B25). 

The model for the line-of-sight velocity distribution of interlopers assumes that they consist of three distinct populations: (i) a population of `splash-back' galaxies associated with the host halo of the primary and extending out to a splash-back radius $r_{\rm sp}\sim 2 r_{\rm vir}(M,z)$ from the primary, (ii) a roughly uniform background population of kinematically decoupled interlopers, and (iii) a kinematically coupled population of interlopers that are infalling towards the halo but located outside of the splash-back radius. The line-of-sight kinematics of the splash-back galaxies is modeled in the same way as the true satellites, while that of the large-scale infalling population is modelled using a data-driven approach based on a set of tertiary galaxies in annular conical volumes at larger projected distances from the primary (see \paperII for details). 

\section{The satellite abundance likelihood}
\label{App:AppB}

In computing the likelihood of satellite kinematics, $\calL_{\rm SK}$, the number of secondaries associated with each primary is used as a conditional. However, that information can also be exploited to put additional constraints on the galaxy-halo connection, especially that of the satellites. Hence, \Basilisk maximizes the {\it combined} likelihood of both
$\calL_{\rm SK}$ and
\begin{equation}
    \calL_{\rm Ns} = \calL(\bD_{\rm Ns}|\btheta) = \prod_{i=1}^{N_{\rm Ns}} P(N_{{\rm sec},i}|L_{{\rm pri},i}, \, z_{{\rm pri},i}, \, \btheta)
\end{equation}
The latter expresses the probabilities for each of the $N_{\rm Ns}$ primaries to have its corresponding number of secondaries, given its luminosity and redshift. As mentioned in the main text, to limit the computational cost, we only use a randomly sampled subset of $N_{\rm Ns} \sim \mathcal{O}(N_+)$ of all $N_{\rm pri}$ primaries. 

Using the halo mass as a latent variable, the probability for a primary of luminosity $L$ at redshift $z$ to have $N_\rms$ secondaries is
\begin{equation} 
P(N_\rms |L,z) = \int \rmd M \, P(M|L,z) \, P(N_\rms|M,L,z)
\end{equation}
with $P(N_\rms|M,L,z)$ given by equation~(\ref{PNsec}) and, using Bayes' theorem,
\begin{equation} 
P(M|L,z) = \frac{P(L|M,z)  \, n(M,z)}{\int \rmd M \, P(L|M,z) \, n(M,z)}
\end{equation}
which contains the same elements as equation~(\ref{ProbMass}).

\section{Sensitivity to CLF parametrization}
\label{App:C}

In \papIII, we explored different extensions of the CLF parametrization, and used Bayesian evidence modeling to test which of these extensions is preferred by the SDSS data. Specifically, we allowed the scatter ($\sigma_\rmc$) in central galaxy luminosity, the faint-end slope ($\alpha_\rmc$) of satellite CLF and the high luminosity cut-off scale of the satellite CLF with respect of central median luminosity ($\Delta_\rms = \log (L_\rms^*/\bar{L}_\rmc$), to have additional degrees of freedom. We found that, in terms of the Bayes factor, the model where $\sigma_\rmc$, $\alpha_\rmc$, and $\Delta_\rms$, all vary linearly with $\log M$ is the optimum model. That is the CLF parametrization that we have used as our fiducial choice throughout the entirety of this analysis.
\begin{figure}
\centering
\includegraphics[width=0.5\textwidth]{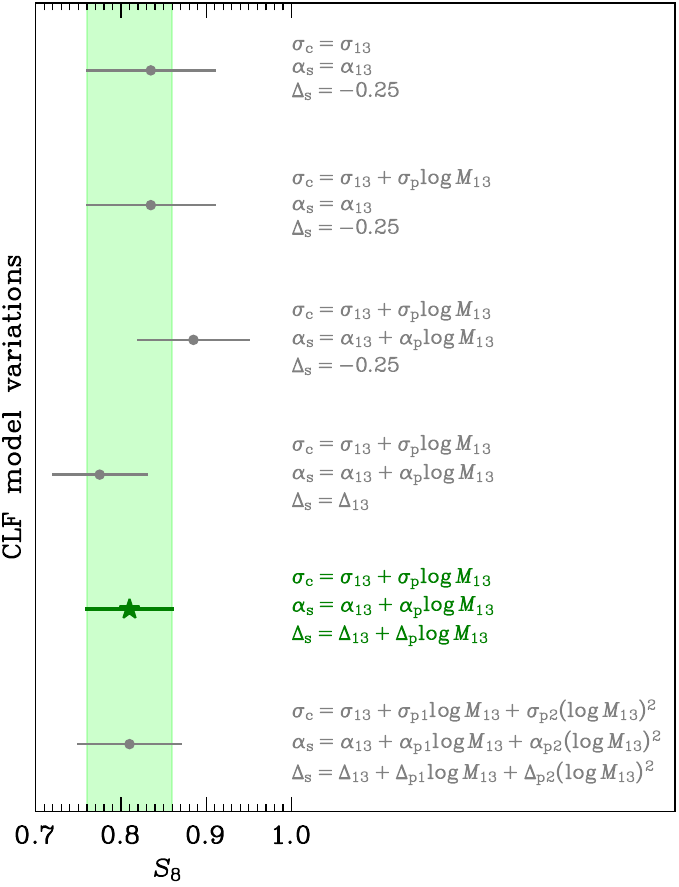}
\caption{Constraints on the $S_8$ parameter for different variations of the assumed galaxy-halo connection model. The differences in the CLF models are indicated in the text corresponding to each constraint. The green star with errorbars corresponds to our fiducial model adopted throughout the main text, which is identical to the one shown in Fig.~\ref{fig:Cosmo_Constraint}. Note that the $S_8$ inference is robust to changes in  the (flexibility) of the CLF model.}
\label{fig:clf_cosmology}
\end{figure}

However, it is worth exploring how the cosmology results depend on the assumed galaxy-halo connection model. Thus, we repeat our analysis for the fiducial B25 baryonic model, for each of the CLF parametrization versions tested in \papIII. For each parametrization, we run \Basilisk on a $3\times 3$ cosmology grid. For each cosmology, we first find the best fit satellite radial profile and then use that to constrain the CLF parameters by running MCMC to convergence. We combine the CLF posteriors from the 9 different cosmologies to use that as the galaxy-halo connection prior in the cosmological evidence modeling. Going through this entire machinery for each variation in CLF parametrization, the resulting $S_8$ parameter inferences, along with the $1 \, \sigma$ confidence interval, are shown in Fig.~\ref{fig:clf_cosmology}. To the right of each $S_8$ constraint we indicate the specific details of the adopted CLF parametrization. The green-shaded band indicates the inference for our fiducial CLF model, which is identical to the green star and error-bar in the top panel of Fig.~\ref{fig:Cosmo_Constraint}. We find that our cosmological inference is remarkably robust to changes in the CLF parametrization. For comparison, in \papIII we show that, for a fixed cosmology, adding some of these additional degrees of freedom to the CLF model results in a drastic improvement of the fit to the SDSS data with significant changes in the inferred galaxy-halo connection. The fact that the cosmological constraints appear to be insusceptible to these different CLF parametrizations suggests that the degeneracy between the cosmological inference and that of the galaxy-halo connection is fairly limited, at least within the method of analysis adopted here.


\label{lastpage}

\end{document}